\documentclass[a4paper,11pt]{article}
\pdfoutput=1 
\usepackage[symbol]{footmisc} 
\usepackage{jheppub} 

\usepackage{color, colortbl}
\usepackage[T1]{fontenc}
\usepackage[utf8]{inputenc}
\usepackage{epsfig,latexsym}
\usepackage{amsmath}
\usepackage[dvipsnames]{xcolor}
\usepackage{adjustbox}
\usepackage{tensor}
\usepackage{graphicx}
\usepackage{verbatim}
\usepackage{mathrsfs}
\usepackage{amssymb}
\usepackage{multirow}
\usepackage{epsfig}
\usepackage{color,colordvi}
\usepackage{appendix}
\usepackage{slashed}
\usepackage{array}
\usepackage{cancel}
\usepackage{float}
\usepackage[font=footnotesize, justification=justified]{caption}
\usepackage{epsf}
\usepackage{tikz}
\usetikzlibrary{external,shapes.geometric,arrows,arrows.meta}
\usepackage{physics}
\usepackage{MnSymbol}
\usepackage{rotating}
\usepackage[normalem]{ulem}
\usepackage{tcolorbox}

\usepackage{bbold}

\allowdisplaybreaks

\DeclareMathAlphabet{\mathpzc}{OT1}{pzc}{m}{it}

\renewcommand{\theequation}{\thesection.\arabic{equation}} \csname
@addtoreset\endcsname{equation}{section}

\newcolumntype{x}[1]{>{\centering\arraybackslash\hspace{0pt}}p{#1}}

\newcommand{\beq}{\begin{equation}}
\newcommand{\eeq}{\end{equation}}
\renewcommand{\[}{\left[}
\renewcommand{\]}{\right]}
\renewcommand{\(}{\left(}
\renewcommand{\)}{\right)}

\newcommand{\be}{\begin{eqnarray}}
\newcommand{\ee}{\end{eqnarray}}
\newcommand{\bea}{\begin{eqnarray}}
\newcommand{\eea}{\end{eqnarray}}
\newcommand{\bi}{\begin{itemize}}
\newcommand{\ei}{\end{itemize}}
\newcommand{\ben}{\begin{enumerate}}
\newcommand{\een}{\end{enumerate}}
\def\bes{\begin{equation*}}
\def\ees{\end{equation*}}
\def\bead{\begin{aligned}}
\def\eead{\end{aligned}}
\def\bmat{\left(\begin{matrix}}
\def\emat{\end{matrix}\right)}

\def\Re{\text{Re}}
\def\Im{\text{Im}}

\def\cD{{\cal D}}

\def\cH{{\cal H}}

\def\cL{{\cal L}}

\def\cO{{\cal O}}

\def\PQ{\text{PQ}}
\def\notPQ{\cancel{\text{PQ}}}
\def\aSMEFT{\text{aSMEFT}}
\def\aSMEFTPQ{\text{aSMEFT}_{\text{PQ}}}
\def\aSMEFTnotPQ{\text{aSMEFT}_{\cancel{\text{PQ}}}}
\def\aLEFT{\text{aLEFT}}
\def\aLEFTPQ{\text{aLEFT}_{\text{PQ}}}
\def\aLEFTnotPQ{\text{aLEFT}_{\cancel{\text{PQ}}}}
\def\s0{\slashed{0}}

\newcommand{\wtilde}{\widetilde}
\newcommand{\hc}{\text{h.c.}}

\allowdisplaybreaks

\newcommand\numberthis{\addtocounter{equation}{1}\tag{\theequation}}

\title{
Hilbert series for ALP EFTs
}

\author[a,b]{Christophe Grojean,}
\emailAdd{christophe.grojean@desy.de}

\author[a,b]{Jonathan Kley,}
\emailAdd{jonathan.kley@desy.de}

\author[a,c]{Chang-Yuan Yao}
\emailAdd{chang.yuan.yao@desy.de}

\affiliation[a]{Deutsches Elektronen-Synchrotron DESY, Notkestr. 85, 22607 Hamburg, Germany}

\affiliation[b]{Institut für Physik, Humboldt-Universität zu Berlin, 12489 Berlin, Germany}

\affiliation[c]{School of Physics, Nankai University, Tianjin 300071, China}

\abstract{Axions and axion-like particles (ALPs) are ubiquitous in popular attempts to solve supercalifragilisticexpialidocious puzzles of Nature. A widespread and vivid experimental programme spanning a vast range of mass scales and decades of couplings strives to find evidence for these elusive but theoretically well-motivated particles. In the absence of clear guiding principle, effective field theories (EFTs) prove to be an efficient tool in this  experimental quest. Hilbert series technologies are a privileged instrument of the EFT toolbox to enumerate and classify operators. In this work, we compute explicitly the Hilbert series capturing the interactions of a generic ALP to the Standard Model particles above and below the electroweak symmetry scale, which allow us to build bases of operators up to dimension 8. In particular, we revealed a remarkable structure of the Hilbert series that isolates the shift-symmetry breaking and preserving interactions. In addition, with the Hilbert series method, we enumerate the sources of CP violation in terms of CP-even, CP-odd and CP-violating operators. Furthermore, we provide an ancillary file of the Hilbert series up to dimension 15 to supplement our findings, which can be used for further analysis and exploration.
}

\begin{document} 
\begin{flushright}
DESY-23-098\\
HU-EP-23/39
\end{flushright}
\maketitle
\flushbottom

\section{Introduction}\label{sec:Intro}

Axions and axionlike particles (ALPs) are well-motivated new physics candidates appearing in many models beyond the Standard Model (SM). The axion was first introduced as a -- by now the most well-known -- solution to the strong CP problem of QCD~\cite{Peccei:1977hh,Peccei:1977ur,Weinberg:1977ma,Wilczek:1977pj,Kim:1979if,Shifman:1979if,Dine:1981rt,Zhitnitsky:1980tq} in the Peccei--Quinn (PQ) mechanism. In addition, it can also act as a candidate for dark matter~\cite{Preskill:1982cy,Abbott:1982af,Dine:1982ah,Marsh:2015xka,DiLuzio:2020wdo} solving two shortcomings of the SM at once. One of the most important properties of the axion is that it benefits from a shift symmetry which is due to its Goldstone nature under the spontaneously broken $U(1)_{\text{PQ}}$ symmetry in the PQ mechanism. Thanks to this shift symmetry, the axion acquires special properties. It is, for instance, protected from receiving a potential which ensures in the case of the QCD axion that the strong CP problem will still be solved by going to the minimum of the potential as long as any shift-breaking effects beyond QCD instantons are small enough. As the shift symmetry is a vital property of ALPs, it is important to study how explicit breaking effects can introduce deviations to the ALP construction from the exact shift symmetric case. Above the scale of spontaneous PQ-breaking, the shift-breaking terms can be understood as terms which break the global $U(1)_{\text{PQ}}$ symmetry. Henceforth, we will therefore also refer to these effects as PQ-breaking effects. 

Explicit breaking effects to the PQ symmetry appear in many well-motivated new physics scenarios. First of all, quantum gravity does not allow for exact global symmetries~\cite{Hawking:1987mz,Giddings:1988cx,Kamionkowski:1992mf,Banks:2010zn} leading to Planck scale suppressed operators\footnote{These effects can become much larger than naively expected when heavy particles are present in the UV above the scale of spontaneous PQ-breaking. Integrating them out can significantly lower the scale of gravity-induced breaking effects \cite{Bonnefoy:2022vop} worsening the so-called axion quality problem.} that break the shift symmetry of the ALP. Besides that, it can also be interesting from a model building point of view to allow for some breaking of the shift symmetry. Examples are the relaxion~\cite{Graham:2015cka,Espinosa:2015eda}, where the breaking term is needed to scan the Higgs mass parameter, or model building for collider anomalies~\cite{Franceschini:2016gxv}. Therefore, it is important to understand how these effects can be captured in a generic way. Due to the vast landscape of theories in which axions can appear, it is convenient to work with an effective field theory (EFT) of the axion coupled to the SM degrees of freedom. In this language, one can capture the interactions between axions and the SM particles in a bottom-up approach while being mostly agnostic about the details of the UV physics~\cite{Srednicki:1985xd,Georgi:1986df}.

A first step towards understanding PQ-breaking effects in an EFT language has been laid out in Ref.~\cite{Bonnefoy:2022rik}, where the leading shift-breaking effects have been captured in Jarlskog-like flavor invariants. Here, we want to carry on this analysis further and explicitly check how these PQ-breaking effects are encoded in operators at higher mass dimensions. One of the main points of Ref.~\cite{Bonnefoy:2022rik} is that, in the presence of shift-breaking effects, the same operators coupling the ALP to fermions describe both shift-breaking and shift-preserving interactions. Therefore, it is difficult to give the interactions their appropriate power counting which can be very different as the scales of spontaneous and explicit PQ-breaking are usually well-separated. In this paper, we want to study if a similar mixing between the shift-breaking and shift-preserving sector also appears in the fermionic sector at higher orders in the EFT expansion. For that we build an operator basis for a pseudoscalar with and without a shift symmetry coupled to the SM degrees of freedom. 

The operator basis we derive could prove useful for phenomenological as well as theoretical studies. Most analyses study the leading dimension-5 interactions of the ALPs, while some also consider effects from higher dimensional operators at dimension~6~\cite{Draper:2012xt,Bauer:2017ris,Bauer:2018uxu,Davoudiasl:2021haa,Brivio:2021fog} and dimension~7~\cite{Bauer:2016zfj,Bauer:2017ris,Bauer:2018uxu}. In particular, the analyses at dimension~7 use an incomplete basis, which may lead to the omission of contributions from other operators that could alter the results of the study. On the more theoretical side, the operators at dimension~8 are of interest for discussion of positivity in the context of the ALP EFT. Furthermore, it could be of interest for matching calculations~\cite{Quevillon:2021sfz} to have a complete set of operators beyond the leading interactions. An important probe for new physics are low-energy experiments looking for small corrections in high-precision experiments and exotic decays of mesons involving ALPs (see e.g. Refs.~\cite{Bjorkeroth:2018dzu,Bauer:2020jbp,MartinCamalich:2020dfe,Calibbi:2020jvd,Bauer:2021mvw}). In order to have a complete effective description of such effects below the electroweak (EW) scale, we will also derive an operator basis for the so called low-energy effective field theory (LEFT) extended with an ALP.

To simplify the procedure of building an operator basis, we will borrow the so-called Hilbert series from representation theory that counts all combinations of objects transforming as singlets under a given group. Here, we will use it to count the singlets under the Lorentz and gauge group as well as to address equation of motion (EOM) and integration by parts (IBP) redundancies that are a big nuisance when building an operator basis. In the context of EFTs, these tools have been developed in Refs.~\cite{Lehman:2015via,Lehman:2015coa,Henning:2015daa,Henning:2015alf,Henning:2017fpj} to build operator bases for EFTs addressing EOM and IBP redundancies with the help of ideas from conformal representation theory. Since then, these tools have proven helpful in many analyses of different EFTs~\cite{Lehman:2015via,Lehman:2015coa,Henning:2015daa,Henning:2015alf,Kobach:2017xkw,Henning:2017fpj,Ruhdorfer:2019qmk,Marinissen:2020jmb,Banerjee:2020bym,Graf:2020yxt,Kondo:2022wcw,Graf:2022rco,Sun:2022aag,Bijnens:2022zqo} to build operator bases and to study different aspects of these EFTs like their behavior under CP. Therefore, in this paper, we will first calculate the full Hilbert series for the constructions of operator bases. To further investigate the CP violation effects in the EFTs, we will implement CP in the Hilbert series language, and the Hilbert series counting CP-even, CP-odd and CP-violating couplings will be calculated.

The main focus of this paper is to demonstrate how the Hilbert series can provide a clear and concise understanding of the separation of the shift-symmetric and shift-breaking sectors in the ALP EFTs. We highlight that these two sectors can be distinctly categorized above mass dimension~5, without any observed mixing between them. We will furthermore show that, making a change of basis that is often considered in the literature and convenient to work in in the presence of shift-breaking effects, one has to consider more seemingly shift-breaking operators with completely constrained Wilson coefficients.

The paper is organized as follows. In Section~\ref{sec:HS}, we briefly introduce the Hilbert series in the context of constructing an EFT operator basis, and explain how to implement a shift symmetry in the calculation. In addition, we also discuss how to include CP in the Hilbert series framework. In Section~\ref{sec:aSMEFT}, we extend the SMEFT with both shift-symmetric and non-shift-symmetric axions, the Hilbert series and the operator counting are present, we identify a remarkable Peccei--Quinn breaking isolation property of the Hilbert series, and the complete and non-redundant operator bases up to dimension~8 are constructed. In addition, we discuss the shift-symmetric limit at mass dimension~5 and beyond, and discuss conditions beyond dimension~5 that have to be considered if one changes the operator basis from the derivatively coupled to the Yukawa-like ALP interactions at dimension-5. Furthermore, we consider the effect of CP transformations on the EFT and count the CP violating couplings. In Section~\ref{sec:aLEFT}, we investigate the axion-extended LEFT with and without a shift symmetry, we calculate the Hilbert series and construct the operator basis up to dimension~8. We discuss again CP violation in the EFT. Finally, we conclude the paper in Section~\ref{sec:Conclusions} and outline potential future directions for our research. In order to provide more information, we show the complete operator bases up to dimension~8 for both axion-extended SMEFT and LEFT in App.~\ref{app:aSMEFTOpBasis} and App.~\ref{app:aLEFTOpBasis} respectively. In App.~\ref{app:HS}, we show additional results for the Hilbert series, and operator counting is also performed for higher dimensions. Finally, in App.~\ref{app:ShiftFieldRedef}, we give details about the change of operator basis from the derivatively coupled to the Yukawa basis, and list all the relevant operators along their constrained Wilson coefficient that have to be considered in the Yukawa basis up to dimension~8.

\section{Hilbert series for EFT operators}\label{sec:HS}
The Hilbert series serves the purpose of systematically counting operators based on their order in fields and derivatives, although it does not provide an explicit construction of these operators. Knowing the number of independent operators is extremely helpful for the construction of an operator basis.
In this section, we will briefly review the tools for the Hilbert series that will be used throughout this paper and have been developed for EFTs in Refs.~\cite{Lehman:2015via,Lehman:2015coa,Henning:2015daa,Henning:2015alf,Henning:2017fpj}.

\subsection{Hilbert series and operator redundancies}
The Hilbert series is a mathematical tool that allows one to determine the number of independent invariants in a theory by considering the power series representation. In the context of operator basis construction, it is a generating function designed to count the number of gauge and Lorentz invariant operators associated with a specific field content $\{\phi_i\}$ and derivatives $\cD$, referred to as spurions. The general form of the Hilbert series is given by
\begin{equation}
\cH(\cD,\{\phi_i\}) = \sum_{r_1,\dots,r_n}\sum_{k} c_{\mathbf{r}\, k} \ \phi_1^{r_1}\dots \phi_n^{r_n} \cD^k,
\label{eq:hs}
\end{equation}
where $c_{\mathbf{r}\, k}\equiv c_{r_1,\dots,r_n, k}$ counts the number of independent operators with $k$ derivatives $\cD$ and $r_i$ fields $\phi_i$. In order to construct an operator basis at given mass dimension, it is helpful to obtain the Hilbert series and interpret the number $c_{\mathbf{r},k}$ as a guiding factor.
The calculation of the Hilbert series of the operator basis can be accomplished by utilizing the group characters. When dealing with compact Lie groups, the group characters are orthonormal when integrate over the group's Haar measure, i.e., 
\begin{equation}
 \int d\mu_G (g)\, \chi_{\mathbf{R}}(g)\, \chi_{\mathbf{R'}}^*(g) = \delta_{\mathbf{R}, \mathbf{R'}}\,,  
\end{equation}
where $\chi_{\mathbf{R}}(g)$ is the character of representation $\mathbf{R}$ of a group $G$ with $g\in G$, and $d\mu_G$ is the Haar measure. 
Therefore, by considering all possible tensor products of the spurions, multiplying their characters, the orthonormality of the group characters allows one to project these products onto the singlets of the group, which results in a complete set of group invariants. The generating function that yields all possible tensor products of spurions is called the plethystic exponential (PE)~\cite{Feng:2007ur,Henning:2017fpj}
\begin{equation}
\text{PE}\left[\phi_{\mathbf{R}}\, \chi_{\mathbf{R}}(z)\right] = \exp\left(\sum_{r=1}^\infty \frac{1}{r} (\pm 1)^{r+1}\phi_{\mathbf{R}}^r\, \chi_{\mathbf{R}}(z^r) \right)\,,
\label{eq:PE}
\end{equation}
where $z=\{z_1,\dots,z_{n}\}$ are the complex parameters of the maximal torus $U(1)^n\subset G$ with $n=\text{rank}\,G$, and $z^r=\{z_1^r,\dots,z_{n}^r\}$, the plus and minus signs correspond to the bosonic and fermionic spurions respectively. For operator basis construction, the PE should include the complete field content of the theory, and the full PE is defined as $\text{PE}[\{\phi_{i}\}]=\prod_i \text{PE}[\phi_{i}]$ without showing characters explicitly. The Hilbert series will be obtained after the group integration, projecting out all of the singlets under the group $G$
\begin{equation}
    \cH(\{\phi_i\})=\int d\mu_G\, \text{PE}[\{\phi_i\}]\,.
\end{equation}
Using the procedure described above, it is possible to calculate the Hilbert series for all potential invariants associated with the symmetry group of a theory. However, it is important to note that these invariants do not form an independent operator basis because the EOM~\cite{Arzt:1993gz} and IBP redundancies are not directly taken into account during the calculation of the Hilbert series. In Ref.~\cite{Henning:2017fpj}, a comprehensive exploration is carried out to examine the structure and implications of the conformal group. It turns out that an operator basis is well organized by the conformal group, which in turn enables one to effectively address redundancies arising from EOM and IBP. To capture the correct degrees of freedom, the so-called single particle module is introduced as a building block which reads as follows for a scalar field $a$
\begin{equation}
\label{eq:single_particle}
    R_{a}=\begin{pmatrix}
        a \\
        \partial_{\mu_1} a \\
        \partial_{\{\mu_1}\partial_{\mu_2\}} a \\
        \vdots
    \end{pmatrix}\,,
\end{equation}
where only the symmetrized combinations of the derivatives are included, as the other combinations just yield a field strength. To avoid EOM redundancies, we have to remove the terms of the form $\(\partial^2 a, \partial_{\mu_1} \partial^2 a,\partial_{\mu_1} \partial_{\mu_2} \partial^2 a, \dots\)$ from the module. This is done by imposing a so-called shortening condition after which only the traceless part of the module remains. We can obtain the character for this single particle module that captures the right degrees of freedom by summing over the characters of its symmetrized and traceless components, i.e.~\cite{Henning:2017fpj}
\begin{equation}
\label{eq:chi_a}
    \chi_a\(\cD,x\) = \sum_{n=0}^{\infty} \cD^{n+d_a} \chi_{\text{Sym}^{n}\(\frac{1}{2},\frac{1}{2}\)} (x) - \sum_{n=2}^{\infty} \cD^{n+d_a} \chi_{\text{Sym}^{n-2}\(\frac{1}{2},\frac{1}{2}\)} (x) = \cD\(1-\cD^2\) P(\cD,x) \, ,
\end{equation}
where $\chi_{\text{Sym}^n\(\frac{1}{2},\frac{1}{2}\)}$ is the character of the $n$th tensor product of the symmetrized $\(\frac{1}{2},\frac{1}{2}\)$ representation of the Lorentz group\footnote{Note that the the Lorentz group is non-compact, and its characters are not orthonormal, which make it difficult to project out the Lorentz invariant by using group integral method in calculation of Hilbert series. However, we can conveniently work in Euclidean space, the Lorentz group $SO(4)$ is isomorphic to $[SU(2)_L\otimes SU(2)_R]/Z_2$, which is a compact group. Then the covariant derivative $\cD$ transforms in the fundemental $\(\frac{1}{2},\frac{1}{2}\)$ representation of the Lorentz group.}, that the covariant derivative, counted by the spurion $\cD$, lives in. $d_a=1$ is the scaling dimension of the $a$, which accounts for the additional factor of $\cD$ in Eq.~\eqref{eq:chi_a}. We have defined the generating function of symmetric products of the vector representation
\begin{equation}
 P(\cD,x) = \sum_{n=0}^{\infty} \cD^n \chi_{\text{Sym}^n\(\frac{1}{2},\frac{1}{2}\)} (x)\, = \frac{1}{(1-\cD x_1)(1-\cD x_1^{-1})(1-\cD x_2)(1-\cD x_2^{-1})},
\end{equation}
which will also appear later in the Hilbert series. $x$ are the maximal torus coordinates of the Lorentz group that will be integrated over in the Hilbert series. The same procedure can be applied to the SM fermions and field strengths to remove EOM redundancies (for details see Ref.~\cite{Henning:2017fpj}).

Furthermore, when dealing with IBP redundancy, it has been discovered that the single particle modules align perfectly with unitary conformal representations of free fields. The local operators can then be constructed by combining single particle modules through tensor products of the unitary conformal representations, which can further be decomposed into irreducible conformal representations. The irreducible representations in the tensor product form a set of independent operators with both the IBP and EOM redundancies eliminated. By performing an additional integral over the conformal group, the independent Lorentz and gauge invariant operators will be obtained, and thus an operator basis can be constructed~\cite{Henning:2017fpj}. The final expression of the Hilbert series can be organized as follows
\begin{equation}
\cH(\cD,\{\phi_i\}) = \int d\mu_{\text{Lorentz}}\int d\mu_{\text{gauge}}\frac{1}{P} \prod_i\text{PE}\left[\frac{\phi_i}{\cD^{d_i}}\chi_i\right]+\Delta\cH(\cD,\{\phi_i\})\,,
\label{eq:hs2}
\end{equation}
where $\{\phi_i\}$ corresponds to all spurions in the theory, and the character $\chi_i$ should be understood as the character of the single particle module $R_{\phi_i}$. The calculation of the conformal character of $R_{\phi_i}$ is weighted with the scaling dimension $d_i$, therefore, each spurion $\phi_i$ in the PE is weighted by $\cD^{-d_i}$. The factor $1/P$ as well as the $d\mu_{\text{Lorentz}}$ are the remnants of the Haar measure for the conformal group after the integral of the dilatations, and $P$ is the generating function for symmetric products of the vector representation that we have defined above, which plays an important role in eliminating IBP redundancies.

One subtlety arises when using the shortened conformal characters to remove the EOM redundancies. Conformal characters are only unitary if their scaling dimensions satisfies a lower bound, its unitarity bound. The characters we use saturate this bound leading to problems with orthogonality of these characters~\cite{Henning:2017fpj,Ruhdorfer:2019qmk}. The additional terms in the Hilbert series arising due to these issues are removed by $\Delta \cH$ in Eq.~\eqref{eq:hs2}, its specific form for a general EFT can be found in Ref.~\cite{Henning:2017fpj}. These terms can be interpreted with tools from differential geometry~\cite{Graf:2020yxt,Bijnens:2022zqo} and usually only appear at mass dimension four and less. We will discuss this in details in the next section where we calculate the Hilbert series.

In order to calculate the Hilbert series defined in Eq.~\eqref{eq:hs2}, the conformal characters corresponding to various single particle modules and the Haar measures for different groups are needed. They have been extensively discussed in previous papers~\cite{Henning:2017fpj,Henning:2015alf}. This paper does not aim to replicate the aforementioned results. Instead, our calculations rely on the formulas presented in Ref.~\cite{Henning:2015alf}, specifically addressing the characters of typical fields and Haar measures.

\subsection{Implementing the ALP shift symmetry}
After introducing how we can obtain the Hilbert series for operators removing IBP and EOM redundancies, we will now discuss how to implement the shift-symmetric character of the axion. The Hilbert series for a shift-symmetric theory was first discussed in Ref.~\cite{Henning:2017fpj}, where a general treatment was introduced within the framework of non-linear realizations. This approach has been further applied to construct operator bases, such as the operator basis for the shift-symmetric scalar coupled to gravity~\cite{Ruhdorfer:2019qmk}, and for the $\cO(N)$ nonlinear sigma model~\cite{Bijnens:2022zqo}. This paper will provide a detailed discussion of the Hilbert series for the SMEFT and LEFT extended with a shift-symmetric axion. The Hilbert series will serve as a guiding tool in constructing the operator basis for axion EFTs. In this section, we will briefly summarize how to implement a shift-symmetric scalar in the Hilbert series using conformal characters.

As the axion arises as the Goldstone boson of the spontaneously broken $U(1)_{\text{PQ}}$, we can adopt the machinery for non-linearly realized symmetries \`a la CCWZ \cite{Coleman:1969sm,Callan:1969sn} developed in Ref.~\cite{Henning:2017fpj} to impose its properties in the Hilbert series. For a spontaneously broken symmetry $G \to H \subset G$, the Goldstone degrees of freedom $\pi^i(x)$ can be parameterized using the following matrix field
\begin{equation}
    \xi(x) = e^{\frac{i \pi^i(x) X^i}{f_{\pi}}}\,,
\end{equation}
where $X^i$ are the broken generators living in the coset space $G/H$ and $f_{\pi}$ is the pion decay constant. To write down the EFT of the pions of the spontaneously broken symmetry, one usually defines the Cartan form
\begin{equation}
    w_{\mu} \equiv \xi^{-1} \partial_{\mu} \xi = u_{\mu}^i X^i + v_{\mu}^a T^a \equiv u_{\mu} + v_{\mu}
\end{equation}
decomposing the degrees of freedom along the broken generators $X^i$ and the unbroken generators $T^a$. Due to the simplicity of the symmetry breaking pattern $U(1)_{\text{PQ}} \to \slashed{0}$, the discussion simplifies drastically for us. There exists only one broken generator and we can simply write
\begin{equation}
    \xi = e^{i \frac{a}{f}}, \qquad w_{\mu} = u_{\mu} = i \frac{\partial_{\mu} a}{f}\,.
\end{equation}
In the following, instead of working with the Cartan form $u_{\mu}$, we will work with the simplified expression $u_{\mu} \sim \partial_{\mu} a$ for the ALP. In order to implement this derivative coupling, we have to remove the scalar itself as a building block from the Hilbert series, amounting to removing the first entry from the single particle module in Eq.~\eqref{eq:single_particle}. This yields
\begin{equation}
\label{eq:singe_particle_a}
    R_{\partial a} = \begin{pmatrix} \partial_{\mu_1} a \\ \partial_{\{ \mu_1} \partial_{ \mu_2 \}} a \\ \partial_{\{ \mu_1} \partial_{ \mu_2} \partial_{ \mu_3 \}} a \\ \vdots \end{pmatrix} \, .
\end{equation}
To remove the first entry in the single particle module in Eq.~\eqref{eq:single_particle} with the help of characters, we have to apply another shortening condition on top of the previous one that eliminates EOM redundancy. By summing over the characters of the remaining elements of the scalar single particle module, we obtain the character of a shift-symmetric singlet scalar~\cite{Henning:2017fpj}
\begin{equation}
\label{eq:chi_da}
\begin{split}
        \chi_{\partial a}\(\cD,x\) & = \sum_{n=1}^{\infty} \cD^{n+d_{a}} \chi_{\text{Sym}^n\(\frac{1}{2},\frac{1}{2}\)} (x) - \sum_{n=2}^{\infty} \cD^{n+d_{a}} \chi_{\text{Sym}^{n-2}\(\frac{1}{2},\frac{1}{2}\)} (x) \\
        &  = \cD^{d_{a}}\left(-1 + \sum_{n=0}^{\infty} \cD^n \chi_{\text{Sym}^n\(\frac{1}{2},\frac{1}{2}\)} (x) - \sum_{n=2}^{\infty} \cD^n \chi_{\text{Sym}^{n-2}\(\frac{1}{2},\frac{1}{2}\)} (x)\right) \\
        & =\cD\left(\( 1 - \cD^2 \) P\(\cD,x\)- 1\right)\, .
\end{split}
\end{equation}
Together with the characters for all other building blocks in the EFT (see Ref.~\cite{Henning:2017fpj}) this completes the discussion of the ingredients for the Hilbert series. We will fix the exact spurion content and some other conventions in the following section.

\subsection{CP in the Hilbert series} \label{sec:CPHilbertSeries}

In order to capture the CP properties of the operators and systematically classify them into CP-even, CP-odd, and CP-violating classes, CP transformations have to be incorporated in the Hilbert series formalism. The first comprehensive discussion can be found in Ref.~\cite{Henning:2017fpj}, and the techniques have been further used to construct Hilbert series of various theories~\cite{Graf:2020yxt,Kondo:2022wcw,Bijnens:2022zqo}. In this subsection, we will present an overview of the necessary ingredients for integrating CP into the Hilbert series framework.

C and P transformations split both the Lorentz group and gauge group into two disconnected groups, i.e.,
\begin{equation}
\begin{split}
&\widetilde{\text{Lorentz}}=\text{Lorentz}\rtimes \Gamma_{\mathcal{P}}=\{\text{Lorentz},\ \text{Lorentz}\rtimes \mathcal{P}\}\equiv \{\wtilde{\text{Lorentz}}_{+},\ \wtilde{\text{Lorentz}}_{-}\}\,,\\
&\widetilde{\text{gauge}}=\text{gauge}\rtimes \Gamma_{\mathcal{C}}=\{\text{gauge},\ \text{gauge}\rtimes \mathcal{C}\}\equiv \{\wtilde{\text{gauge}}_{+},\ \wtilde{\text{gauge}}_{-}\}\,.
\end{split}
\end{equation}
The CP-even Hilbert series can be calculated by averaging the two Hilbert series from the $\widetilde{\text{Lorentz}}_+\times \widetilde{\text{gauge}}_+$ and $\widetilde{\text{Lorentz}}_-\times \widetilde{\text{gauge}}_-$ branches.\footnote{We can denote these two branches as $C^+P^+$ and $C^-P^-$, and there are another two branches, namely $C^+P^-$ and $C^-P^+$. However, since we only care about the combined effects of CP, the two branches given here are enough for our analysis. If one wants to analyze the properties of the Hilbert series under the single parity, another two branches should also be involved. See Ref.~\cite{Graf:2020yxt} for details.} They correspond to the invariants under $\text{SO}(4)\times\text{SU}(3)\times\text{SU}(2)\times U(1)$ and $(\text{SO}(4)\times\text{SU}(3)\times\text{SU}(2)\times U(1))\mathcal{C}\mathcal{P}$ respectively. The calculation of these two branches follows a slightly modified version of Eq.~\eqref{eq:hs2} given by\footnote{We have omitted the $\Delta\cH$ terms in these two Hilbert series, which could appear in the Hilbert series of $\aSMEFTPQ$ and $\aLEFTPQ$ defined in next subsection at dimension~5. We will not show them explicitly here, and they will be added to the Hilbert series in the ancillary file of this paper.}
\begin{align}
\cH_+(\cD,\{\check{\phi_i}\}) =& \int d\mu_{\widetilde{\text{Lorentz}}_+}(x)\int d\mu_{\widetilde{\text{gauge}}_+}(z)\frac{1}{P_+(\cD,x)} \prod_i\text{PE}\left[\frac{\check{\phi_i}}{\cD^{d_i}}\chi_{i}^+(\cD,x,z)\right]\,,\label{eq:hsp}\\
\cH_-(\cD,\{\check{\phi_i}\}) =& \int d\mu_{\widetilde{\text{Lorentz}}_-}(\tilde{x})\int d\mu_{\widetilde{\text{gauge}}_-}(\tilde{z})\frac{1}{P_-(\cD,\tilde{x})} \prod_i\text{PE}'\left[\frac{\check{\phi_i}}{\cD^{d_i}}\chi_i^-(\cD,x,z)\right]\,,
\label{eq:hsm}
\end{align}
where we have explicitly include the parameter $x\equiv(x_1,x_2)$ for the Lorentz group, and $z\equiv(z_{c,1},z_{c,2},z_W,z_Y)$ for the gauge groups $\text{SU}(3)_c\times\text{SU}(2)_W\times\text{U}(1)_Y$.

Since the spurions $\phi$ and $\phi^\dagger$ transform to each other under CP, we introduce the direct sum $\check{\phi}=\phi\oplus \phi^\dagger$ as a building block to simplify the counting of the operators. For a real singlet (pseudo-)scalar $a$, transforming trivially under CP, the spurion $a$ itself works as a building block. The character $\chi^+$ of the new spurion $\check{\phi}$ is simply the sum of the characters of spurion and its conjugate. The plus branch of the gauge and Lorentz group is the part of the group including C- and P-transformations that is unchanged by those transformations. Therefore, the group measures and $P$ function of $\cH_+$ are the same as those of the full Hilbert series $\cH$. For $\cH_-$, the symmetries are from the minus branches, the group measures and characters should be updated with the ``folding'' technique~\cite{Henning:2017fpj, Graf:2020yxt}. Applying the folding technique, one finds, for instance, that the Haar measure and character of the negative branch of the Lorentz group correspond to those of $\widetilde{\text{Lorentz}}_- = \text{Sp}(2)$ and $\widetilde{SU(3)_-} = \text{Sp}(2)$ for the color part of the gauge group. A detailed discussion will not be provided in this paper, but a summary of these updated expressions can be found in Ref.~\cite{Kondo:2022wcw}. It is worth mentioning that when the odd branch is involved, the characters for even power and odd power in the PE are different, which is indicated by the notation $\text{PE}'$. The $\chi_i^-(\cD,x,z)$ used in odd and even powers are given by
\begin{equation}
\label{eq:char_minus}
\text{odd-power:~} \chi_i^{P^-}(\cD,\tilde{x})\chi_i^{C^-}(\tilde{z})\,,\quad \text{even-power:~} \chi_i^{P^+}(\cD,\bar{x})\chi_i^{C^+}(\bar{z})\,,
\end{equation}
where $\chi_i^{P^\pm}$ is the character corresponding to the Lorentz group, and $\chi_i^{C^\pm}$ is the character related to the gauge groups. After the folding is applied, some group parameters become redundant. Therefore, reduced parameters $\tilde{x}\equiv x_1$ and $\tilde{z}\equiv (z_{c,1},z_W)$ in the odd powers of the characters are introduced. For the even powers, we find that $\chi^-(\cD,x,z)=\chi^+(\cD,\bar{x},\bar{z})$ and it is convenient to introduce the notation $\bar{x}\equiv(x_1,1)$ and $\bar{z}\equiv(z_{c,1},1,z_W,1)$. For the SMEFT particle content, all of the odd-power terms in the PE vanish because none of the states is invariant under CP transformations~\cite{Kondo:2022wcw}. This case was discussed in Ref.~\cite{Kondo:2022wcw} and the expressions for all the measures and characters for the SMEFT particle content can be found there. However, for a singlet scalar extension to the SMEFT, the odd-power terms of the singlet scalar are non-vanishing~\cite{Henning:2017fpj, Graf:2020yxt}. For the ALP-dependent part of the Hilbert series the odd-power characters of the the non-shift-symmetric and shift-symmetric axions are given by
\begin{align}
\label{eq:chi_a_minus}
\chi_a^-\(\cD,x\) =& \chi_a^{P^-}\(\cD,\tilde{x}\) = -\cD\(1-\cD^2\) P_-(\cD,\tilde{x}) \, ,\\
\chi_{\partial a}^-\(\cD,x\) =& \chi_{\partial a}^{P^-}\(\cD,\tilde{x}\) = -\cD\(\(1-\cD^2\) P_-(\cD,\tilde{x})-1\) \, ,
\end{align}
where $P_-(\cD,\tilde{x}) = \frac{1}{(1-\cD x_1)(1-\cD x_1^{-1})(1-\cD^2)}$ and an overall minus sign is introduced to capture the pseudo-scalar nature of the axion under CP. The even-power characters are given in Eq.~\eqref{eq:chi_a} and Eq.~\eqref{eq:chi_da} respectively with $P(\cD,x)=P_+(\cD,\bar{x})=P(\cD,\bar{x})$.

As already mentioned above, the CP-even Hilbert series is an average of $\cH_+$ and $\cH_-$, and the CP-odd Hilbert series can be obtained by $\cH-\cH_{\text{even}}$ with $\cH=\cH_{+}$. We summarize them as follows,
\begin{equation}
  \cH=\cH_{\text{even}}+\cH_{\text{odd}}=\cH_+\,,\quad \cH_{\text{even}}=\frac{1}{2}(\cH_++\cH_-)\,,\quad \cH_{\text{odd}}=\frac{1}{2}(\cH_+-\cH_-)\,.
\end{equation}
Once the plus and minus branches of the Hilbert series are calculated according to Eq.~\eqref{eq:hsp} and Eq.~\eqref{eq:hsm}, the CP-even and CP-odd splitting will be straightforward.

From the above discussion, the CP-odd operators can be identified. However there is no one-to-one correspondence between the CP-odd operators and CP-violating sources. Since the CP-violating effects are captured by both the CP property of the operator and the Wilson coefficient in front of it in the EFT Lagrangian. The redefinitions of the fermion fields can possibly remove the CP phases of an operator, leading to a vanishing CP-violating effect. In this paper, we will follow the definition in Ref.~\cite{Kondo:2022wcw}, where the CP-violating operators are considered as those CP-odd operators where the CP phases cannot be removed by rephasing degree of freedom of the SM Lagrangian. As we know, there are four $U(1)$ symmetries corresponding to lepton family number and baryon number, under which the CP phases in the SM Lagrangian are invariant. If a CP-odd operator at higher mass dimension is not invariant under at least one of the four $U(1)$ transformations, the CP phases of this single operator can be removed by the $U(1)$ transformation. Consequently, 
the CP-violating operators should be captured after applying four more $U(1)$ symmetries.

In practice, the CP-violating Hilbert series can be calculated by performing the additional integrals correspond to the $U(1)$ symmetries of the SM Lagrangian. We will discuss the axion-extended SMEFT and LEFT in this paper, which will be defined in the next subsection. For the SMEFT extended theory, the four $U(1)$ symmetries are just baryon and lepton family numbers $U(1)_{L_i} \times U(1)_B$ as mentioned above. For the LEFT extended theory, more $U(1)$ symmetries are generated by the breaking of  $SU(2)_W$. They correspond to the up and down quark family numbers $U(1)_{u_i}$ and $U(1)_{d_i}$, and charged lepton family number $U(1)_{e_i}$. There is no $U(1)$ symmetry in the neutrino sector due to the Majorana nature of the mass term. Therefore, in total, we have to perform $N_u+N_d+N_e$ of $U(1)$ integrals depending on the number of fermion flavors. It is worth mentioning that according to Eq.~\eqref{eq:char_minus}, the calculation of the $\cH_-$ branch does not involve the $U(1)$ characters. Therefore, we only need to perform the additional $U(1)$ integrals in the $\cH_+$ branch. We define the CP-violating Hilbert series in form of
\begin{equation}
\cH_{\text{CPV}}=\(U(1)~\text{inv.~}\cH_{+}\)-\cH_{-}\,.
\end{equation}

\subsection{Conventions}
This paper will discuss axion EFTs at different energy scales, we consider both SMEFT and LEFT extended with axion, and they are refereed to as aSMEFT and aLEFT respectively. Depending on whether there is a shift symmetry for the axion, four types of EFTs are shown explicitly along with corresponding spurions as follows

\begin{itemize}
\item $\aSMEFTPQ$: SMEFT extended with a shift-symmetric axion

\qquad$\{\cD, \partial a, Q, Q^\dagger, L, L^\dagger, H, H^\dagger, u, u^\dagger, d, d^\dagger, e, e^\dagger, B_L, B_R, W_L, W_R, G_L, G_R\}$\,,

\item $\aSMEFTnotPQ$: SMEFT extended with a non-shift-symmetric axion

\qquad$\{\cD, a, Q, Q^\dagger, L, L^\dagger, H, H^\dagger, u, u^\dagger, d, d^\dagger, e, e^\dagger, B_L, B_R, W_L, W_R, G_L, G_R\}$\,,

\item $\aLEFTPQ$: LEFT extended with a shift-symmetric axion

\qquad$\{\cD, \partial a, u_L^{\vphantom\dagger}, u_L^\dagger, u_R^{\vphantom\dagger}, u_R^\dagger, d_L^{\vphantom\dagger}, d_L^\dagger, d_R^{\vphantom\dagger}, d_R^\dagger, e_L^{\vphantom\dagger}, e_L^\dagger, e_R^{\vphantom\dagger}, e_R^\dagger, \nu_L^{\vphantom\dagger}, \nu_L^\dagger, F_L, F_R, G_L, G_R\}$\,,

\item $\aLEFTnotPQ$: LEFT extended with a non-shift-symmetric axion

\qquad$\{\cD, a, u_L^{\vphantom\dagger}, u_L^\dagger, u_R^{\vphantom\dagger}, u_R^\dagger, d_L^{\vphantom\dagger}, d_L^\dagger, d_R^{\vphantom\dagger}, d_R^\dagger, e_L^{\vphantom\dagger}, e_L^\dagger, e_R^{\vphantom\dagger}, e_R^\dagger, \nu_L^{\vphantom\dagger}, \nu_L^\dagger, F_L, F_R, G_L, G_R\}$\,,
\end{itemize}
where we have followed the convention in Ref.~\cite{Henning:2015alf} to only use the $\(\frac{1}{2},0\)$ representation left-handed fermions to form the Hilbert series, and the superscript ``$c$'' for the right-handed conjugate fields $u^c_{(R)}, d^c_{(R)}$ and $e^c_{(R)}$ are omitted when we calculate the Hilbert series. The field strength $X=F,B,W,G$ is redefined to extract the chiral components $X^{\mu\nu}_{L,R}=\frac{1}{2}(X^{\mu\nu}\pm i \tilde{X}^{\mu\nu})$ that transform as $(1,0)$ and $(0,1)$ under the Lorentz group respectively. The representations and charges of these SM fields under Lorentz group $SO(4)$ and gauge groups $SU(3)_c\otimes SU(2)_{W}\otimes U(1)_Y$ are shown explicitly in Ref.~\cite{Henning:2015alf}, we will not show them here. The axion field transforms as a singlet under all groups.  

With the spurions shown for $\aSMEFTnotPQ$, as we mentioned previously, we can use the Haar measures and characters in Ref.~\cite{Henning:2015alf} to calculate the Hilbert series easily based on Eq.~\eqref{eq:hs2}. The only difference compared to the SMEFT lies in the inclusion of the axion field $a$, which can be easily handled by adding additional PE of the spurion $a$ in the calculation of the Hilbert series. For $\aSMEFTPQ$, we only need to change the spurion ``$a$'' to ``$\partial a$'', and adopt the character in Eq.~\eqref{eq:chi_da}, the calculation of Hilbert series will also be straightforward. 

For the axion in LEFT, all left-handed and right-handed fermion fields are completely independent below the electroweak~(EW) scale as a UV completion could be chiral (LEFT captures both SMEFT- and HEFT-like UV completions). 
The $SU(2)_W$ symmetry is broken, and $U(1)_\text{em}$ is generated with the charge given by $Q=Y+T_3$. The calculation of the Hilbert series of $\aLEFTPQ$ and $\aLEFTnotPQ$ follows the same fashion, except that we don't need to integrate over the $SU(2)$ group, which makes the calculation even easier. Similarly, in the calculation, we need to take care of the spurions ``$\partial a$'' and ``$a$'', and use the corresponding shift and non-shift-symmetric character respectively.

Throughout this paper we will use a grading in the mass dimension of the fields to obtain the Hilbert series at each order in the EFT expansion. For this, we will rescale the spurions with their mass dimensions $\phi \to \epsilon\phi$ for scalars, $\psi \to \epsilon^{3/2} \psi$ for fermions, $X \to \epsilon^2 X$ for field strengths, $\partial a\to \epsilon^2 \partial a$ and $\cD \to \epsilon\cD$ for the covariant derivative. We define the graded Hilbert series as $\cH(\epsilon) = \sum_i \epsilon^i \cH_i$. It should be noted that the calculation of the full Hilbert series is impossible. However, in the construction of the operator basis, it suffices to focus only on a specific mass dimension. 

The calculation of the Hilbert series can be simplified by working only with the corresponding mass dimension of the integrand in Eq.~\eqref{eq:hs2}. This means that we can expand $\text{PE}/P$ to the desired mass dimension first and then perform the integration. Nevertheless, expanding the integrand itself becomes challenging at higher mass dimensions. To address this, a FORM code called ECO (Efficient Counting of Operators)~\cite{Marinissen:2020jmb} has been developed specifically for efficient Hilbert series calculation. For this project, we developed our own Mathematica code to generate the Hilbert series that can be used efficiently at higher mass dimensions. It allowed us to compute the Hilbert series up to dimension~15 for all axion EFTs. Our code is designed with a broader scope in mind and we intend to publish it in a forthcoming paper~\cite{Grojean:2023}, making it readily available whenever the explicit form of the Hilbert series is needed. 

\section{aSMEFT}\label{sec:aSMEFT}
We are now ready to calculate the Hilbert series and construct all operators of the SMEFT extended with a light pseudoscalar. We will start with an ALP, i.e., a pseudoscalar with a shift symmetry, stemming from its Goldstone boson nature under the spontaneously broken global $U(1)_{\text{PQ}}$ symmetry.

\subsection{\texorpdfstring{$\aSMEFTPQ$}{aSMEFTPQ}}\label{sec:aSMEFT_SS}
Using the tools we have introduced in the last section, we can calculate the Hilbert series for the given spurions that define the EFT at low energies. For the explicit shift symmetry we will work with a derivatively coupled pseudoscalar $\partial a$ as mentioned in the last section. Evaluating Eq.~\eqref{eq:hs2} for the given spurions, we find the Hilbert series for one generation of fermions up to mass dimension~8 to be\footnote{Note that $\cH_5^{\PQ}$ here only corresponds to the first term in Eq.~\eqref{eq:hs2} and we still have to add $\Delta \cH$, which is non-trivial here, to get the correct full result. We also remove the pure SMEFT operators from our Hilbert series to only capture the axion coupled operators.}
\begin{align*}
\label{eq:HSaSMEFTSS}
\cH_5^{\PQ}  \,=\, & \partial a\, Q Q^{\dagger} + \partial a\, u u^{\dagger} + \partial a\, d d^{\dagger} + \partial a\, L L^{\dagger} + \partial a\, e e^{\dagger}  + \partial a\, H H^{\dagger} \cD  \\
&- \partial a\, B_L \cD - \partial a\, B_R \cD - \partial a\, \cD^3 \, , \\[3pt]
\cH_6^{\PQ}  \,=\, &  (\partial a)^2 H H^{\dagger} \, , \\[3pt]
\cH_7^{\PQ}  \,=\, & \partial a\, Q Q^{\dagger} B_L + \partial a\, Q Q^{\dagger} B_R + \partial a\, Q Q^{\dagger} G_L + \partial a\, Q Q^{\dagger} G_R + \partial a\, Q Q^{\dagger} W_L + \partial a\, Q Q^{\dagger} W_R \\
&+ \partial a\, u u^{\dagger} B_L + \partial a\, u u^{\dagger} B_R + \partial a\, u u^{\dagger} G_L + \partial a\, u u^{\dagger} G_R + \partial a\, d d^{\dagger} B_L + \partial a\, d d^{\dagger} B_R \\
& + \partial a\, d d^{\dagger} G_L + \partial a\, d d^{\dagger} G_R + \partial a\, L L^{\dagger} B_L + \partial a\, L L^{\dagger} B_R + \partial a\, L L^{\dagger} W_L + \partial a\, L L^{\dagger} W_R \\
& + \partial a\, e e^{\dagger} B_L + \partial a\, e e^{\dagger} B_R + 2 \partial a\, Q Q^{\dagger} H H^{\dagger} + \partial a\, u u^{\dagger} H H^{\dagger} + \partial a\, d d^{\dagger} H H^{\dagger} \\
& + 2 \partial a\, L L^{\dagger} H H^{\dagger} + \partial a\, e e^{\dagger} H H^{\dagger} + \partial a\, B_L H H^{\dagger} \cD + \partial a\, B_R H H^{\dagger} \cD + \partial a\, W_L H H^{\dagger} \cD \\
& + \partial a\, W_R H H^{\dagger} \cD + \partial a\, H^2 H^{\dagger2} \cD + 2 \partial a\, Q u H \cD + 2 \partial a\, Q^{\dagger} u^{\dagger} H^{\dagger} \cD + 2 \partial a\, Q d H^{\dagger} \cD \stepcounter{equation}\tag{\theequation}\\
& + 2 \partial a\, Q^{\dagger} d^{\dagger} H \cD + 2 \partial a\, L e H^{\dagger} \cD + 2 \partial a\, L^{\dagger} e^{\dagger} H \cD\, ,  \\[3pt]
\cH_8^{\PQ}  \,=\, & (\partial a)^4 + (\partial a)^2 Q Q^{\dagger} \cD + (\partial a)^2 u u^{\dagger} \cD + (\partial a)^2 d d^{\dagger} \cD + (\partial a)^2 L L^{\dagger} \cD + (\partial a)^2 e e^{\dagger} \cD \\
& + (\partial a)^2 B_L^2 + (\partial a)^2 B_L B_R + (\partial a)^2 B_R^2 + (\partial a)^2 G_L^2 + (\partial a)^2 G_L G_R + (\partial a)^2 G_R^2 \\
& + (\partial a)^2 W_L^2 + (\partial a)^2 W_L W_R + (\partial a)^2 W_R^2 + \partial a\, Q d^{\dagger2} L^{\dagger} + \partial a\, Q^{\dagger} d^2 L + \partial a\, u d^{\dagger} L^{\dagger2} \\
& + \partial a\, u^{\dagger} d L^2 + 2 (\partial a)^2 H H^{\dagger} \cD^2 + 2 \partial a\, L^2 H^2 \cD + 2 \partial a\, L^{\dagger2} H^{\dagger2} \cD + (\partial a)^2 H^2 H^{\dagger2} \\
& + (\partial a)^2 Q u H + (\partial a)^2 Q^{\dagger} u^{\dagger} H^{\dagger} + (\partial a)^2 Q d H^{\dagger} + (\partial a)^2 Q^{\dagger} d^{\dagger} H + (\partial a)^2 L e H^{\dagger} \\
& + (\partial a)^2 L^{\dagger} e^{\dagger} H\, . \\
\end{align*}
We can interpret the Hilbert series as follows. Every term corresponds to an operator with the field content given by the spurions and the multiplicity given by the prefactor. For the first term in $\cH_5^{\PQ}$, for instance, we expect one operator with the field content $\partial_{\mu} a, Q, Q^{\dagger}$. With this information, it is simple to build a gauge and Lorentz invariant operator $\partial_{\mu} a \, \bar{Q} \gamma^{\mu} Q$.

The negative terms appearing at dimension~5 do not correspond to non-redundant operators and they are canceled exactly by other terms in $\Delta \cH$ as mentioned in the last section. They correspond to co-closed but not co-exact forms and can be calculated immediately from the expression of $\Delta \cH$ given in Ref.~\cite{Henning:2017fpj}. Evaluating the expression, we find for the terms that are relevant for the ALP EFT\footnote{Note that the $\cD^4$ term that usually appears in $\cH_0$ and is canceled by a term in $\Delta \cH$ does not appear here because we only keep terms in the Hilbert series which include at least on ALP field $a$.}
\begin{equation} \label{eq:DeltaH}
\Delta \cH = \partial a\, B_L \cD + \partial a\, B_R \cD + \partial a \, \cD^3 \, ,
\end{equation}
which exactly cancel the negative terms in $\cH_0$. The form of the terms is similar to those found in the discussion of the QCD chiral Lagrangian in Ref.~\cite{Graf:2020yxt}, where a more involved case of a non-linearly realized symmetry is analyzed. The only difference is due to the different counting of mass dimension of the spurions capturing the Goldstone degrees of freedom. As mentioned in the last section, in the case of the ALP we have simplified $u_{\mu} = i \frac{\partial_{\mu}a}{f}$ where we put the $1/f$ suppression into the Wilson coefficient. Hence, in our case $\[ \partial a \] = 2$, whereas usually $\[ u_{\mu} \] = 1$. Therefore terms like $u \, B_L \cD \sim \partial a \, B_L \cD$ appear at dimension~5 here. Note that a term of the form of the last term in Eq.~\eqref{eq:DeltaH} only appears for scalars that are singlets under the gauge group~\cite{Henning:2017fpj}, while the other terms can also appear for non-Abelian gauge groups if the scalar transforms in the adjoint representation.
 
We should also note that the Hilbert series in Eq.~\eqref{eq:HSaSMEFTSS} only takes one flavor of fermions into account. Sometimes this is not enough for constructing an operator basis with multiple flavors because some operator structures are only non-vanishing for multiple flavors of fermions. In order to construct an operator basis for a general number of flavors $N_f$, we need to capture the dependence of the Hilbert series on $N_f$. This can be easily realized by making $N_f$ copies of the corresponding fermions' PE in the calculation of the Hilbert series, which is equivalent to simply adding a factor of $N_f$ in front of the fermionic part of the PE~(c.f. Eq.~\eqref{eq:PE}). Indeed, looking at the Hilbert series at dimension~8 for generic $N_f$, we find the following terms
\begin{equation}
\cH_8^{\PQ} \supset  \frac{1}{3} N_f^2 \(N_f^2-1\) \partial a \, d^{\dagger 3} e + \frac{1}{3} N_f^2 \(N_f^2-1\) \partial a \, d^3 e^{\dagger} \, ,
\end{equation}
which evidently vanishes for $N_f = 1$. The reason why these terms only appear for $N_f >1 $ is the antisymmetric color structure of the down quarks in that operator which only gives rise to a non-vanishing operator if there are at least two different flavors of down quarks (c.f. operator $\cO_{\partial aed}$ in Tab.~\ref{tab:aSMEFT_SS_dim8}).

The Hilbert series in Eq.\eqref{eq:HSaSMEFTSS} does not quite give the correct number of non-redundant operators and some further adjustments have to be performed. First of all, the redundant operator $\cO_{\partial aH} = \partial^{\mu} a \, \Bigl(H^{\dagger} i \overleftrightarrow{D}_{\hspace{-2pt}\mu} H \Bigr)$ corresponding to the term $\partial a \, H H^{\dagger} \cD$ in $\cH_5^{\PQ}$ is not removed automatically.\footnote{We can check that this does not happen again at higher mass dimensions using Eq.~\eqref{eq:separation}. Without imposing a shift symmetry any operator that exactly gives an EOM redundant operator upon using IBP will be removed because one derivative is no longer fixed to the ALP by demanding derivatively coupled ALP interactions.} This is because it can be removed by a global hypercharge transformation on the Higgs field which is not captured in our Hilbert series approach. In general, all derivative couplings of the ALP to SM particles are only defined up to redefinitions by exact global symmetries~\cite{Georgi:1986df}. This is also relevant for the conservation of baryon and lepton number. We did not impose the condition $\partial_{\mu} j^{\mu} = 0$ for conserved currents in our approach. This applies to the operators of type $\partial_{\mu} a \, \bar{\psi} \gamma^{\mu} \psi$ where flavor diagonal parts of the Wilson coefficients can be removed by moving the derivative to the fermions by IBP and using the conservation of baryon and lepton family number  $\partial_{\mu}j_{B}^{\mu} = \partial_{\mu}j_{L_i}^{\mu} = 0$ \cite{Bonilla:2021ufe}.\footnote{To be precise, baryon number and lepton family number are both only conserved classically and are anomalous at loop level. Taking the anomalies into account shifts the Wilson coefficients of the operators $\cO_{a\tilde{B}}$ and $\cO_{a\tilde{W}}$ \cite{Bonilla:2021ufe}. } In principal we could also implement a shortening condition through conformal characters that removes redundancies of the form $\partial_{\mu} j^{\mu} = 0$~\cite{Henning:2017fpj}. Then, one would have to implement the conserved currents explicitly into the Hilbert series, so it is easier to just remove these redundancies by hand at the end, as we do here.

Secondly, there are operators of the form $a F \wtilde{F}$ at mass dimension~5 which do not appear in the Hilbert series. This is due to the fact that we use $\partial a$ and $F$ as a building block because they have nice transformation properties under gauge and Lorentz transformations. One can easily show by moving the derivative from the first field strength by IBP to the ALP that this operator is shift-symmetric. Then however, the gauge field $A_{\mu}$ appears by itself and we would have to use the gauge fields themselves as a building block which is practically unfeasible with the Hilbert series, and would also be inconvenient from the perspective of constructing the operators.
Alternatively, as is well-known, after shifting the ALP by a constant $a \to a + c$, chiral transformations can remove the shift in front of this type of operator. It will prove useful that we also build an operator basis for an ALP without a shift symmetry in Section~\ref{sec:aSMEFT_nonSS}, where we use just $a$ as a building block and notice that after taking the shift-symmetric limit the $a F \wtilde{F}$ type operators will remain. In this way we can systematically include this type of operator in our construction.

For most terms in the Hilbert series it is straightforward to build Lorentz and gauge invariant operators from the spurion content and get the correct number of independent operators as indicated by the Hilbert series. There is one exception that is a bit more involved, the operators of type $(\partial a)^2 X^2$ at dimension~8. From the Hilbert series in Eq.~\eqref{eq:HSaSMEFTSS}, we can read of that we should expect three non-redundant operators $X_L^2 (\partial a)^2$, $X_L X_R (\partial a)^2$ and $X_R^2 (\partial a)^2$. However, one can naively build 4 operators
\begin{equation}
    \partial_{\mu} a \partial^{\mu}a \, B_{\nu\rho} B^{\nu\rho}, \quad \partial_{\mu} a \partial^{\mu}a \, B_{\nu\rho} \wtilde{B}^{\nu\rho}, \quad \partial_{\mu} a \partial^{\nu}a \, B^{\mu\rho} B_{\nu\rho}, \quad \partial_{\mu} a \partial^{\nu}a \, B^{\mu\rho} \wtilde{B}_{\nu\rho}
\end{equation}
sharing a complicated relation that renders one of the operators redundant. We can use the Schouten identity (see e.g. Ref.~\cite{Chala:2021cgt})
\begin{equation}
g_{\mu\nu} \epsilon_{\alpha\beta\gamma\delta} + g_{\mu\alpha} \epsilon_{\beta\gamma\delta\nu} + g_{\mu\beta} \epsilon_{\gamma\delta\nu\alpha} + g_{\mu\gamma} \epsilon_{\delta\nu\alpha\beta} + g_{\mu\delta} \epsilon_{\nu\alpha\beta\gamma} = 0
\end{equation}
to relate the two operators with a dual field strength. Contracting the indices in the identity with a generic rank-2 tensor $T^{\mu\nu}$ (which we identify with $\partial^{\mu}a\partial^{\nu}a$) and an anti-symmetric rank-2 tensor $X_{\mu\nu}$ yields
\begin{equation}
T_{\mu\nu} X^{\mu\rho} \wtilde{X}_{\hphantom{\nu}\rho}^{\nu} = \frac{1}{4} T_{\mu}^{\mu} X_{\nu\rho} \wtilde{X}^{\nu\rho}
\end{equation}
explaining the number of operators of that type in the Hilbert series. Our complete basis up to mass dimension~8 can be found in Tabs.~\ref{tab:aSMEFT_SS_dim5},~\ref{tab:aSMEFT_SS_dim6},~\ref{tab:aSMEFT_SS_dim7} and~\ref{tab:aSMEFT_SS_dim8} in App.~\ref{app:aSMEFTSSOpBasis}.

There are several ways to cross-check our results. As a sanity check for our implementation of the Hilbert series we can use the $a^0$ terms to compare our results for the Hilbert series with Ref.~\cite{Henning:2015alf} and the operators with the SMEFT operator basis up to dimension~8~\cite{Weinberg:1979sa,Grzadkowski:2010es,Lehman:2014jma,Murphy:2020rsh}. 

Furthermore, some results for the ALP EFT are available in the literature up to dimension~7. Our results at dimension~5 are consistent with the usual basis at dimension~5 (see e.g. Ref.~\cite{Georgi:1986df}). The results at dimension~6 are consistent with Ref.~\cite{Bauer:2017ris,Bauer:2018uxu,Brivio:2021fog,Bonilla:2021ufe}. Some results at dimension~7 can be found in Ref.~\cite{Bauer:2016zfj,Bauer:2017ris} and are consistent with our operator basis. More results for higher-dimensional operators can be found in Ref.~\cite{Brivio:2017ije} where the authors match some of the operators in the chiral electroweak EFT extended with an axion to the one in a linear realization. All operators they find are either in our basis or equivalent to operators in our basis due to field redefinitions and IBP.

To get a feeling for how the number of operators $\#\,\cO_{d_i}^{\PQ}$ behaves as a function of the mass dimension $d_i$, we set all field spurions in the Hilbert series with full flavor dependence to unity such that only the dependence on $N_f$ remains. By rephasing all fermionic spurions with lepton number and baryon number transformations respectively, i.e. $\ell \to \epsilon_{L} \ell,\  \ell^{\dagger} \to \epsilon_{L}^{-1} \ell^{\dagger}$ and $q \to \epsilon_{B}^{1/3} q,\  q^{\dagger} \to \epsilon_{B}^{-1/3} q^{\dagger}$, we can in addition obtain the number of operators that break lepton and baryon number at each mass dimension. Then for each power of $\epsilon_{B,L}$ in the following expressions, baryon number and lepton number are violated by one unit. After taking care of the caveats we have mentioned above by hand, we find\footnote{After all terms have been simplified, we make the replacement $\epsilon_i^{-n} \to \epsilon_i^n$, such that we count all operators in the same way that break baryon and lepton number by $|n|$ units with respect to the conserving case.}
\begin{align}
& \#\,\cO_5^{\PQ} \,=\, 2 - N_f + 5 N_f^2 \, , \nonumber\\
& \#\,\cO_6^{\PQ} \,=\, 1 \, , \nonumber \\
& \#\,\cO_7^{\PQ} \,=\, 5 + 39 N_f^2 \, , \nonumber \\ 
& \#\,\cO_8^{\PQ} \,=\, \left(13+11 N_f^2\right)+\left(-\frac{2 N_f^2}{3}+\frac{8 N_f^4}{3}\right) \epsilon_B \epsilon_L+\left(4 N_f^2+2 N_f^4\right) \epsilon_L^2 \, , \\
& \#\,\cO_9^{\PQ} \,=\, \left(74+\frac{1799 N_f^2}{4}-\frac{N_f^3}{2}+\frac{847 N_f^4}{4}\right)+46 N_f^4 \epsilon_B \epsilon_L+\left(N_f+N_f^2\right) \epsilon_L^2 \, , \nonumber \\ 
& \#\,\cO_{10}^{\PQ} \,=\, \left(74+\frac{431 N_f^2}{2}+\frac{91 N_f^4}{2}\right)+\left(8 N_f^3+106 N_f^4\right) \epsilon_B \epsilon_L+\left(-N_f+75 N_f^2+N_f^3+121 N_f^4\right) \epsilon_L^2 \, . \nonumber
\end{align}
One can also easily get the total number of operators by setting $\epsilon_{B,L}\to 1$.
Beyond mass dimension~10 the expressions become too long to be presented here. We give more results in App.~\ref{app:HSaSMEFT} and the Hilbert series up to mass dimension~15 with full spurion and flavor dependence in an ancillary Mathematica notebook.\footnote{In the notebook, we have taken care of the caveats mentioned above by hand and have also added $\Delta \cH$ such that the negative terms coming from co-closed but not co-exact forms are canceled, so that the correct number of non-redundant operators can be obtained easily.} One notices that baryon and lepton number breaking operators only appear at mass dimension~8. This is due to the derivative coupling of the ALP that only allows for baryon and lepton number breaking terms through the coupling of $\partial_{\mu} a$ to 4-fermion operators. 

\begin{figure}[t]
  \includegraphics[width=\linewidth]{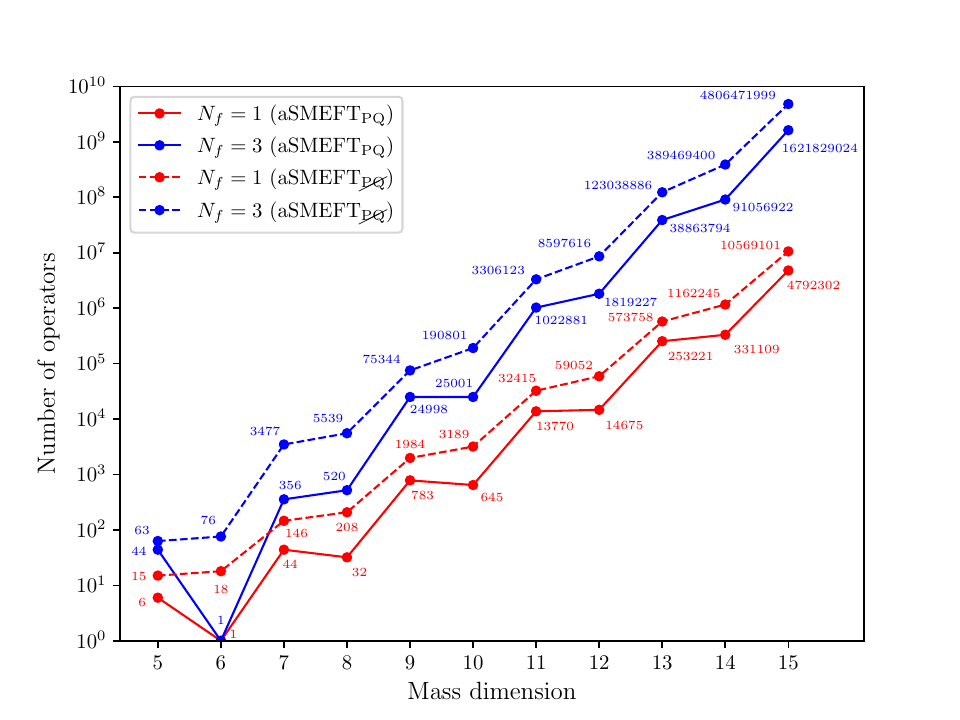}
  \caption{The number of operators in the aSMEFT with and without a shift symmetry for the ALP plotted against the mass dimension for $N_f=1$ and $N_f=3$ number of flavors.}
  \label{fig:aSMEFT_NoOpsVsMassDim}
\end{figure}

In Fig.~\ref{fig:aSMEFT_NoOpsVsMassDim}, we have plotted the number of operators of the ALP EFT up to mass dimension~15 for one and three flavors of fermions. One can see the usual growth of operators with mass dimension where some unusual features appear for the ALP EFT due to the derivative nature of the shift-symmetric ALP couplings. The shift symmetry, for instance, only allows for one coupling at dimension~6. At higher mass dimensions on the other hand, the multiplicity of operators increases in the same manner as one is accustomed to from other EFTs.\footnote{There is an interesting asymptotic scaling behavior~\cite{Melia:2020pzd} for the number of operators in effective field theories that should also apply to the plot in Fig.~\ref{fig:aSMEFT_NoOpsVsMassDim} for large mass dimensions beyond the drop at dimension~6.}

\subsection{\texorpdfstring{$\aSMEFTnotPQ$}{aSMEFTnotPQ}}\label{sec:aSMEFT_nonSS}
 In this section, we will analyze the difference between an ALP and a generic pseudoscalar. Hence, we relax the assumption of a shift symmetry for the pseudoscalar which no longer necessarily has to be connected to the spontaneous breaking of a PQ symmetry. However, there can still be such a connection by assuming that the spontaneously broken global symmetry is only approximate. Then, all shift-breaking operators are understood as small corrections to those that break the shift symmetry and it is important to understand the limit of taking the shift symmetry.

As the shift symmetry is now relaxed, we can work with $a$ itself as a building block for the Hilbert series. As in the previous section, we can evaluate the Hilbert series from Eq.~\eqref{eq:hs2} using the appropriate characters and Haar measures for the given field spurions and symmetries. For one generation of fermions we obtain up to mass dimension~7
\begin{equation}
\begin{split}
\cH_5^{\notPQ} = & a^5 + a B_L^2 + a B_R^2 + a G_L^2 + a G_R^2 + a W_L^2 + a W_R^2 + a^3 H H^{\dagger} + a H^2 H^{\dagger2} + a Q u H \\
&+ a Q^{\dagger} u^{\dagger} H^{\dagger} + a Q d H^{\dagger} + a Q^{\dagger} d^{\dagger} H + a L e H^{\dagger} + a L^{\dagger} e^{\dagger} H \\
= & a \, \cH_{4}^{\text{SM}} + a^5 + a^3 H H^{\dagger} \, , \\[3pt]
\cH_6^{\notPQ} = & a \, \cH_5^{\notPQ} + a H^2 L^2 + a H^{\dagger2} L^{\dagger2} + a^2 H H^{\dagger} \cD^2 \, , \\[3pt]
\cH_7^{\notPQ} = & a \, \cH_6^{\notPQ} + a \, \cH_6^{\text{SMEFT}} + a Q Q^{\dagger} B_L \cD + a Q Q^{\dagger} B_R \cD + a Q Q^{\dagger} G_L \cD + a Q Q^{\dagger} G_R \cD \\
& + a Q Q^{\dagger} W_L \cD + a Q Q^{\dagger} W_R \cD + a u u^{\dagger} B_L \cD + a u u^{\dagger} B_R \cD + a u u^{\dagger} G_L \cD + a u u^{\dagger} G_R \cD \\
& + a d d^{\dagger} B_L \cD + a d d^{\dagger} B_R \cD + a d d^{\dagger} G_L \cD + a d d^{\dagger} G_R \cD + a L L^{\dagger} B_L \cD + a L L^{\dagger} B_R \cD \\
& + a L L^{\dagger} W_L \cD + a L L^{\dagger} W_R \cD + a e e^{\dagger} B_L \cD + a e e^{\dagger} B_R \cD + 2 a Q Q^{\dagger} H H^{\dagger} \cD \\
& + a u u^{\dagger} H H^{\dagger} \cD + a d d^{\dagger} H H^{\dagger} \cD + 2 a L L^{\dagger} H H^{\dagger} \cD + a e e^{\dagger} H H^{\dagger} \cD + a B_L H H^{\dagger} \cD^2 \\
& + a B_R H H^{\dagger} \cD^2 + a W_L H H^{\dagger} \cD^2 + a W_R H H^{\dagger} \cD^2 + a H^2 H^{\dagger2} \cD^2 + 2 a Q u H \cD^2  \\
& + 2 a Q^{\dagger} u^{\dagger} H^{\dagger} \cD^2 + 2 a Q d H^{\dagger} \cD^2 + 2 a Q^{\dagger} d^{\dagger} H \cD^2 + 2 a L e H^{\dagger} \cD^2 + 2 a L^{\dagger} e^{\dagger} H \cD^2 \, .
\end{split}
\end{equation}
At dimension~5, $\cH_5^{\PQ}$ is absent and we find
\begin{equation}
    \cH_5^{\notPQ} = a \cH_4^{\notPQ} + a \cH_4^{\text{SM}}\,,
\end{equation}
which is the well-known result that, at dimension~5, the fermionic operators with the derivatively coupled ALP are redundant by the EOM~\cite{Brivio:2017ije,Bauer:2020jbp,Chala:2020wvs,Bonilla:2021ufe}.
Allowing for lepton number breaking, we find the following relation at the level of the Hilbert series
\begin{equation} \label{eq:aSMEFTnonSS_HS6}
        \cH_6^{\notPQ} = a \, \cH_5^{\notPQ} + a \, \cH_5^{\text{SMEFT}} + \cH_6^{\PQ} \( \partial a \to a \cD \) \, .
\end{equation}
Here, $\cH_i^{\notPQ}$ is the Hilbert series of the ALP EFT with $a$ as a building block, $\cH_i^{\text{SMEFT}}$ is the Hilbert series of the SMEFT and $\cH_6^{\PQ}$ is the Hilbert series of the ALP EFT with $\partial a$ as a building block, each at mass dimension~$i$. The expression in the bracket is understood as replacing the spurion $\partial a$ of the derivatively coupled ALP with the ALP spurion $a$ and the spurion of the covariant derivative $\cD$.

This relation at the level of the Hilbert series should hold true to any mass dimension beyond dimension~5.\footnote{We have checked that this separation appears up to mass dimension~15 and believe that it holds also true for all higher order operators. To get a mixing of the two sector one needs an EOM relation between two of the operators which implies that the effects of one operator is already captured by the other. In particular such an EOM relation has to arise upon moving the derivative from the axion to the rest of the operator using IBP. As the operators become more and more complicated for higher mass dimensions, it is less and less likely that through this procedure a structure is obtained which exactly resembles the EOM of an SM particle as at dimension~5.}  In general, we conjecture that the Hilbert series fulfills the following condition at mass dimension~$n$
\begin{tcolorbox}[top=2mm,bottom=3mm]
\begin{equation}
\label{eq:separation}
        \cH_n^{\notPQ} = a \, \cH_{n-1}^{\notPQ} + a \, \cH_{n-1}^{\text{SMEFT}} + \cH_n^{\PQ} (\partial a \to a \cD)\,, \qquad n>5
\end{equation} 
\end{tcolorbox}
\noindent which we have verified to hold true up to $n=15$. From now on we will refer to this relation as the \emph{Peccei--Quinn breaking isolation condition} or \emph{shift-breaking isolation condition}. We want to emphasize the importance of this equation. It states that above dimension~5 the EFT splits into a part generated by simply multiplying the operators at the previous mass dimension -- which immediately follows from the singlet scalar nature of the ALP -- and a second part which is exactly the EFT built with a derivatively coupled, i.e. explicitly shift-invariant ALP. This separation of the shift-breaking and shift-symmetric sectors of the ALP EFT can be captured with the Hilbert series in a concise way. We will explore the implications of this further in Section~\ref{sec:ShiftSym}. 

The operator construction for the shift-breaking case is trivial as the ALP is a singlet both under the Lorentz and the gauge group. Then, one can multiply any gauge and Lorentz invariant operator to receive a new Lorentz invariant operator. Looking again at Eq.~\eqref{eq:separation}, it can be seen that, by this construction, one can obtain the complete operator basis after adding the derivatively coupled terms that we have constructed in Section~\ref{sec:aSMEFT_SS}.

Our complete basis for an ALP without a shift symmetry coupled to the SM particles at mass dimension~5 can be found in Tab.~\ref{tab:aSMEFT_nonSS_dim5}. The operator bases up to mass dimension~8 can be constructed easily with the shift-breaking isolation condition, see App.~\ref{app:aSMEFTnonSSOpBasis} for details. For completeness, we also give the renormalizable part of the Lagrangian here, as a potential can be generated for the ALP once the assumption of a shift symmetry for the ALP is loosened. The renormalizable Lagrangian is given by
\begin{equation} \label{eq:aSMEFTnonPQreno}
\cL_{\leq4}^{a} = \frac{1}{2} \partial_{\mu} a \, \partial^{\mu} a - \frac{m_{a,0}^2}{2} \, a^2 + C_{a^3} \, a^3 + C_{a^4} \, a^4 + C_{aH^2} \, a \, |H|^2 + C_{a^2H^2} \, a^2 |H|^2 \, .
\end{equation}
There are some results available in the literature. We have cross-checked our operator basis with those of Refs.~\cite{Gripaios:2016xuo,Franceschini:2016gxv} and find agreement.

As before, we can obtain a formula for the number of operators at each mass dimension by setting all spurions to unity. We find
\begin{align}
 \#\,\cO_5^{\notPQ} \,=\,& 9 + 6 N_f^2 \, , \nonumber \\
 \#\,\cO_6^{\notPQ} \,=\,& \left(10+6 N_f^2\right)+\left(N_f+N_f^2\right) \epsilon_L^2 \, , \nonumber \\
 \#\,\cO_7^{\notPQ} \,=\,& \left(30+\frac{315 N_f^2}{4}+\frac{N_f^3}{2}+\frac{107 N_f^4}{4}\right)+\left(\frac{2 N_f^2}{3}+N_f^3+\frac{19 N_f^4}{3}\right) \epsilon_B \epsilon_L+\left(N_f+N_f^2\right) \epsilon_L^2 \, , \nonumber \\
 \#\,\cO_8^{\notPQ} \,=\,& \left(43+\frac{359 N_f^2}{4}+\frac{N_f^3}{2}+\frac{107 N_f^4}{4}\right)+\left(3 N_f+\frac{41 N_f^2}{3}+N_f^3+\frac{37 N_f^4}{3}\right) \epsilon_L^2\\
 &+\left(2 N_f^3+16 N_f^4\right) \epsilon_B \epsilon_L\, . \nonumber
\end{align}
Note that lepton number violating terms already appear at dimension~6 for an ALP without a shift symmetry, whereas for an ALP with a shift symmetry lepton and baryon number violating terms only appear at dimension~8 because the ALP no longer has to be derivatively coupled, and can for instance simply multiply the Weinberg operator of the SMEFT to give a lepton number violating operator at dimension~6. One can again find more results beyond dimension~8 in App.~\ref{app:HSaSMEFT} and the Hilbert series up to mass dimension~15 with full spurion content and flavor dependence in the ancillary Mathematica notebook.

In Fig.~\ref{fig:aSMEFT_NoOpsVsMassDim}, we have plotted the number of operators against the mass dimension for the SMEFT extended with an ALP without a shift symmetry. When the shift symmetry is relaxed, it is trivial to build new singlets under the gauge and Lorentz group by just multiplying by $a$ as can also be seen in the structure of the Hilbert series in Eq.~\eqref{eq:separation}. Comparing the number of operators at dimension~5 between the explicitly shift-symmetric and non-shift-symmetric Lagrangian in Fig~\ref{fig:aSMEFT_NoOpsVsMassDim}, one can see that the difference $63-44=19$ corresponds exactly to the number of shift-breaking invariants from Ref.~\cite{Bonnefoy:2022rik} (13) plus the number of conditions that have to be imposed on the bosonic shift symmetry breaking operators (6) (c.f. Tab.~\ref{tab:aSMEFT_nonSS_dim5}). We will discuss the shift-symmetric limit in more details in Section~\ref{sec:ShiftSym}.

\subsection{Taking the shift-symmetric limit}\label{sec:ShiftSym}
In the aSMEFT there is a subtlety in how to properly identify the shift-symmetric and shift-breaking couplings of the ALP to the fermions at dimension~5. This is because, as we saw in Section~\ref{sec:aSMEFT_nonSS}, the dimension-5 operators coupling the ALP derivatively to the SM particles become redundant due to the fermion EOM once the shift symmetry is relaxed. After removing the EOM redundancy, one can use the apparently non-shift-symmetric interactions between the ALP and the SM particles to describe a shift-symmetric ALP given that the couplings of the interactions follow a set of 13 relations \cite{Bonnefoy:2022rik} (see also Refs.~\cite{Brivio:2017ije,Bauer:2020jbp,Chala:2020wvs,Bonilla:2021ufe} where the constraints on the couplings were formulated as matrix relations in flavor space). This leads to difficulties in the EFT picture because the same operator has to capture physics corresponding to the shift-breaking and shift-symmetric sector which usually arise at very different scales. Furthermore, taking the shift-symmetric limit in the EFT where one uses $a$ as a building block requires some care.

We will first present the results at dimension~5 known in the literature and will then check if similar new relations arise at higher mass dimensions using our results from the previous section. We consider the following Lagrangian
\begin{equation} \label{eq:aSMEFTLag45}
\cL^a = - \bar{L} Y_e H e - \bar{Q} Y_u \tilde{H} u - \bar{Q} Y_d H d + \frac{a}{f} \( \bar{L} C_{ae} H e + \bar{Q} C_{au} \tilde{H} u + \bar{Q} C_{ad} H d \) + \hc  \, .
\end{equation}
On a first glance, this Lagrangian does not look shift invariant. However, after shifting the ALP $a \to a+c$, one can perform field redefinitions on the fermion fields that allow to remove the shift at $\cO\(\frac{1}{f}\)$ from the Lagrangian by imposing the following relations on the dimension-5 Wilson coefficients\footnote{Another way to derive these results can be found in App.~\ref{app:ShiftFieldRedef}, where we start in the derivatively coupled basis and make a change of basis to go to the Yukawa basis.}
\begin{equation}\label{eq:Dim5YukShiftRelations}
C_{au} = i \( c_Q Y_u - Y_u c_u \), \ C_{ad} = i \( c_Q Y_d - Y_d c_d \), \ C_{ae} = i \( c_L Y_e - Y_e c_e \) \, .
\end{equation}
Then, the apparently non-shift-symmetric Lagrangian can be made shift-symmetric. Furthermore, these matrix relations can be cast into order parameters which allow to implement the different power countings of the shift-breaking and shift-conserving sector in a straightforward way \cite{Bonnefoy:2022rik}. From now on, we will refer to the Lagrangian in Eq.~\eqref{eq:aSMEFTLag45} as the Yukawa basis given that the relations in Eq.~\eqref{eq:Dim5YukShiftRelations} are fulfilled.

We will now explore if similar relations exist in the non-shift-symmetric EFT at higher mass dimensions using the operator basis we have derived above. The first observation we want to make is based on the Peccei--Quinn breaking isolation condition in Eq.~\eqref{eq:separation}. For $n>5$ we have found previously
\begin{equation} 
        \cH_n^{\notPQ} = a \, \cH_{n-1}^{\notPQ} + a \, \cH_{n-1}^{\text{SMEFT}} + \cH_n^{\PQ} \( \partial a \to a \cD \) \, .
\end{equation}
The Hilbert series which is obtained by imposing the shift symmetry explicitly,  $\cH_n^{\PQ}$, appears fully in the Hilbert series of the theory where $a$ itself is used as a spurion. 

This implies that beyond dimension-5 no further EOM redundancies appear and all operator structures stay non-redundant in the presence of shift-breaking interactions. Therefore, if one decides to work in the operator basis with derivatively coupled interactions at dimension-5, all shift-symmetric couplings are exactly captured by the derivative interactions. 

One has to be more careful when working in the Yukawa basis, which is the more natural basis in the presence of shift-breaking effects as we will comment on below. Here, one has to take care while removing the EOM redundancy at dimension~5 when higher-order operators are considered in the EFT.\footnote{We thank Quentin Bonnefoy for pointing this out and Pham Ngoc Hoa Vuong for sharing calculations with us convincing us of the importance of those terms.} In particular, one should use field redefinitions instead of simply plugging in the SM EOM to remove the derivatively coupled operators at dimension~5 and keep all terms that are generated by these field redefinitions up to the considered order in the EFT. We have done this carefully in App.~\ref{app:ShiftFieldRedef} and find that the field redefinition removing the EOM redundancy at dimension~5 indeed generates more (seemingly shift-breaking) operators with fully constrained Wilson coefficients that restore the shift symmetry.

This is important as the spontaneous and explicit breaking of PQ usually arise at very different scales. Working in the Yukawa basis, these effects are captured by the same operators and it is not straightforward how to implement the correct power counting for both sectors. As was pointed out in Ref.~\cite{Bonnefoy:2022rik}, one way around this is to consider flavor invariants acting as order parameters for the ALP shift symmetry. In this language it is possible to consistently implement the power counting of the theory.

One also has to keep these relations in mind while taking the shift-symmetric limit going from the EFT without a shift-symmetry to the EFT with a shift-symmetry in the Yukawa basis. Instead of setting all non-derivatively coupled operators to zero, one should set them to the constrained form that is found applying the appropriate field redefinitions. Note that the Yukawa basis is in some sense the more natural basis to perform this limit because the EOM redundancy at dimension-5 requires the derivatively coupled fermionic operator to be absent from the operator basis in favor of the Yukawa-like operator. We list all the relations that have to be imposed in App.~\ref{app:ShiftRelations}.

These additional constrained interactions are crucial for explicit calculations in the Yukawa basis. If the additional terms are not included, one will run into results in the shift-symmetric EFT in the Yukawa basis which are not shift-invariant. E.g. if two insertions of the dimension-5 ALP-Yukawa couplings are considered, one must also add the diagram with the constrained interaction of the dimension-6 ALP-Yukawa coupling. Note that up to dimension-7 only the ALP-Yukawa operators with higher powers of the ALPs have to be considered, and only starting at dimension-8 more operators with constrained Wilson coefficients are generated by the field redefinition if additional operators generated by applying the field redefinitions to SMEFT operators are ignored.

The analysis presented here can also be understood from an amplitudes point of view~\cite{Bertuzzo:2023} by imposing the Adler's zero condition~\cite{Adler:1964um,Adler:1965ga}. Here, special care has to be taken in how to impose the Adler's zero condition leading to the well-known conditions on the dimension-5 couplings and more relations at higher mass dimensions, consistent with our analysis with the Hilbert series and field redefinitions. In the amplitudes approach, these relations can be understood from fundamental properties of amplitudes like analyticity and regularity of the amplitude in the limit of soft ALP momenta (for details see Ref.~\cite{Bertuzzo:2023}).

For the aLEFT, that we will construct in the next section, one can perform a similar analysis which works in the same way. Therefore, we will skip the discussion of the shift symmetric limit there.

\subsection{CP violation in the aSMEFT}\label{sec:aSMEFTCPV}

{\arraycolsep=4pt
\begin{table}[ht!]
    \centering
\scalebox{0.95}{
$\begin{array}{|c|ccc|ccc|}
\hline
 \multirow{ 2}{*}{\text{Dim.}} & \multicolumn{3}{c|}{\aSMEFTPQ} & \multicolumn{3}{c|}{\aSMEFTnotPQ} \\ \cline{2-7}
  & \text{CP-even} & \text{CP-odd} & \text{CP-violating} & \text{CP-even} & \text{CP-odd} & \text{CP-violating} \\ \hline
 \multirow{1.6}{*}{5} & 6 & 0 & 0 & 6 & 9 & 9 \\[-2.5mm]
  & 29 & 15 & 9 & 30 & 33 & 27 \\ 
 \multirow{1.6}{*}{6} & 1 & 0 & 0 & 11 & 7 & 6 \\[-2.5mm]
  & 1 & 0 & 0 & 40 & 36 & 24 \\ 
 \multirow{1.6}{*}{7} & 26 & 18 & 18 & 60 & 86 & 81 \\[-2.5mm]
  & 189 & 167 & 128 & 1647 & 1830 & 1062 \\ 
 \multirow{1.6}{*}{8} & 22 & 10 & 6 & 123 & 85 & 61 \\[-2.5mm]
  & 271 & 249 & 33 & 2872 & 2667 & 912 \\ 
 \multirow{1.6}{*}{9} & 427 & 356 & 332 & 942 & 1042 & 945 \\[-2.5mm]
  & 12662 & 12336 & 6807 & 37345 & 37999 & 20476 \\ 
 \multirow{1.6}{*}{10} & 356 & 289 & 134 & 1678 & 1511 & 979 \\[-2.5mm]
  & 12702 & 12299 & 1733 & 95929 & 94872 & 21555 \\ 
 \multirow{1.6}{*}{11} & 7053 & 6717 & 5926 & 15978 & 16437 & 13942 \\[-2.5mm]
  & 513504 & 509377 & 235519 & 1651318 & 1654805 & 702019 \\ 
 \multirow{1.6}{*}{12} & 7491 & 7184 & 2812 & 29909 & 29143 & 16295 \\[-2.5mm]
  & 910536 & 908691 & 60630 & 4301474 & 4296142 & 759162 \\ 
 \multirow{1.6}{*}{13} & 127404 & 125817 & 104553 & 285800 & 287958 & 227861 \\[-2.5mm]
  & 19442371 & 19421423 & 7978922 & 61499879 & 61539007 & 22689934 \\ 
 \multirow{1.6}{*}{14} & 166364 & 164745 & 54104 & 583011 & 579234 & 279807 \\[-2.5mm]
  & 45535198 & 45521724 & 2494107 & 194761001 & 194708399 & 25144913 \\ 
 \multirow{1.6}{*}{15} & 2400015 & 2392287 & 1868885 & 5279487 & 5289614 & 3909730 \\[-2.5mm]
  & 810986291 & 810842733 & 284971909 & 2403111000 & 2403360999 & 764583481 \\[1mm] \hline
\end{array}$}
\caption{Number of CP-even, CP-odd and CP-violating operators for $\aSMEFTPQ$~(left) and $\aSMEFTnotPQ$~(right) from dimension 5 to 15. In each dimension, the two rows correspond to $N_f=1$ and $N_f=3$ respectively.}
\label{tab:aSMEFT_CP_counting}
\end{table}
}

Adding a $\mathcal{C}$- and $\mathcal{P}$- transformation to the Hilbert series as described in Section~\ref{sec:CPHilbertSeries} allows us to count the number of CP-odd and CP-even parameters in the effective Lagrangian. In Tab.~\ref{tab:aSMEFT_CP_counting} we show the results for the $\aSMEFT$ with and without a shift symmetry for the ALP. Furthermore, we show the number of CP-violating couplings which are the number of CP-odd couplings that cannot be removed after using the freedom of performing rephasings on all fermion fields that leave the renormalizable part of the Lagrangian invariant, i.e. $U(1)_{L_i}^3 \times U(1)_B$ rephasings for the $\aSMEFT$. In this analysis we turn on one operator at a time, such that all possible rephasings can be used for each operator. 

In the bosonic sector this counting is straightforward, because all bosonic operators in our basis are eigenstates of CP and those operators which transform with a sign under CP are CP-violating. In the fermionic sector it is not so simple to identify all CP-violating couplings because a flavor transformation can be performed on top of the CP transformation to remove CP-violating parameters. Therefore, it is advisable to use flavor invariants to characterize CP-violating parameters as was first done in Refs.~\cite{Jarlskog:1985ht,Jarlskog:1985cw} for the SM and has recently been extended to EFTs like the SMEFT~\cite{Bonnefoy:2021tbt,Bonnefoy:2023bzx}. The flavor invariants keep track of exactly this additional freedom that we have just described.

In the $\aSMEFTnotPQ$,  the dimension-5 CP-violating Hilbert series is given by\footnote{We have redefined $\check{\phi}$ to $\phi$ to simplify the notation. All the SM spurions in the Hilbert series should be understood as a direct sum of the field and its conjugate. Since the axion $a$ transforms trivially under the $\mathcal{C}$, it can be considered as a building block in the CP-extended Hilbert series. For instance, the term $aLHe$ should be written as $a\check{L}\check{H}\check{e}$, which can be understood as CP-odd operator $aL^{\dagger}He^{\dagger}+aLH^{\dagger}e$ or CP-even operator $aL^{\dagger}He^{\dagger}-aLH^{\dagger}e$, depending on which Hilbert series it belongs to. In Eq.~\eqref{eq:aSMEFT_PQ_dim5_CPV}, the term $3aLHe$ corresponds to 3 CP-violating operators $aL_i^{\dagger}He_i^{\dagger}+aL_iH^{\dagger}e_i$ for $i=1,2,3$.}
\begin{equation}
\label{eq:aSMEFT_PQ_dim5_CPV}
\cH_{5,\text{CPV}}^{\notPQ}=a^5+a B^2+a W^2+a G^2+a^3 H^2 +a H^4 +3 a L H e+9 a Q H u +9 a Q H d\,,
\end{equation}
from which we can count the number of CP-violating parameters in each operator. In terms of the flavor invariants, all fermionic couplings are described by 3 generic complex $3\times 3$ matrices $C_{ae}$, $C_{au}$ and $C_{ad}$. Following the discussion in Refs.~\cite{Bonnefoy:2021tbt,Bonnefoy:2023bzx}, we can find the following flavor invariants that capture all primary sources\footnote{We denote all CP-odd couplings as primary, which can form a flavor invariant quantity at leading order in the EFT. For instance, the complex $3\times 3$ matrix $C_{ae}$ has 9 CP-odd parameters. However, there only exist three flavor invariant CP-odd quantities at the leading order in the EFT expansion as can be seen in Eq.~\eqref{eq:aSMEFTCPVLep}. All other CP-odd parameters can only appear at subleading orders in the EFT expansion. This changes in the quark sector due to the existence of the CKM matrix which being charged under rephasings can give rise to more primary sources of CP violation. See also Ref.~\cite{Bonnefoy:2021tbt}.} of CP violation in the leptonic sector of the EFT
\begin{equation} \label{eq:aSMEFTCPVLep}
    \Re\Tr\(C_{ae} Y_e^{\dagger}\), \quad \Re\Tr\(X_e C_{ae} Y_e^{\dagger}\), \quad \Re\Tr\(X_e^2 C_{ae} Y_e^{\dagger}\) \,,
\end{equation}
where $X_e = Y_e Y_e^\dagger$ and later also $X_{u,d} = Y_{u,d} Y_{u,d}^\dagger$. We can see that the number of flavor invariants exactly match the corresponding term $+3aLHe$ in the Hilbert series. Setting these invariants to zero gives sufficient and necessary conditions for CP conservation. 

In the quark sector we have
\begin{equation}
\begin{split}
    L_{0000}\(C_{au} Y_u^\dagger\), \ L_{1000}\(C_{au} Y_u^\dagger\), \ & L_{0100}\(C_{au} Y_u^\dagger\), \ L_{1100}\(C_{au} Y_u^\dagger\), \ L_{0110}\(C_{au} Y_u^\dagger\), \\
     L_{2200}\(C_{au} Y_u^\dagger\), \ L_{0220}&\(C_{au} Y_u^\dagger\), \ L_{1220}\(C_{au} Y_u^\dagger\), \ L_{0122}\(C_{au} Y_u^\dagger\) \,,
\end{split}
\end{equation}
where we have defined $L_{abcd}(C)=\Re\Tr\(X_u^a X_d^b X_u^c X_d^d C\)$ and similar relations hold true in the down sector with $C_{au} Y_u^\dagger \to C_{ad} Y_d^\dagger$. These 18 flavor invariants are encoded in the Hilbert series by the terms $+9aQHu+9aQHd$. In total, we have 21 CP-odd flavor invariants for $N_f=3$ which have to vanish for CP to be conserved in the fermionic sector of the theory and 6 CP-odd operators in the bosonic sector which can be easily identified from Tab.~\ref{tab:aSMEFT_nonSS_dim5}. This is consistent with the counting in Tab.~\ref{tab:aSMEFT_CP_counting}. For higher dimensional operators, the CP-even, CP-odd and CP-violating Hilbert series are shown in App.~\ref{app:HSaSMEFT}.

In the dimension-5 $\aSMEFTPQ$, all couplings are described by 5 hermitian matrices in the derivatively coupled basis. Interestingly, no primary sources of CP violation can be written down for the leptonic sector because there exist no quantities charged under rephasings of the lepton fields in the SM Lagrangian. This is consistent with the Yukawa basis where we have to impose shift symmetry on the invariants in Eq.~\eqref{eq:aSMEFTCPVLep}. Because these CP invariants are identical to those capturing the shift symmetry, there are no remaining primary sources of CP violation in the leptonic sector of the EFT as was already noted in Ref.~\cite{Bonnefoy:2022rik}.
This is also captured by the CP-violating Hilbert series
\begin{equation}
\label{eq:aSMEFT_ss_CPV}
\cH_{5,\text{CPV}}^{\PQ}= 3\partial a\, Q^2 + 3\partial a\, u^2 + 3\partial a\, d^2\,,
\end{equation}
where the leptonic sector is absent due to the application of lepton family number rephasings.

In the quark sector we find the following CP-violating invariants for the couplings $C_{\partial a Q}$
\begin{equation}
    \tilde{L}_{1100}\(C_{\partial a Q} \), \ \tilde{L}_{2200}\(C_{\partial a Q} \), \ \tilde{L}_{1122}\(C_{\partial a Q} \),
\end{equation}
where $\tilde{L}_{abcd}(C)=\Im\Tr\(X_u^a X_d^b X_u^c X_d^d C\)$ 
and similar relations for $C_{\partial a u}$ and $C_{\partial a d}$, where we have to replace $C_{\partial a Q} \to Y_u C_{\partial a u} Y_u^\dagger$ and $C_{\partial a Q} \to Y_d C_{\partial a d} Y_d^\dagger$ respectively. The number of flavor invariants matches precisely the number of CP violating couplings counted by the Hilbert series in Eq.~\eqref{eq:aSMEFT_ss_CPV}.

We can again compare this to the Yukawa basis. There, we have found 18 flavor-invariant quantities at leading order in the EFT for the ALP-Yukawa couplings in the shift-breaking form. According to the discussion in Ref.~\cite{Bonnefoy:2022rik}, 9 of those have to be set to zero, in order to obtain a shift-symmetric Lagrangian giving agreement between the two bases. We can furthermore compare with the counting in Tab.~\ref{tab:aSMEFT_CP_counting}. In the $\aSMEFTPQ$, all CP-violating couplings are forbidden in the bosonic sector and, as we just counted, there are 9 CP-odd flavor invariants for $N_f=3$. This is consistent with the counting at dimension-5 in Tab.~\ref{tab:aSMEFT_CP_counting}.

It is also interesting to understand why there are exactly 9 CP-odd and 1 CP-even parameters which capture the shift-breaking interactions in the quark sector of the aSMEFT at dimension 5. One can first notice that this is exactly the same number of parameters as there are physical parameters in the quark sector of the renormalizable Lagrangian, but with opposite CP parity due to the ALP being a pseudoscalar. One can easily verify that the same is true in the lepton sector and still holds true if one adds more fermions to the theory, like sterile neutrinos with a Yukawa coupling or a Majorana mass term.

This correspondence between the number of physical parameters in the dimension-4 Lagrangian and the number of shift-breaking parameters in the ALP EFT can be understood as follows. Shifting the ALP $a \to a +c$ and trying to remove the shift with field redefinitions, as was done in the last section, one has the freedom to remove a parameter at dimension-5 for each physical parameter present at dimension 4. Furthermore, if there is a degeneracy in the mass spectrum at dimension 4, one also has more freedom to remove parameters at dimension-5, preserving this correspondence even for degenerate spectra. This however does not mean that if CP conservation is imposed at dimension-4 that the single CP-even shift-breaking coupling at dimension 5 will automatically vanish. The parameters at dimension 4 and dimension 5 are of course independent parameters that are only connected by the field redefinition and the same behavior under flavor transformations. As we just mentioned, it is exactly this behavior under flavor transformations that allows us to remove more parameters in the case of a degenerate spectrum at dimension 4. Setting the phase of the CKM matrix to zero does not increase the flavor symmetry and therefore the independent parameter at dimension 5 cannot be removed.

Therefore, imposing CP conservation on the EFT with shift-breaking operators does not yield a shift-symmetric EFT as one might expect. This can be seen in a straightforward way from the invariants in Ref.~\cite{Bonnefoy:2022rik}. 

\section{aLEFT}\label{sec:aLEFT}
With the Hilbert series implemented for the SMEFT extended with an ALP, it is fairly straightforward to also construct it for the EFT below the electroweak scale. As discussed in Section~\ref{sec:HS}, the main difference is the different particle content where the heavy particles of the SM, the $W,Z,t$ and $h$, now have been integrated out and the fact that the left-handed fermions are no longer related through their appearance in $SU(2)$ doublets. Since the gauge group below the EW scale is only $SU(3)_c \times U(1)_{\text{em}}$, both a linear and a non-linear realization of the EW symmetry can be captured in the LEFT. Hence, we can capture effects of HEFT-like ALP couplings to the SM particles~\cite{Brivio:2017ije,Bonnefoy:2020gyh,Altmannshofer:2022ckw} in the LEFT extended with an ALP. Furthermore, the effective description of these interactions is of importance for experiments operating at these scales, for instance meson decays giving flavor constraints on ALP couplings to fermions~\cite{Bauer:2021mvw}. Performing the construction and matching to the QCD chiral Lagrangian~\cite{Georgi:1986df,GrillidiCortona:2015jxo,Bauer:2020jbp,Bauer:2021wjo} beyond leading order would complete a full EFT description beyond leading order at all scales.

\subsection{\texorpdfstring{$\aLEFTPQ$}{aLEFTPQ}}\label{sec:aLEFT_SS}
We start again by calculating the Hilbert series for the effective operators by mass dimension. We present here the Hilbert series up to mass dimension~7
\begin{align}
\cH_{5}^{\aLEFTPQ} \,=\, & \partial a\, u_L u_L^{\dagger} + \partial a\, u_R u_R^{\dagger} + \partial a\, d_L d_L^{\dagger} + \partial a\, d_R d_R^{\dagger} + \partial a\, \nu_L \nu_L^{\dagger} + \partial a\, e_L e_L^{\dagger} + \partial a\, e_R e_R^{\dagger} \nonumber\\
&- \partial a\, F_L \cD - \partial a\, F_R \cD - \partial a\, \cD^3 \, , \nonumber\\
\cH_{6}^{\aLEFTPQ} \,=\, & 0 \, , \nonumber\\
\cH_{7}^{\aLEFTPQ} \,=\, & (\partial a)^2 u_L u_R + (\partial a)^2 u_L^{\dagger} u_R^{\dagger} + (\partial a)^2 d_L d_R + (\partial a)^2 d_L^{\dagger} d_R^{\dagger} + (\partial a)^2 \nu_L^2 + (\partial a)^2 \nu_L^{\dagger2}\nonumber\\
& + (\partial a)^2 e_L e_R + (\partial a)^2 e_L^{\dagger} e_R^{\dagger} + \partial a\, u_L u_L^{\dagger} F_L + \partial a\, u_L u_L^{\dagger} F_R + \partial a\, u_L u_L^{\dagger} G_L \\
& + \partial a\, u_L u_L^{\dagger} G_R + \partial a\, u_R u_R^{\dagger} F_L + \partial a\, u_R u_R^{\dagger} F_R + \partial a\, u_R u_R^{\dagger} G_L + \partial a\, u_R u_R^{\dagger} G_R \nonumber\\
& + \partial a\, d_L d_L^{\dagger} F_L + \partial a\, d_R d_R^{\dagger} F_L + \partial a\, d_L d_L^{\dagger} F_R + \partial a\, d_R d_R^{\dagger} F_R + \partial a\, d_L d_L^{\dagger} G_L \nonumber\\
& + \partial a\, d_R d_R^{\dagger} G_L + \partial a\, d_L d_L^{\dagger} G_R + \partial a\, d_R d_R^{\dagger} G_R + \partial a\, \nu_L \nu_L^{\dagger} F_L + \partial a\, \nu_L \nu_L^{\dagger} F_R \nonumber\\
& + \partial a\, e_L e_L^{\dagger} F_L + \partial a\, e_R e_R^{\dagger} F_L + \partial a\, e_L e_L^{\dagger} F_R + \partial a\, e_R e_R^{\dagger} F_R \, ,\nonumber
\end{align}
and more results can be found in App.~\ref{app:HSaLEFT} and in the ancillary Mathematica notebook.

As for the SMEFT extended with an axion, it is mostly straightforward to identify the independent Lorentz and gauge invariant operator structures from the spurions in the Hilbert series. Due to the increased number of independent fermions below the electroweak scale the number of operators increases quickly in the aLEFT. In particular, it can be challenging to identify the many 4-fermion operators coupled to $\partial a$ at dimension~8. Using Fierz identities from Refs.~\cite{Liao:2012uj,Liao:2016qyd}, one can show that the operators in Tab.~\ref{tab:aLEFT_SS_dim8} are the only non-redundant ones.

Something else that one has to keep in mind is that the flavor symmetries below the EW scale change with respect to what we have discussed before as the left-handed fermions no longer come together in doublets. Therefore, all mass terms can be diagonalized and the misalignment captured by the Cabibbo--Kobayashi--Maskawa~(CKM) matrix that one finds above the EW scale is shifted to dimension~6 in the $1/m_W$ expansion allowing for more possible rephasings for the quarks. Instead of only applying baryon number rephasings on the quark fields, we can rephase each flavor of quarks by themselves. In the lepton sector, the charged leptons keep their lepton family number rephasing properties while the neutrinos do not benefit from any flavor symmetries due to the lepton number breaking term $\bar{\nu}_L\nu_L^c + \hc$ that we allow for in the most generic Lagrangian. In total we expect $N_u+N_d+N_e$ conserved currents that enable us to remove operators via imposing $\partial_{\mu} j^{\mu} = 0$ after integrating by parts. As in the aSMEFT, we have to impose these conditions by hand after calculating the Hilbert series. As discussed in the last section, we have to add $aF\wtilde{F},aG\wtilde{G}$ to the operator basis by hand as well.

Our complete basis up to mass dimension~8 for an ALP coupled derivatively to all particles in the SM below the EW scale can be found in Tabs.~\ref{tab:aLEFT_SS_dim5},~\ref{tab:aLEFT_SS_dim7} and~\ref{tab:aLEFT_SS_dim8} in App.~\ref{app:aLEFTSSOpBasis}. The operator basis at mass dimension~5 in the aLEFT is consistent with the operators used in Ref.~\cite{Bauer:2020jbp}. As before, we can use the ALP-independent terms in the Hilbert series as a sanity check for our implementation of the Hilbert series and compare them with the known results for the operator bases up to dimension~8 in the LEFT~\cite{Jenkins:2017jig,Liao:2020zyx,Murphy:2020cly}.

\begin{figure}[t]
  \includegraphics[width=\linewidth]{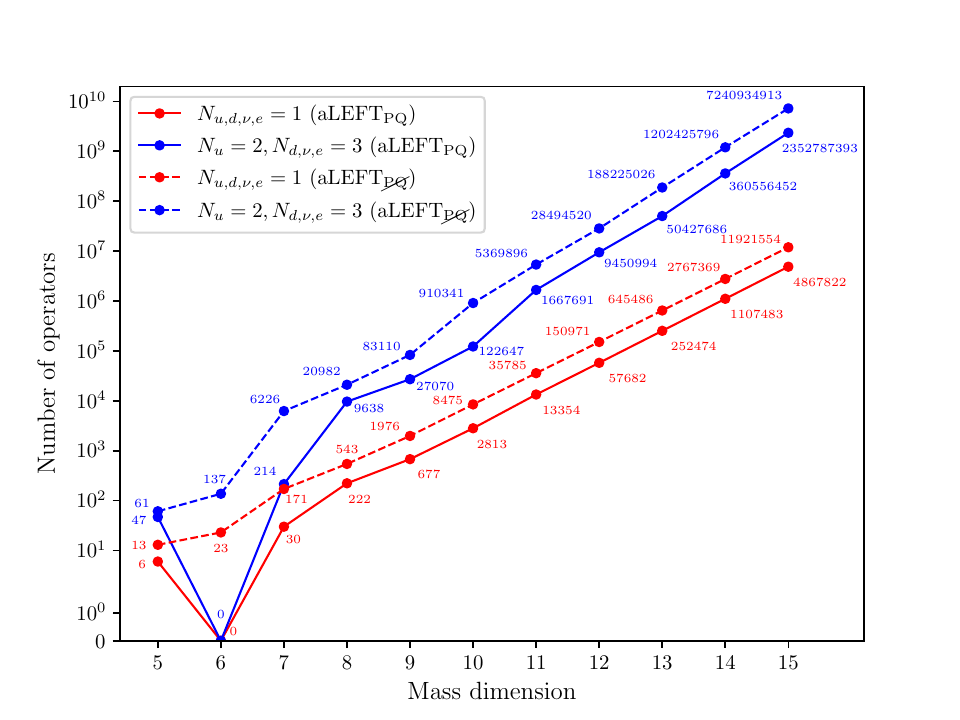}
  \caption{The number of operators in the aLEFT with and without a shift symmetry for the ALP plotted against the mass dimension for $N_u=N_d=N_{\nu}=N_e=1$ and $N_u=2,N_d=N_{\nu}=N_e=3$ number of flavors. Note that the y-axis has a linear scaling between 0 and 1 to accommodate for the 0 at mass dimension~6.}
\label{fig:aLEFT_NoOpsVsMassDim}
\end{figure}

With the Hilbert series calculated, we will now repeat the analysis of the flavor dependence we performed for the aSMEFT. As before, we count the number of independent operators for generic number of flavors using lepton and baryon number rephasings to single out the lepton and baryon number breaking operators. Using $N_{\nu}=N_e = N_d$\footnote{We keep $N_u$ independent here, since we want to take the limit $N_{\nu}=N_e = N_d=3$, $N_u=2$ later amounting to the usual flavor content of the EFT below the EW scale.} to keep the expressions more concise, we find
\begin{equation}
\begin{split}
 \#\,\cO_{5}^{\aLEFTPQ} \,=\,& 2-2N_d+5 N_d^2-N_u+2 N_u^2 \, , \\
 \#\,\cO_{6}^{\aLEFTPQ} \,=\,& 0 \, , \\
 \#\,\cO_{7}^{\aLEFTPQ} \,=\,& \left(18 N_d^2+10 N_u^2\right)+\left(N_d+N_d^2\right) \epsilon_L^2 \, , \\
 \#\,\cO_{8}^{\aLEFTPQ} \,=\,& \left(7+23 N_d^2+36 N_d^4+16 N_d^3 N_u+14 N_u^2+52 N_d^2 N_u^2+8 N_u^4\right)\\
 &+\left(-\frac{4 N_d^2}{3}+\frac{16 N_d^4}{3}+16 N_d^3 N_u+16 N_d^2 N_u^2\right) \epsilon_B \epsilon_L\\
 &+\left(-N_d+\frac{7 N_d^2}{3}+\frac{26 N_d^4}{3}+16 N_d^3 N_u+4 N_d^2 N_u^2\right) \epsilon_L^2 \, .
\end{split}
\end{equation}

As before, shift symmetry protects from lepton number breaking. The lepton number breaking, gauge-invariant neutrino mass term $\bar{\nu}_L^c \nu_L + \hc$ that can be written down below the EW scale allows for lepton number breaking already at dimension~7.

\subsection{\texorpdfstring{$\aLEFTnotPQ$}{aLEFTnotPQ}}\label{sec:aLEFT_nonSS}

We will now proceed by calculating the Hilbert series for the LEFT extended with a generic scalar field $a$ that can but no longer necessarily has to be connected to the spontaneous breaking of a PQ symmetry. The first two terms in the expansion of the Hilbert series in the mass dimension of the operators are given by
\begin{equation}
\begin{split}
\cH_{5}^{\aLEFTnotPQ} \,=\, & a^5 + a^2 u_L u_R + a^2 u_L^{\dagger} u_R^{\dagger} + a^2 d_L d_R + a^2 d_L^{\dagger} d_R^{\dagger} + a^2 \nu_L^2 + a^2 \nu_L^{\dagger2} + a^2 e_L e_R \\
& + a^2 e_L^{\dagger} e_R^{\dagger} + a F_L^2 + a F_R^2 + a G_L^2 + a G_R^2 \, , \\
\cH_{6}^{\aLEFTnotPQ} \,=\, & a^6 + a^3 u_L u_R + a^3 u_L^{\dagger} u_R^{\dagger} + a^3 d_L d_R + a^3 d_L^{\dagger} d_R^{\dagger} + a^3 \nu_L^2 + a^3 \nu_L^{\dagger2} + a^3 e_L e_R \\
& + a^3 e_L^{\dagger} e_R^{\dagger} + a^2 F_L^2 + a^2 F_R^2 + a^2 G_L^2 + a^2 G_R^2 + a u_L u_R F_L + a u_L u_R G_L\\
& + a u_L^{\dagger} u_R^{\dagger} F_R  + a u_L^{\dagger} u_R^{\dagger} G_R + a d_L d_R F_L + a d_L^{\dagger} d_R^{\dagger} F_R + a d_L d_R G_L\\
& + a d_L^{\dagger} d_R^{\dagger} G_R + a e_L e_R F_L + a e_L^{\dagger} e_R^{\dagger} F_R \, . \\
\end{split}
\end{equation}
From these, one can already see again that the same structure appears as for the SMEFT. Firstly, due to the ALP being a singlet pseudoscalar operators can simply be built by multiplying operators at the previous mass dimension with $a$. Secondly, beyond dimension~5 we find a similar PQ-breaking isolation condition that we have found for the aSMEFT
\begin{tcolorbox}[top=2mm,bottom=3mm]
\begin{equation}
\label{eq:separation_aLEFT}
    \cH_{n}^{\aLEFTnotPQ} = a \, \cH_{n-1}^{\aLEFTnotPQ} + a \, \cH_{n-1}^{\text{LEFT}} + \cH_{n}^{\aLEFTPQ} \( \partial a \to a \cD \) \, 
\end{equation}
\end{tcolorbox}
\noindent Therefore the same discussion about shift symmetry that we will present in the following section also applies to the aLEFT. 

In the aLEFT, there exists a subtlety in how to count the couplings of a single ALP to the fermion mass terms. This can be seen by studying the renormalizable part of the Lagrangian 
\begin{equation} \label{eq:aLEFTnonPQreno}
\begin{split}
    \cL_{\leq4}^{a} = & \frac{1}{2} \partial_{\mu} a \partial^{\mu} a - \frac{m_{a,0}^2}{2} a^2 + C_{a^3} a^3 + C_{a^4} a^4 + \frac{a}{f} \( \bar{u}_L C_{au}^{SR} u_R + \bar{d}_L C_{ad}^{SR} d_R + \bar{e}_L C_{ae}^{SR} e_R \right. \\
    & \left. + \bar{\nu}_L C_{a\nu}^{SR} \nu_L^c +\hc \) \, ,
\end{split}
\end{equation}
that we have neglected up to this point. Here, one can see that the operators that couple one ALP field to the fermions naively already appear at dimension~4, instead of dimension~5 in the SMEFT analysis. If one however performs a matching of the aSMEFT to the aLEFT one obtains
\begin{equation}
   \{ C_{au}^{SR}, C_{ad}^{SR}, C_{ae}^{SR} \} = \frac{v}{\sqrt{2}} \{C_{au}, C_{ad}, C_{ae} \} \quad\text{and}\quad C_{a\nu}^{SR} = \frac{v^2}{2} C_{aLH}
\end{equation}
after expanding the Higgs around its vacuum expectation value $v$ and the operators can be identified with dimension-5 and dimension-6 operators in the aSMEFT.

The operator basis at dimension~5 can be found in Tab.~\ref{tab:aLEFT_nonSS_dim5}. Based on the shift-breaking isolation condition in Eq.~\eqref{eq:separation_aLEFT}, the higher dimensional operator bases can be constructed easily, see App.~\ref{app:aLEFTnonSSOpBasis} for details.

We can once more compare the numbers in Fig.~\ref{fig:aLEFT_NoOpsVsMassDim} to see if we can understand the results in terms of the invariants from Ref.~\cite{Bonnefoy:2022rik}. As in the dimension-5 aSMEFT, the derivatively coupled ALP interactions with the fermions become redundant in the presence of the dimension-5 ALP Yukawa couplings. We have just discussed that the Yukawa ALP operators already naively appear at mass dimension~4 in the aLEFT, so we have to be careful in our discussion. Looking at Tab.~\ref{tab:aLEFT_nonSS_dim5}, we can see that the same number of ALP-fermion interactions appear at dimension~4 and dimension~5 in the aLEFT because one operator can just be obtained by multiplying the other one by $a$. Hence, we can rely on the numbers at dimension~5 to understand the counting. 

The difference at dimension~5 in Fig.~\ref{fig:aLEFT_NoOpsVsMassDim} can be explained as follows. According to Ref.~\cite{Bonnefoy:2022rik}, the 13 conditions for shift invariance reduce to 8 invariants below the electroweak scale at leading order. The remaining 4 invariants that are present due to the correlations induced by the left-handed quark doublet get shifted to higher mass dimensions in the $1/m_W^2$ expansion after integrating out the $W$ from the aSMEFT\footnote{If one starts from a HEFT-like scenario, these correlations will not be there in the first place which is both captured in the aLEFT in higher-dimensional operators upon matching to a HEFT-/SMEFT-like scenario.} and one more invariant is removed because the top quark is integrated out. With respect to Ref.~\cite{Bonnefoy:2022rik} we also allow for lepton number breaking here which allows for a neutrino mass term. This implies that there are an additional 3 relations in the fermionic sector.
In total, we have to subtract the 11 conditions obtained from the fermionic sector at dimension~4 (which are counted in the same way at dimension~5), the 1 condition obtained from removing the operator $a^5$ and the 2 conditions obtained from removing $aFF$ and $aGG$ from the 61 terms at dimension~5 which yields exactly the 47 terms in Fig.~\ref{fig:aLEFT_NoOpsVsMassDim}.

We have also once more performed the counting of operators for each mass dimension by setting all spurions to unity and applying the same procedure as before to single out the lepton and baryon number violating terms. We find
\begin{align*}
\label{eq:nOaLEFTnotPQ}
\#\,\cO_{5}^{\aLEFTnotPQ} \,=\,& \left(5+4 N_d^2+2 N_u^2\right)+\left(N_d+N_d^2\right) \epsilon_L^2 \, , \\
\#\,\cO_{6}^{\aLEFTnotPQ} \,=\,& \left(5+10 N_d^2+6 N_u^2\right)+2 N_d^2 \epsilon_L^2 \, , \\
\#\,\cO_{7}^{\aLEFTnotPQ} \,=\,& \left(7+\frac{131 N_d^2}{4}+\frac{3 N_d^3}{2}+\frac{87 N_d^4}{4}+10 N_d^3 N_u+19 N_u^2+32 N_d^2 N_u^2+5 N_u^4\right)\\
&+\left(-\frac{4 N_d^2}{3}-2 N_d^3+\frac{10 N_d^4}{3}-4 N_d^2 N_u+10 N_d^3 N_u+10 N_d^2 N_u^2\right) \epsilon_B \epsilon_L\\
&+\left(N_d+3 N_d^2+2 N_d^3+6 N_d^4+10 N_d^3 N_u+N_d N_u^2+3 N_d^2 N_u^2\right) \epsilon_L^2\\
&+\left(-\frac{N_d^2}{6}+\frac{N_d^4}{6}\right) \epsilon_L^4 \, , \numberthis\\
\#\,\cO_{8}^{\aLEFTnotPQ} \,=\,&\left(14+\frac{335 N_d^2}{4}-\frac{N_d^3}{2}+\frac{303 N_d^4}{4}+34 N_d^3 N_u+53 N_u^2+110 N_d^2 N_u^2+17 N_u^4\right)\\
&+\left(-\frac{4 N_d^2}{3}+2 N_d^3+\frac{34 N_d^4}{3}+4 N_d^2 N_u+34 N_d^3 N_u+34 N_d^2 N_u^2\right) \epsilon_B \epsilon_L\\
&+\left(4 N_d+10 N_d^2-3 N_d^3+19 N_d^4+34 N_d^3 N_u-N_d N_u^2+9 N_d^2 N_u^2\right) \epsilon_L^2\\
&+\left(-\frac{N_d^2}{6}+\frac{N_d^4}{6}\right) \epsilon_L^4 \, .
\end{align*}
Again, we only show the leading results here and the remaining results with full spurion and flavor dependence can be found in the ancillary notebook.
Note that due to the operator $a \bar{\nu}\nu_L^c + \hc$, a lepton-number violating term can already be written down at dimension~4 (it has the same dependence on the number of flavors as the corresponding term at mass dimension~5 quoted at the end of $\#\,\cO_{5}^{\aLEFTnotPQ}$ in Eq.~\eqref{eq:nOaLEFTnotPQ}), which is the operator that captures the effects of the derivatively coupled operator $\partial_{\mu} a \, \bar{\nu}_L \gamma^{\mu} \nu_L$ that does not violate lepton number. This only makes sense if the coefficient of $a \bar{\nu}\nu_L^c + \hc$ is proportional to the renormalizable spurion of lepton number breaking $m_{\nu}$ which is indeed the case as one can check from the usual relations one expects at dimension~5.

\subsection{CP violation in the aLEFT} \label{sec:aLEFTCPV}

{\arraycolsep=4pt
\begin{table}[ht!]
    \centering
\scalebox{0.95}{
$\begin{array}{|c|ccc|ccc|}
\hline
 \multirow{ 2}{*}{\text{Dim.}} & \multicolumn{3}{c|}{\aLEFTPQ} & \multicolumn{3}{c|}{\aLEFTnotPQ} \\ \cline{2-7}
  & \text{CP-even} & \text{CP-odd} & \text{CP-violating} & \text{CP-even} & \text{CP-odd} & \text{CP-violating} \\ \hline
\multirow{1.6}{*}{5} & 6 & 0 & 0 & 6 & 7 & 7 \\[-2.5mm]
 & 30 & 17 & 3 & 30 & 31 & 17 \\
\multirow{1.6}{*}{6} & 0 & 0 & 0 & 12 & 11 & 11 \\[-2.5mm]
 & 0 & 0 & 0 & 69 & 68 & 32 \\
\multirow{1.6}{*}{7} & 15 & 15 & 15 & 68 & 103 & 85 \\[-2.5mm]
 & 107 & 107 & 49 & 2995 & 3231 & 634 \\
\multirow{1.6}{*}{8} & 116 & 106 & 72 & 294 & 249 & 173 \\[-2.5mm]
 & 4830 & 4808 & 698 & 10620 & 10362 & 1467 \\
\multirow{1.6}{*}{9} & 370 & 307 & 205 & 951 & 1025 & 709 \\[-2.5mm]
 & 13691 & 13379 & 1860 & 41320 & 41790 & 6120 \\
\multirow{1.6}{*}{10} & 1444 & 1369 & 901 & 4312 & 4163 & 2521 \\[-2.5mm]
 & 61565 & 61082 & 8224 & 455647 & 454694 & 33450 \\
\multirow{1.6}{*}{11} & 6836 & 6518 & 3759 & 17727 & 18058 & 10168 \\[-2.5mm]
 & 836128 & 831563 & 53634 & 2683815 & 2686081 & 163719 \\
\multirow{1.6}{*}{12} & 28965 & 28717 & 15483 & 75775 & 75196 & 38924 \\[-2.5mm]
 & 4726245 & 4724749 & 271917 & 14249141 & 14245379 & 763605 \\
\multirow{1.6}{*}{13} & 126851 & 125623 & 63572 & 321876 & 323610 & 158051 \\[-2.5mm]
 & 25222133 & 25205553 & 1305402 & 94093443 & 94131583 & 3848880 \\
\multirow{1.6}{*}{14} & 554379 & 553104 & 262485 & 1385189 & 1382180 & 630296 \\[-2.5mm]
 & 180283648 & 180272804 & 6861666 & 601237390 & 601188406 & 19339749 \\
\multirow{1.6}{*}{15} & 2436838 & 2430984 & 1084823 & 5956959 & 5964595 & 2569894 \\[-2.5mm]
 & 1176447813 & 1176339580 & 35693696 & 3620363967 & 3620570946 & 98145863 \\ \hline
\end{array}$}
\caption{Number of CP-even, CP-odd and CP-violating operators for $\aLEFTPQ$~(left) and $\aLEFTnotPQ$~(right) from dimension 5 to 15. In each dimension, the two rows correspond to $N_{u,d,e,\nu}=1$ and $N_{u}=2,\,N_{d,e,\nu}=3$ respectively.}
\label{tab:aLEFT_CP_counting}
\end{table}
}

In this section we will discuss CP transformations in the $\aLEFT$. The same general discussion as for the aSMEFT in Section~\ref{sec:aSMEFTCPV} applies. As discussed before, in the $\aLEFT$ all fermions are independent fields and are no longer subject to correlations through linear electroweak symmetry breaking. This gives rise to a larger flavor symmetry group and all mass terms can be fully diagonalized. As a consequence, the Lagrangian is invariant under a larger group of rephasings, which for the $\aLEFT$ is $U(1)_{e_i}^3 \times U(1)_{u_i}^2 \times U(1)_{d_i}^3$. Keeping this in mind, we find the following Hilbert series counting the CP-violating couplings in the $\aLEFTnotPQ$ 
\begin{equation} \label{eq:aLEFT_HSCPV_nonSS}
\cH_{5,\text{CPV}}^{\aLEFTnotPQ}=a^5+a F^2+a G^2+3 a^2 e_L e_R+6 a^2 \nu _L^2+2 a^2 u_L u_R+3 a^2 d_L d_R\,.
\end{equation}
The counting of CP-even, CP-odd and CP-violating couplings in the aLEFT up to mass dimension 15 can be found in Tab.~\ref{tab:aLEFT_CP_counting}. 

The number of primary CP-odd invariants decreases in the quark sector and we find the following three invariants in the $\aLEFTnotPQ$ at dimension 4\footnote{If we set $N_{u}=2$ and $N_{d,e,\nu}=3$, there will be only two non-redundant flavor invariants for $f=u$, and in total $2+3\times 3=11$ flavor invariants for fermions $f=u,d,e,\nu$.}
\begin{equation} \label{eq:aLEFT_CPoddInvs}
    \Re\Tr\(C_{af}^{SR} m_f^\dagger\), \quad \Re\Tr\(X_f C_{af}^{SR} m_f^\dagger\), \quad \Re\Tr\( X_f^2 C_{af}^{SR} m_f^\dagger\)
\end{equation}
for each type of fermion $f=u,d,e,\nu$. Here, $X_{u,d,e,\nu} = m_{u,d,e,\nu} m_{u,d,e,\nu}^\dagger$. 

As we have discussed in the last section, the leading ALP-fermion interactions will move to dimension 4 in the aLEFT that can be matched to the dimension-5 interactions in the aSMEFT. We can still use the flavor invariants at dimension 4 to check the results for the dimension-5 ALP-Yukawa operators because they have the same structure in flavor space. We just have to keep track of the CP properties which is different for $a$ and $a^2$ multiplying the fermion bilinear. However, since completely generic as well as symmetric coupling matrices have the same amount of CP-even and CP-odd parameters, the numbers do not change here.

The 11 CP-odd flavor invariants together with the 3 CP-odd bosonic operators at leading order give 14 CP violating parameters that can appear in observables at the leading order in the EFT expansion. Comparing this to the expression in Eq.~\eqref{eq:aLEFT_HSCPV_nonSS}, we find what looks like a mismatch between our counting with flavor invariants and the Hilbert series in the neutrino sector. However, one has to keep in mind that the Hilbert series counts all CP-violating couplings, i.e. all couplings that are CP-odd and cannot possibly be removed by rephasings of the fermion fields. The flavor invariants capture all physical degrees of freedom that can interfere with the SM at leading order. These numbers agree if there is a CP-odd rephasing invariant of the Wilson coefficient corresponding to the flavor invariant at the same order in the EFT power counting. 

The Wilson coefficients of the electrons, for instance, allows for the following CP-odd rephasing invariants $C_{ae,ii}$ at leading order in the EFT corresponding to the flavor invariants $\Re\Tr(X_e^{0,1,2}C_{ae}^{SR} m_e^\dagger)$ that capture the interference of the EFT and the SM. Due to the Majorana nature of the neutrinos, no rephasing invariant exists at leading order in the EFT that only contains the Wilson coefficient of the effective ALP-neutrino operator. Only after using the spurious transformation of the neutrino mass term under rephasings of the neutrino fields, one can build a rephasing invariant quantity which are exactly the flavor invariant shown in Eq.~\eqref{eq:aLEFT_CPoddInvs}. This is not captured by the Hilbert series, as it only counts the number of parameters for each effective operator which cannot possibly be removed by a rephasing. Once the difference between the number of CP-violating parameters (6) and those parameters that can interfere with the SM (3) is taken into account, the numbers in Tab.~\ref{tab:aLEFT_CP_counting} and the counting using the flavor invariants matches again. For operators at higher mass dimensions, similar consideration should be taken for the neutrino-coupled operators.

Turning to the $\aLEFTPQ$, we find
\begin{equation}\label{eq:aLEFT_HSCPV_SS}
    \cH_{5,\text{CPV}}^{\aLEFTPQ} = 3 \partial a \, \nu_L^2 \,.
\end{equation}
All CP-odd bosonic operators are forbidden by the shift symmetry. In the fermion sector all couplings are hermitian matrices whose phases can not interfere with the renormalizable part of the Lagrangian due to the lack of a parameter that is charged under rephasings below the electroweak scale. Therefore, from our flavor invariant analysis we expect no CP-violating parameters that can interfere with the renormalizable part of the Lagrangian. This is compatible with Eq.~\eqref{eq:aLEFT_HSCPV_SS} if the neutrinos are properly taken into account as we just discussed for the $\aLEFTnotPQ$.

\section{Conclusions}\label{sec:Conclusions}
In this paper, we have investigated how shift-breaking effects in ALP EFTs above and below the electroweak scale can be captured beyond the leading interactions at dimension~5. For that we have constructed operator bases up to dimension~8 using the Hilbert series and presented the results for the Hilbert series up to dimension~15, making it possible to extend our operator bases to higher mass dimensions if needed. We provide the full results of the Hilbert series in an ancillary file including the full $N_f$ dependence. These results can be used to perform the operator counting for any specific $N_f$. In addition, we have identified the operators associated with violation of baryon and lepton number in the full expression of the Hilbert series. Furthermore, the CP violation effects are also discussed in the Hilbert series framework, the CP-even, CP-odd and CP-violating Hilbert series are given up to dimension~15 in the ancillary file. The Hilbert series is calculated using our own Mathematica code, which is specifically designed to address general problems. This code will be made publicly available in an upcoming publication~\cite{Grojean:2023}, allowing researchers to use it for a wide range of applications beyond the scope of this study.

At the level of the Hilbert series, we have found what we call the PQ-breaking isolation condition stating that beyond mass dimension~5 the operators describing shift-breaking couplings of the ALP to the SM are clearly isolated from the shift-preserving couplings. This is in stark contrast with the dimension-5 interactions where the two sectors mix due to the EOM redundancy that relate the derivatively coupled operators with fermions to the ALP-Yukawa couplings. We have discussed how to properly take care of the EOM redundancy at dimension-5 when higher order operators are considered in the EFT. Then, more relations beyond the well-known relations at dimension-5 should be imposed when going from the derivatively coupled basis to the Yukawa basis. We have constructed those relations and show them explicitly up to dimension-8.

There are several ways in which one could proceed with our results. First, one can study how the higher-dimensional shift-symmetric operators influence phenomenology. Since at dimension~5 the ALP already receives couplings to all SM particles except the Higgs, and at dimension~7 the ALP is coupled to all SM particles, we do not expect large corrections. The only exception could be specific channels which do not get a contribution in the EFT at lower mass order or which rely on intermediate particles from the SM implying that their amplitudes do not grow (as fast) with energy as a pure contact interaction. Another interesting possibility would be to study the interplay of shift-symmetric and shift-breaking operators if the scale of explicit PQ-breaking is not much larger than the scale of spontaneous breaking. In regards to low-energy experiments, our complete basis for the LEFT extended with an ALP should prove helpful to perform analyses beyond the leading order (c.f. for instance Ref. \cite{Bauer:2020jbp,Bauer:2021mvw}). With the full operator basis at dimension~8 one could study the implications of positivity on ALP EFTs, extending the analysis of Ref.~\cite{Kim:2023pwf} in the scalar sector. Finally, with a complete basis one could extend the efforts of Refs.~\cite{Chala:2020wvs,Bauer:2020jbp,DasBakshi:2023lca} and calculate the renormalization group equations of operators of higher mass dimension and their contributions to the renormalization group equations of operators at dimension~5.\\

\section*{Acknowledgments}
We thank Enrico Bertuzzo and Gabriel Massoni Salla for enlightening discussions on the axion scattering amplitudes that give an alternative way to obtain a basis of the axion interactions to the SM particles. We thank Guilherme Guedes for discussions on operator basis redundancies and Pham Ngoc Hoa Vuong for discussions on shift-breaking effects in ALP EFTs. Furthermore, we thank Emanuele Gendy for collaboration at early stages of this project and Quentin Bonnefoy for numerous discussions on this topic and comments on the manuscript. We thank Katharina Albrecht for cross-checking some of the results in this paper.

This work is supported by the Deutsche Forschungsgemeinschaft under Germany’s Excellence Strategy EXC 2121 “Quantum Universe” -- 390833306, as well as by the grant 491245950. This project also has received funding from the European Union’s Horizon Europe research and innovation programme under the Marie Skłodowska-Curie Staff Exchange grant agreement No 101086085 - ASYMMETRY. This research was supported in part by Perimeter Institute for Theoretical Physics. Research at Perimeter Institute is supported by the Government of Canada through the Department of Innovation, Science and Economic Development and by the Province of Ontario through the Ministry of Research and Innovation. C.Y.Y. is supported in part by the Grants No. NSFC-11975130, No. NSFC-12035008, No. NSFC-12047533, the National Key Research and Development Program of China under Grant No. 2017YFA0402200, the China Postdoctoral Science Foundation under Grant No. 2018M641621 and the Helmholtz-OCPC International Postdoctoral Exchange Fellowship Program.

\section*{Note added}
During the final stages of this project, the papers \cite{Song:2023lxf, Song:2023jqm} appeared on the arXiv. There, the EFT describing the interactions of a pseudoscalar, with and without a shift symmetry,  with the SM particles above the electroweak symmetry breaking scale was also considered. The Young tensor methods was used to find the Lorentz and SM gauge group singlets, which is an orthogonal approach to the Hilbert series methods we have used. 
The main result of our work is the identification of the particular structure of the shift-symmetry of an ALP in the EFT picture. We have cross-checked our results with those in Refs.~\cite{Song:2023lxf, Song:2023jqm}, and find agreement for the number of operators for $N_f = 1,3$, and the bases of operators seem to be equivalent after applying appropriate transformations. 

\clearpage
\appendix

\section{Operator basis for the aSMEFT up to mass dimension~8}
\label{app:aSMEFTOpBasis}

Using the Hilbert series as a guide, we have constructed independent operator bases for aSMEFT, encompassing dimensions up to 8, for both shift-symmetric and non-shift-symmetric theories, which are shown explicitly in the following two subsections.

\subsection{With shift symmetry} \label{app:aSMEFTSSOpBasis}

For aSMEFT with a shift symmetry, the operator bases from dimension~5 to dimension~8 are constructed, they are grouped in Tabs.~\ref{tab:aSMEFT_SS_dim5},~\ref{tab:aSMEFT_SS_dim6},~\ref{tab:aSMEFT_SS_dim7} and~\ref{tab:aSMEFT_SS_dim8} respectively. The operators are classified according to the reduced Hilbert series, we identify fermion with $\psi$, field strength with $X$, and scalar with $H$, and also taking $\partial a$ as a building block, the Hilbert series with $N_f=1$ for each mass dimension will be reduced as
\begin{equation}
\label{eq:HS_aSMEFT_reduce}
\begin{split}
\cH_{5}^{\PQ}\,=\,& 5 \partial a\, \psi^2 + 3 a X^2\,,\qquad
\cH_{6}^{\PQ}\,=\, (\partial a)^2 H^2\,,\\
\cH_{7}^{\PQ}\,=\,& 20 \partial a\, \psi^2 X + 4 \partial a\, X H^2 \cD + 7 \partial a\, \psi^2 H^2 + \partial a\, H^4 \cD + 12 \partial a\, \psi^2 H \cD\,,\\
\cH_{8}^{\PQ}\,=\,& (\partial a)^4 + 5 (\partial a)^2 \psi^2 \cD + 9 (\partial a)^2 X^2 + 4 \partial a\, \psi^4 + 2 (\partial a)^2 H^2 \cD^2 + 4 \partial a\, \psi^2 H^2 \cD \\
& + (\partial a)^2 H^4 + 6 (\partial a)^2 \psi^2 H + [2 \partial a\, \psi^4]\,,
\end{split}
\end{equation}
where for the Hilbert series at dimension~5, we have taken care of the caveats discussed in Section~\ref{sec:aSMEFT_SS}, i.e., the Higgs coupled term and the negative terms are removed, and the $a X^2$ terms are added by hand. As already mentioned in Section~\ref{sec:aSMEFT_SS}, setting $N_f=1$ will lead to vanishing terms. In order to construct an operator basis for general $N_f$, such vanishing terms should also be taken into account, which are given as additional terms in the brackets in Eq.~\eqref{eq:HS_aSMEFT_reduce} and the corresponding operators are marked with $(\star)$ in the tables of operator bases. For instance, the additional terms $[2 \partial a\, \psi^4]$ correspond to the operator $\cO_{\partial aed}+\hc$ in Tab.~\ref{tab:aSMEFT_SS_dim8}.

\begin{table}[h!]
\centering
\begin{tabular}{ |c|c|c|c| }
\hline
\multicolumn{2}{|c|}{$\partial a \, \psi^2$} & \multicolumn{2}{c|}{$a X^2$} \\
\hline
$\cO_{\partial aL}$ & $\partial_{\mu} a \, \(\bar{L} \gamma^{\mu} L\)$ & $\cO_{a\tilde{B}}$ & $a B_{\mu\nu} \wtilde{B}^{\mu\nu}$ \\ 
$\cO_{\partial ae}$ & $\partial_{\mu} a \, \(\bar{e} \gamma^{\mu} e\)$ & $\cO_{a\tilde{W}}$ & $a W_{\mu\nu}^I \wtilde{W}^{I,\mu\nu}$ \\
$\cO_{\partial aQ}$ & $\partial_{\mu} a \, \(\bar{Q} \gamma^{\mu} Q\)$ & $\cO_{a\tilde{G}}$ & $a G_{\mu\nu}^a \wtilde{G}^{a,\mu\nu}$  \\
$\cO_{\partial au}$ & $\partial_{\mu} a \, \(\bar{u} \gamma^{\mu} u\)$ & & \\
$\cO_{\partial ad}$ & $\partial_{\mu} a \, \(\bar{d} \gamma^{\mu} d\)$ & & \\
\hline
\end{tabular}
\caption{Operators in the aSMEFT at mass dimension~5 with $\partial a$ as a building block. Note that $\cO_{\partial aH} = \partial^{\mu} a \, \Bigl(H^{\dagger} i \overset{\leftrightarrow}{D}_{\hspace{-2pt}\mu} H \Bigr)$ is a redundant operator and can be removed via a global hypercharge transformation~\cite{Georgi:1986df,Bauer:2020jbp}. Imposing lepton and baryon number conservation at the level of the renormalizable Lagrangian, 3~(1) flavor diagonal entries of the operators coupling the ALP to leptons (quarks), for instance $\cO_{\partial aL,ii}$ and $\cO_{\partial aQ,11}$, can be removed~\cite{Bonilla:2021ufe}. Furthermore, we have used that the shift in the operators of class $a X^2$ can be removed using anomalous chiral transformations on the fermion fields making the operators shift-symmetric without an explicit derivative on the axion field.}
\label{tab:aSMEFT_SS_dim5}
\end{table}

\begin{table}[h!]
\centering
\begin{tabular}{ |c|c| }
\hline
\multicolumn{2}{|c|}{$(\partial a)^2 H^2$} \\
\hline
$\cO_{\partial a^2H^2}$ & $\partial_{\mu} a \, \partial^{\mu} a \, |H|^2$ \\ 
\hline
\end{tabular}
\caption{Operators in the aSMEFT at mass dimension~6 with $\partial a$ as a building block.}
\label{tab:aSMEFT_SS_dim6}
\end{table}

\begin{table}[h!]
\centering
\begin{tabular}{ |c|c|c|c| }
\hline
\multicolumn{2}{|c|}{$\partial a \, \psi^2 X$} & \multicolumn{2}{c|}{$\partial a \, X H^2 D$} \\
\hline
$\cO_{\partial aLB}$ & $\partial^{\mu} a \, \( \bar{L} \gamma^{\nu} L \) B_{\mu\nu}$ & $\cO_{\partial aHB}$ & $\partial_{\mu} a \, \Bigl(H^{\dagger} i \overleftrightarrow{D}_{\hspace{-2pt}\nu} H \Bigr) B^{\mu\nu}$ \\
$\cO_{\partial aL\tilde{B}}$ & $\partial^{\mu} a \, \( \bar{L} \gamma^{\nu} L \) \wtilde{B}_{\mu\nu}$ & $\cO_{\partial aH\tilde{B}}$ & $\partial_{\mu} a \, \Bigl(H^{\dagger} i \overleftrightarrow{D}_{\hspace{-2pt}\nu} H \Bigr) \wtilde{B}^{\mu\nu}$ \\
$\cO_{\partial aeB}$ & $\partial^{\mu} a \, \( \bar{e} \gamma^{\nu} e \) B_{\mu\nu}$ & $\cO_{\partial aHW}$ & $\partial_{\mu} a \, \Bigl(H^{\dagger} i \overleftrightarrow{D}_{\hspace{-2pt}\nu}^I H \Bigr) W^{I,\mu\nu}$ \\
$\cO_{\partial ae\tilde{B}}$ & $\partial^{\mu} a \, \( \bar{e} \gamma^{\nu} e \) \wtilde{B}_{\mu\nu}$ & $\cO_{\partial aH\tilde{W}}$ & $\partial_{\mu} a \, \Bigl(H^{\dagger} i \overleftrightarrow{D}_{\hspace{-2pt}\nu}^I H \Bigr) \wtilde{W}^{I,\mu\nu}$ \\ \cline{3-4}
$\cO_{\partial aQB}$ & $\partial^{\mu} a \, \( \bar{Q} \gamma^{\nu} Q \) B_{\mu\nu}$ & \multicolumn{2}{c|}{$\partial a \, \psi^2 H^2$} \\ \cline{3-4}
$\cO_{\partial aQ\tilde{B}}$ & $\partial^{\mu} a \, \( \bar{Q} \gamma^{\nu} Q \) \wtilde{B}_{\mu\nu}$ & $\cO_{\partial aLH^2}^{(1)}$ & $\partial_{\mu} a \, \(\bar{L} \gamma^{\mu} L\) |H|^2$ \\
$\cO_{\partial auB}$ & $\partial^{\mu} a \, \( \bar{u} \gamma^{\nu} u \) B_{\mu\nu}$ & $\cO_{\partial aLH^2}^{(2)}$ & $\partial_{\mu} a \, \(\bar{L} \gamma^{\mu} \sigma^I L \) \( H^{\dagger} \sigma^I H \)$ \\
$\cO_{\partial au\tilde{B}}$ & $\partial^{\mu} a \, \( \bar{u} \gamma^{\nu} u \) \wtilde{B}_{\mu\nu}$ & $\cO_{\partial aeH^2}$ & $\partial_{\mu} a \, \(\bar{e} \gamma^{\mu} e\) |H|^2$ \\
$\cO_{\partial adB}$ & $\partial^{\mu} a \, \( \bar{d} \gamma^{\nu} d \) B_{\mu\nu}$ & $\cO_{\partial aQH^2}^{(1)}$ & $\partial_{\mu} a \, \(\bar{Q} \gamma^{\mu} Q\) |H|^2$ \\
$\cO_{\partial ad\tilde{B}}$ & $\partial^{\mu} a \, \( \bar{d} \gamma^{\nu} d \) \wtilde{B}_{\mu\nu}$ & $\cO_{\partial aQH^2}^{(2)}$ & $\partial_{\mu} a \, \(\bar{Q} \gamma^{\mu} \sigma^I Q\) \( H^{\dagger} \sigma^I H \)$ \\
$\cO_{\partial aLW}$ & $\partial^{\mu} a \, \( \bar{L} \gamma^{\nu} \sigma^I L \) W_{\mu\nu}^I$ & $\cO_{\partial auH^2}$ & $\partial_{\mu} a \, \(\bar{u} \gamma^{\mu} u\) |H|^2$ \\
$\cO_{\partial aL\tilde{W}}$ & $\partial^{\mu} a \, \( \bar{L} \gamma^{\nu} \sigma^I L \) \wtilde{W}_{\mu\nu}^I$ & $\cO_{\partial adH^2}$ & $\partial_{\mu} a \, \(\bar{d} \gamma^{\mu} d \) |H|^2$ \\ \cline{3-4}
$\cO_{\partial aQW}$ & $\partial^{\mu} a \, \( \bar{Q} \gamma^{\nu} \sigma^I Q \) W_{\mu\nu}^I$ & \multicolumn{2}{c|}{$\partial a \, \psi^2 H D + \hc$} \\ \cline{3-4}
$\cO_{\partial aQ\tilde{W}}$ & $\partial^{\mu} a \, \( \bar{Q} \gamma^{\nu} \sigma^I Q \) \wtilde{W}_{\mu\nu}^I$ & $\cO_{\partial a eHD}^{(1)}$ & $\partial_{\mu}a \(D^{\mu} \bar{L} \)H e$ \\
$\cO_{\partial aQG}$ & $\partial^{\mu} a \, \( \bar{Q} \gamma^{\nu} T^a Q \) G_{\mu\nu}^a$ & $\cO_{\partial a eHD}^{(2)}$ & $\partial_{\mu}a \, \bar{L} H \(D^{\mu} e \)$ \\
$\cO_{\partial aQ\tilde{G}}$ & $\partial^{\mu} a \, \( \bar{Q} \gamma^{\nu} T^a Q \) \wtilde{G}_{\mu\nu}^a$ & $\cO_{\partial a uHD}^{(1)}$ & $\partial_{\mu}a \(D^{\mu} \bar{Q} \) \wtilde{H} u$ \\
$\cO_{\partial auG}$ & $\partial^{\mu} a \, \( \bar{u} \gamma^{\nu} T^a u \) G_{\mu\nu}^a$ & $\cO_{\partial a uHD}^{(2)}$ & $\partial_{\mu}a \, \bar{Q} \wtilde{H} \(D^{\mu} u \)$ \\
$\cO_{\partial au\tilde{G}}$ & $\partial^{\mu} a \, \( \bar{u} \gamma^{\nu} T^a u \) \wtilde{G}_{\mu\nu}^a$ & $\cO_{\partial a dHD}^{(1)}$ & $\partial_{\mu}a \(D^{\mu} \bar{Q} \)H d$ \\
$\cO_{\partial adG}$ & $\partial^{\mu} a \, \( \bar{d} \gamma^{\nu} T^a d \) G_{\mu\nu}^a$ & $\cO_{\partial a dHD}^{(2)}$ & $\partial_{\mu}a \, \bar{Q} H \(D^{\mu} d \)$ \\ \cline{3-4}
$\cO_{\partial ad\tilde{G}}$ & $\partial^{\mu} a \, \( \bar{d} \gamma^{\nu} T^a d \) \wtilde{G}_{\mu\nu}^a$ & \multicolumn{2}{c|}{$\partial a \, H^4 D$} \\  \cline{3-4}
 & & $\cO_{\partial a H^4}$ & $\partial^{\mu} a \, \Bigl(H^{\dagger} i \overleftrightarrow{D}_{\hspace{-2pt}\mu} H \Bigr) |H|^2$ \\ 
\hline
\end{tabular}
\caption{Operators in the aSMEFT at mass dimension~7 with $\partial a$ as a building block. }
\label{tab:aSMEFT_SS_dim7}
\end{table}

\begin{table}[h!]
\centering
\begin{tabular}{ |c|c|c|c| }
\hline
\multicolumn{2}{|c|}{$(\partial a)^2 X^2$} & \multicolumn{2}{c|}{$(\partial a)^2 \psi^2 D$} \\
\hline
$\cO_{\partial a^2B}^{(1)}$ & $\partial_{\mu}a\partial^{\mu}a \, B_{\nu\rho}B^{\nu\rho}$ & $\cO_{\partial a^2LD}$ & $\partial_{\mu}a\partial_{\nu}a \, \( \bar{L} \gamma^{\mu} \overleftrightarrow{D}^{\nu}L\)$ \\
$\cO_{\partial a^2B}^{(2)}$ & $\partial_{\mu}a\partial^{\nu}a \, B^{\mu\rho}B_{\nu\rho}$ & $\cO_{\partial a^2eD}$ & $\partial_{\mu}a\partial_{\nu}a \, \( \bar{e} \gamma^{\mu} \overleftrightarrow{D}^{\nu}e\)$ \\
$\cO_{\partial a^2\tilde{B}}$ & $\partial_{\mu}a\partial^{\mu}a \, B_{\nu\rho}\wtilde{B}^{\nu\rho}$ & $\cO_{\partial a^2QD}$ & $\partial_{\mu}a\partial_{\nu}a \, \( \bar{Q} \gamma^{\mu} \overleftrightarrow{D}^{\nu}Q\)$ \\ 
$\cO_{\partial a^2W}^{(1)}$ & $\partial_{\mu}a\partial^{\mu}a \, W_{\nu\rho}^I W^{I,\nu\rho}$ & $\cO_{\partial a^2uD}$ & $\partial_{\mu}a\partial_{\nu}a \, \( \bar{u} \gamma^{\mu} \overleftrightarrow{D}^{\nu}u\)$ \\ 
$\cO_{\partial a^2W}^{(2)}$ & $\partial_{\mu}a\partial^{\nu}a \, W^{I,\mu\rho} W_{\nu\rho}^I$ & $\cO_{\partial a^2dD}$ & $\partial_{\mu}a\partial_{\nu}a \, \( \bar{d} \gamma^{\mu} \overleftrightarrow{D}^{\nu}d\)$ \\ \cline{3-4}
$\cO_{\partial a^2\tilde{W}}^{(2)}$ & $\partial_{\mu}a\partial^{\mu}a \, W_{\nu\rho}^I \wtilde{W}^{I,\nu\rho}$ & \multicolumn{2}{c|}{$(\partial a)^2 \psi^2 H + \hc$} \\ \cline{3-4}
$\cO_{\partial a^2G}^{(1)}$ & $\partial_{\mu}a\partial^{\mu}a \, G_{\nu\rho}^a G^{a,\nu\rho}$ & $\cO_{\partial a^2 eH}$ & $\partial_{\mu}a\partial^{\mu}a \, \bar{L} H e$ \\ 
$\cO_{\partial a^2G}^{(2)}$ & $\partial_{\mu}a\partial^{\nu}a \, G^{a,\mu\rho} G_{\nu\rho}^a$ & $\cO_{\partial a^2 uH}$ & $\partial_{\mu}a\partial^{\mu}a \, \bar{Q} H u$ \\
$\cO_{\partial a^2\tilde{G}}$ & $\partial_{\mu}a\partial^{\mu}a \, G_{\nu\rho}^a \wtilde{G}^{a,\nu\rho}$ & $\cO_{\partial a^2 dH}$ & $\partial_{\mu}a\partial^{\mu}a \, \bar{Q} H d$ \\ \hline
 \multicolumn{2}{|c|}{$(\partial a)^4$} & \multicolumn{2}{c|}{$(\partial a)^2 H^2 D^2$} \\ \hline
$\cO_{\partial a^4}$ & $\partial_{\mu} a \partial^{\mu}a \partial_{\nu} a \partial^{\nu}a$ & $\cO_{\partial a^2DH^2}^{(1)}$ & $\partial_{\mu}a\partial^{\mu}a D_{\nu}H^{\dagger} D^{\nu}H$ \\ \cline{1-2}
\multicolumn{2}{|c|}{$(\partial a)^2 H^4$} & $\cO_{\partial a^2DH^2}^{(2)}$ & $\partial_{\mu}a\partial_{\nu}a D^{\mu}H^{\dagger} D^{\nu}H$  \\ \cline{1-2}
$\cO_{\partial a^2H^4}$ & $\partial_{\mu}a \partial^{\mu}a |H|^4$ & & \\ \hline
\multicolumn{4}{|c|}{$\slashed{B}$ and $\slashed{L}$ terms} \\ \hline
\multicolumn{2}{|c|}{$\partial a \, \psi^4 + \hc$} & \multicolumn{2}{c|}{$\partial a \, \psi^2 H^2 D + \hc$} \\ \hline
$\cO_{\partial a Ldu}$ & $\partial_{\mu} a \, \(\bar{L}^cL\) \( \bar{d} \gamma^{\mu} u \)$ & $\cO_{\partial a LHD}^{(1)}$ & $\partial_{\mu} a \, \(\bar{L}^c H \) \(\tilde{H}^{\dagger} D^{\mu} L \)$ \\
 $\cO_{\partial a LQd}$ & $\epsilon^{\alpha\beta\gamma} \partial_{\mu} a \, \(\bar{L} d_{\alpha}\) \( \bar{Q}_{\beta}^c \gamma^{\mu} d_{\gamma} \)$ & $\cO_{\partial a LHD}^{(2)}$ & $\partial_{\mu} a \, \(\bar{L}^c D^{\mu} H \) \(\tilde{H}^{\dagger} L \)$ \\
$\cO_{\partial a ed}~(\star)$ & $\epsilon^{\alpha\beta\gamma} \partial_{\mu} a \, \(\bar{d}_{\alpha}^c d_{\beta} \) \(\bar{e} \gamma^{\mu} d_{\gamma} \)$ & & \\
\hline
\end{tabular}
\caption{Operators in the aSMEFT at mass dimension~8 with $\partial a$ as a building block. Note that the operator $\cO_{\partial a ed}$ marked with $(\star)$ only exists for $N_f \neq 1$ because otherwise all contractions of the antisymmetric color structure will sum to zero (the first current in the operator is symmetric under $\alpha \leftrightarrow \beta$ for one generation of fermions).}
\label{tab:aSMEFT_SS_dim8}
\end{table}

\clearpage

\subsection{Without shift symmetry} \label{app:aSMEFTnonSSOpBasis}

For aSMEFT without a shift symmetry, we can use the shift-breaking isolation condition Eq.~\eqref{eq:separation} to construct the operator basis easily. The operator bases at higher dimensions can be constructed with
\begin{equation}
        \cL_n^{\notPQ} = a \, \cL_{n-1}^{\notPQ} + a \, \cL_{n-1}^{\text{SMEFT}} + \cL_n^{\PQ}
\end{equation}
for $n>5$. We start with the dimension-5 operator basis shown in Tab.~\ref{tab:aSMEFT_nonSS_dim5}, the dimension-6,~7, and 8 operator bases can be constructed successively. For instance, the operators at dimension~6 can be constructed with $\cL_{6}^{\notPQ} = a \, \cL_{5}^{\notPQ} + a \, \cL_{5}^{\text{SMEFT}} + \cL_{6}^{\PQ}$, where the operators in $\cL_{5}^{\text{SMEFT}}$ have been shown in Ref.~\cite{Weinberg:1979sa}, and the dimension-6 shift-symmetric operator basis associated with $\cL_{6}^{\PQ}$ is given in Tab.~\ref{tab:aSMEFT_SS_dim6}. The construction of operator bases of dimension~7 and 8 follows the same manner, and the SMEFT operator bases at dimension~6 and 7 are needed~\cite{Grzadkowski:2010es,Lehman:2014jma}. For completeness, the axion-dependent renormalizable operators can be found in Eq.~\eqref{eq:aSMEFTnonPQreno}.

\begin{table}[h!]
\centering
\begin{tabular}{ |c|c|c|c| }
\hline
\multicolumn{2}{|c|}{$V(a,H)$} & \multicolumn{2}{c|}{$a X^2$} \\
\hline
$\cO_{a^5}$ & $a^5$ & $\cO_{aB}$ & $a B_{\mu\nu} B^{\mu\nu}$ \\
$\cO_{a^3H^2}$ & $a^3 |H|^2$ & $\cO_{a\tilde{B}}$ & $a B_{\mu\nu} \wtilde{B}^{\mu\nu}$ \\
$\cO_{aH^4}$ & $a |H|^4$ & $\cO_{aW}$ & $a W_{\mu\nu}^I W^{I,\mu\nu}$ \\ \cline{1-2}
\multicolumn{2}{|c|}{$a \psi^2 H + \hc$} & $\cO_{a\tilde{W}}$ & $a W_{\mu\nu}^I \wtilde{W}^{I,\mu\nu}$ \\ \cline{1-2}
$\cO_{ae}$ & $a \bar{L} H e$ & $\cO_{aG}$ & $a G_{\mu\nu}^a G^{a,\mu\nu}$ \\
$\cO_{au}$ & $a \bar{Q} \tilde{H} u$ & $\cO_{a\tilde{G}}$ & $a G_{\mu\nu}^a \wtilde{G}^{a,\mu\nu}$ \\
$\cO_{ad}$ & $a \bar{Q} H d$ & & \\
\hline
\end{tabular}
\caption{Operators in the aSMEFT at mass dimension~5 with $a$ as a building block.}
\label{tab:aSMEFT_nonSS_dim5}
\end{table}

\section{Operator basis for the aLEFT up to mass dimension~8}
\label{app:aLEFTOpBasis}

\subsection{With shift symmetry} \label{app:aLEFTSSOpBasis}

By setting $N_{u,d,\nu,e}\to 1$ and restoring the vanishing terms, the reduced Hilbert series for operator basis up to dimension~8 are given by
\begin{align}
\cH_{5}^{\aLEFTPQ}\,=\,& 7 \partial a\, \psi^2 + 2 a X^2\,,\quad
\cH_{6}^{\aLEFTPQ}\,=\, 0\,,\quad
\cH_{7}^{\aLEFTPQ}\,=\, 8 (\partial a)^2 \psi^2 + 22\, \partial a\, \psi^2 X\,,\nonumber\\
\cH_{8}^{\aLEFTPQ}\,=\,& (\partial a)^4 + 7 (\partial a)^2 \psi^2 \cD + 6 (\partial a)^2 X^2 + 32 \partial a\, \psi^2 X \cD + 176 \partial a\, \psi^4 + [ 6 \partial a\, \psi^4]\,.
\end{align}
The dimension-5 and dimension-7 operator bases are given in Tab.~\ref{tab:aLEFT_SS_dim5} and Tab.~\ref{tab:aLEFT_SS_dim7} respectively, and dimension-8 operator basis is collected in Tabs.~\ref{tab:aLEFT_SS_dim8},~\ref{tab:aLEFT_SS_dim8_2},~\ref{tab:aLEFT_SS_dim8_3}. We can see that the numbers of operators precisely match the numbers present in the reduced Hilbert series.

\begin{table}[h!]
\centering
\begin{tabular}{ |c|c|c|c| }
\hline
\multicolumn{2}{|c|}{$\partial a \, \psi^2$} & \multicolumn{2}{c|}{$\partial a \, \psi^2$~(cont.)} \\ \hline
$\cO_{\partial ae}^{VL}$ & $\partial_{\mu} a \, \(\bar{e}_L \gamma^{\mu} e_L\)$ & $\cO_{\partial ad}^{VL}$ & $\partial_{\mu} a \, \(\bar{d}_L \gamma^{\mu} d_L\)$ \\
$\cO_{\partial ae}^{VR}$ & $\partial_{\mu} a \, \(\bar{e}_R \gamma^{\mu} e_R\)$ & $\cO_{\partial ad}^{VR}$ & $\partial^{\mu} a \, \(\bar{d}_R \gamma^{\mu} d_R\)$ \\ \cline{3-4}
$\cO_{\partial a\nu}^{VL}$ & $\partial_{\mu} a \, \(\bar{\nu}_L \gamma^{\mu} \nu_L\)$ & \multicolumn{2}{c|}{$a X^2$} \\ \cline{3-4}
$\cO_{\partial au}^{VL}$ & $\partial_{\mu} a \, \(\bar{u}_L \gamma^{\mu} u_L\)$ & $\cO_{a\tilde{F}}$ & $a F_{\mu\nu} \wtilde{F}^{\mu\nu}$ \\
$\cO_{\partial au}^{VR}$ & $\partial_{\mu} a \, \(\bar{u}_R \gamma^{\mu} u_R\)$ & $\cO_{a\tilde{G}}$ & $a G_{\mu\nu}^a \wtilde{G}^{a,\mu\nu}$ \\
\hline
\end{tabular}
\caption{Operators in the aLEFT at mass dimension~5 with $\partial a$ as a building block. Note that imposing lepton and baryon number conservation, 3 flavor diagonal entries of the operators coupling the ALP to leptons and quarks, for instance $\cO_{\partial ae,ii}^{VL}$, $\cO_{\partial au,ii}^{VL}$ and $\cO_{\partial ad,ii}^{VL}$, can be removed~\cite{Bonilla:2021ufe}. Furthermore, we have used that the shift in the operators of class $a X^2$ can be removed using anomalous chiral transformations on the fermion fields making the operators shift-symmetric without an explicit derivative on the axion field.}
\label{tab:aLEFT_SS_dim5}
\end{table}

\begin{table}[h!]
\centering
\begin{tabular}{ |c|c|c|c| }
\hline
\multicolumn{2}{|c|}{$(\partial a)^2 \psi^2 + \hc$} & \multicolumn{2}{c|}{$\partial a\, X \psi^2$~(cont.)} \\
\hline
$\cO_{\partial a^2e}^{SR}$ & $\partial_{\mu} a \partial^{\mu} a \, \(\bar{e}_L e_R\)$ & $\cO_{\partial au\tilde{F}}^{VR}$ & $\partial^{\mu} a \, \(\bar{u}_R \gamma^{\nu} u_R\) \wtilde{F}_{\mu\nu}$ \\
$\cO_{\partial a^2\nu}^{SR} \( \slashed{L} \)$ & $\partial_{\mu} a \partial^{\mu} a \, \(\bar{\nu}_L \nu_L^c\)$ & $\cO_{\partial adF}^{VL}$ & $\partial^{\mu} a \, \(\bar{d}_L \gamma^{\nu} d_L\) F_{\mu\nu}$ \\
$\cO_{\partial a^2u}^{SR}$ & $\partial_{\mu} a \partial^{\mu} a \, \(\bar{u}_L u_R\)$ & $\cO_{\partial adF}^{VR}$ & $\partial^{\mu} a \, \(\bar{d}_R \gamma^{\nu} d_R\) F_{\mu\nu}$ \\
$\cO_{\partial a^2d}^{SR}$ & $\partial_{\mu} a \partial^{\mu} a \, \(\bar{d}_L d_R\)$ & $\cO_{\partial ad\tilde{F}}^{VL}$ & $\partial^{\mu} a \, \(\bar{d}_L \gamma^{\nu} d_L\) \wtilde{F}_{\mu\nu}$ \\ \cline{1-2}
\multicolumn{2}{|c|}{$\partial a \, X \psi^2$} & $\cO_{\partial ad\tilde{F}}^{VR}$ & $\partial^{\mu} a \, \(\bar{d}_R \gamma^{\nu} d_R\) \wtilde{F}_{\mu\nu}$ \\ \cline{1-2}
$\cO_{\partial aeF}^{VL}$ & $\partial^{\mu} a \, \(\bar{e}_L \gamma^{\nu} e_L\) F_{\mu\nu}$ & $\cO_{\partial auG}^{VL}$ & $\partial^{\mu} a \, \(\bar{u}_L \gamma^{\nu} T^a u_L\) G_{\mu\nu}^a$ \\
$\cO_{\partial aeF}^{VR}$ & $\partial^{\mu} a \, \(\bar{e}_R \gamma^{\nu} e_R\) F_{\mu\nu}$ & $\cO_{\partial auG}^{VR}$ & $\partial^{\mu} a \, \(\bar{u}_R \gamma^{\nu} T^a u_R\) G_{\mu\nu}^a$ \\
$\cO_{\partial ae\tilde{F}}^{VL}$ & $\partial^{\mu} a \, \(\bar{e}_L \gamma^{\nu} e_L\) \wtilde{F}_{\mu\nu}$ & $\cO_{\partial au\tilde{G}}^{VL}$ & $\partial^{\mu} a \, \(\bar{u}_L \gamma^{\nu} T^a u_L\) \wtilde{G}_{\mu\nu}^a$ \\
$\cO_{\partial ae\tilde{F}}^{VR}$ & $\partial^{\mu} a \, \(\bar{e}_R \gamma^{\nu} e_R\) \wtilde{F}_{\mu\nu}$ & $\cO_{\partial au\tilde{G}}^{VR}$ & $\partial^{\mu} a \, \(\bar{u}_R \gamma^{\nu} T^a u_R\) \wtilde{G}_{\mu\nu}^a$ \\
$\cO_{\partial a\nu F}^{VL}$ & $\partial^{\mu} a \, \(\bar{\nu}_L \gamma^{\nu} \nu_L\) F_{\mu\nu}$ & $\cO_{\partial adG}^{VL}$ & $\partial^{\mu} a \, \(\bar{d}_L \gamma^{\nu} T^a d_L\) G_{\mu\nu}^a$ \\
$\cO_{\partial a\nu \tilde{F}}^{VL}$ & $\partial^{\mu} a \, \(\bar{\nu}_L \gamma^{\nu} \nu_L\) \wtilde{F}_{\mu\nu}$ & $\cO_{\partial adG}^{VR}$ & $\partial^{\mu} a \, \(\bar{d}_R \gamma^{\nu} T^a d_R\) G_{\mu\nu}^a$ \\
$\cO_{\partial au F}^{VL}$ & $\partial^{\mu} a \, \(\bar{u}_L \gamma^{\nu} u_L\) F_{\mu\nu}$ & $\cO_{\partial ad\tilde{G}}^{VL}$ & $\partial^{\mu} a \, \(\bar{d}_L \gamma^{\nu} T^a d_L\) \wtilde{G}_{\mu\nu}^a$ \\
$\cO_{\partial au F}^{VR}$ & $\partial^{\mu} a \, \(\bar{u}_R \gamma^{\nu} u_R\) F_{\mu\nu}$ & $\cO_{\partial ad\tilde{G}}^{VR}$ & $\partial^{\mu} a \, \(\bar{d}_R \gamma^{\nu} T^a d_R\) \wtilde{G}_{\mu\nu}^a$ \\
$\cO_{\partial au\tilde{F}}^{VL}$ & $\partial^{\mu} a \, \(\bar{u}_L \gamma^{\nu} u_L\) \wtilde{F}_{\mu\nu}$ & &
\\[2pt]
\hline
\end{tabular}
\caption{Operators in the aLEFT at mass dimension~7 with $\partial a$ as a building block. The lepton number violating operator $\cO_{\partial a^2\nu}^{LR}$ is marked with $(\slashed{L})$.
}
\label{tab:aLEFT_SS_dim7}
\end{table}

\begin{table}[h!]
\centering
\begin{tabular}{ |c|c|c|c| }
\hline
\multicolumn{2}{|c|}{$\partial a \, \psi^4 + \hc$} & \multicolumn{2}{c|}{$\partial a \, \psi^4 + \textrm{h.c.}$~(cont.)} \\ \hline
$\cO_{\partial aee}^{VL,SR}$ & $\partial_{\mu}a \, \( \bar{e}_L \gamma^{\mu} e_L \) \(\bar{e}_L e_R\) $ & $\cO_{\partial aeu}^{VL,TR}$ & $\partial_{\mu}a \, \( \bar{e}_L \gamma_{\nu} e_L \) \(\bar{u}_L \sigma^{\mu\nu} u_R\)$ \\
$\cO_{\partial aee}^{VR,SR}$ & $\partial_{\mu}a \, \( \bar{e}_R \gamma^{\mu} e_R \) \(\bar{e}_L e_R\) $ & $\cO_{\partial aeu}^{VR,SR}$ & $\partial_{\mu}a \, \( \bar{e}_R \gamma^{\mu} e_R \) \(\bar{u}_L u_R\)$ \\
$\cO_{\partial auu}^{VL1,SR}$ & $\partial_{\mu}a \, \( \bar{u}_L \gamma^{\mu} u_L \) \(\bar{u}_L u_R\) $ & $\cO_{\partial aeu}^{VR,TR}$ & $\partial_{\mu}a \, \( \bar{e}_R \gamma_{\nu} e_R \) \(\bar{u}_L \sigma^{\mu\nu} u_R\)$ \\
$\cO_{\partial auu}^{VL8,SR}$ & $\partial_{\mu}a \, \( \bar{u}_L \gamma^{\mu} T^a u_L \) \(\bar{u}_L T^a u_R\) $ & $\cO_{\partial aed}^{VL,SR}$ & $\partial_{\mu}a \, \( \bar{e}_L \gamma^{\mu} e_L \) \(\bar{d}_L d_R\)$ \\
$\cO_{\partial auu}^{VR1,SR}$ & $\partial_{\mu}a \, \( \bar{u}_R \gamma^{\mu} u_R \) \(\bar{u}_L u_R\) $ & $\cO_{\partial aed}^{VL,TR}$ & $\partial_{\mu}a \, \( \bar{e}_L \gamma_{\nu} e_L \) \(\bar{d}_L \sigma^{\mu\nu} d_R\)$ \\
$\cO_{\partial auu}^{VR8,SR}$ & $\partial_{\mu}a \, \( \bar{u}_R \gamma^{\mu} T^a u_R \) \(\bar{u}_L T^a u_R\) $ & $\cO_{\partial aed}^{VR,SR}$ & $\partial_{\mu}a \, \( \bar{e}_R \gamma^{\mu} e_R \) \(\bar{d}_L d_R\)$ \\
$\cO_{\partial add}^{VL1,SR}$ & $\partial_{\mu}a \, \( \bar{d}_L \gamma^{\mu} d_L \) \(\bar{d}_L d_R\) $ & $\cO_{\partial aed}^{VR,TR}$ & $\partial_{\mu}a \, \( \bar{e}_R \gamma_{\nu} e_R \) \(\bar{d}_L \sigma^{\mu\nu} d_R\)$ \\
$\cO_{\partial add}^{VL8,SR}$ & $\partial_{\mu}a \, \( \bar{d}_L \gamma^{\mu} T^a d_L \) \(\bar{d}_L T^a d_R\) $ & $\cO_{\partial aue}^{VL,SR}$ & $\partial_{\mu}a \, \( \bar{u}_L \gamma^{\mu} u_L \) \(\bar{e}_L e_R\)$ \\
$\cO_{\partial add}^{VR1,SR}$ & $\partial_{\mu}a \, \( \bar{d}_R \gamma^{\mu} d_R \) \(\bar{d}_L d_R\) $ & $\cO_{\partial aue}^{VL,TR}$ & $\partial_{\mu}a \, \( \bar{u}_L \gamma_{\nu} u_L \) \(\bar{e}_L \sigma^{\mu\nu} e_R\)$ \\
$\cO_{\partial add}^{VR8,SR}$ & $\partial_{\mu}a \, \( \bar{d}_R \gamma^{\mu} T^a d_R \) \(\bar{d}_L T^a d_R\)$ & $\cO_{\partial aue}^{VR,SR}$ & $\partial_{\mu}a \, \( \bar{u}_R \gamma^{\mu} u_R \) \(\bar{e}_L e_R\)$ \\
$\cO_{\partial a\nu e}^{VL,SR}$ & $\partial_{\mu}a \, \( \bar{\nu}_L \gamma^{\mu} \nu_L \) \(\bar{e}_L e_R\)$ & $\cO_{\partial aue}^{VR,TR}$ & $\partial_{\mu}a \, \( \bar{u}_R \gamma_{\nu} u_R \) \(\bar{e}_L \sigma^{\mu\nu} e_R\)$ \\
$\cO_{\partial a\nu e}^{VL,TR}$ & $\partial_{\mu}a \, \( \bar{\nu}_L \gamma_{\nu} \nu_L \) \(\bar{e}_L \sigma^{\mu\nu} e_R\)$ & $\cO_{\partial ade}^{VL,SR}$ & $\partial_{\mu}a \, \( \bar{d}_L \gamma^{\mu} d_L \) \(\bar{e}_L e_R\)$ \\
$\cO_{\partial aud}^{VL1,SR}$ & $\partial_{\mu}a \, \( \bar{u}_L \gamma^{\mu} u_L \) \(\bar{d}_L d_R\) $ & $\cO_{\partial ade}^{VL,TR}$ & $\partial_{\mu}a \, \( \bar{d}_L \gamma_{\nu} d_L \) \(\bar{e}_L \sigma^{\mu\nu} e_R\)$ \\
$\cO_{\partial aud}^{VL8,SR}$ & $\partial_{\mu}a \, \( \bar{u}_L \gamma^{\mu} T^a u_L \) \(\bar{d}_L T^a d_R\) $ & $\cO_{\partial ade}^{VR,SR}$ & $\partial_{\mu}a \, \( \bar{d}_R \gamma^{\mu} d_R \) \(\bar{e}_L e_R\)$ \\
$\cO_{\partial aud}^{VL1,TR}$ & $\partial_{\mu}a \, \( \bar{u}_L \gamma_{\nu} u_L \) \(\bar{d}_L \sigma^{\mu\nu}  d_R\) $ & $\cO_{\partial ade}^{VR,TR}$ & $\partial_{\mu}a \, \( \bar{d}_R \gamma_{\nu} d_R \) \(\bar{e}_L \sigma^{\mu\nu} e_R\)$ \\
$\cO_{\partial aud}^{VL8,TR}$ & $\partial_{\mu}a \, \( \bar{u}_L \gamma_{\nu} T^a u_L \) \(\bar{d}_L \sigma^{\mu\nu} T^a d_R\) $ & $\cO_{\partial a\nu u}^{VL,SR}$ & $\partial_{\mu}a \, \( \bar{\nu}_L \gamma^{\mu} \nu_L \) \(\bar{u}_L u_R\)$ \\
$\cO_{\partial aud}^{VR1,SR}$ & $\partial_{\mu}a \, \( \bar{u}_R \gamma^{\mu} u_R \) \(\bar{d}_L d_R\) $ & $\cO_{\partial a\nu u}^{VL,TR}$ & $\partial_{\mu}a \, \( \bar{\nu}_L \gamma_{\nu} \nu_L \) \(\bar{u}_L \sigma^{\mu\nu} u_R\)$ \\
$\cO_{\partial aud}^{VR8,SR}$ & $\partial_{\mu}a \, \( \bar{u}_R \gamma^{\mu} T^a u_R \) \(\bar{d}_L T^a d_R\) $ & $\cO_{\partial a\nu d}^{VL,SR}$ & $\partial_{\mu}a \, \( \bar{\nu}_L \gamma^{\mu} \nu_L \) \(\bar{d}_L d_R\)$ \\
$\cO_{\partial aud}^{VR1,TR}$ & $\partial_{\mu}a \, \( \bar{u}_R \gamma_{\nu} u_R \) \(\bar{d}_L \sigma^{\mu\nu}  d_R\) $ & $\cO_{\partial a\nu d}^{VL,TR}$ & $\partial_{\mu}a \, \( \bar{\nu}_L \gamma_{\nu} \nu_L \) \(\bar{d}_L \sigma^{\mu\nu} d_R\)$ \\ \cline{3-4}
$\cO_{\partial aud}^{VR8,TR}$ & $\partial_{\mu}a \, \( \bar{u}_R \gamma_{\nu} T^a u_R \) \(\bar{d}_L \sigma^{\mu\nu} T^a d_R\) $ & \multicolumn{2}{c|}{$(\partial a)^2 X^2$} \\ \cline{3-4}
$\cO_{\partial adu}^{VL1,SR}$ & $\partial_{\mu}a \, \( \bar{d}_L \gamma^{\mu} d_L \) \(\bar{u}_L u_R\)$ & $\cO_{\partial a^2F}^{(1)}$ & $\partial_{\mu}a\partial^{\mu}a \, F_{\nu\rho}F^{\nu\rho}$ \\
$\cO_{\partial adu}^{VL8,SR}$ & $\partial_{\mu}a \, \( \bar{d}_L \gamma^{\mu} T^a d_L \) \(\bar{u}_L T^a u_R\) $ & $\cO_{\partial a^2F}^{(2)}$ & $\partial_{\mu}a\partial^{\nu}a\, F^{\mu\rho} F_{\nu\rho}$ \\
$\cO_{\partial adu}^{VL1,TR}$ & $\partial_{\mu}a \, \( \bar{d}_L \gamma_{\nu} d_L \) \(\bar{u}_L \sigma^{\mu\nu}  u_R\) $ & $\cO_{\partial a^2\tilde{F}}$ & $\partial_{\mu}a\partial^{\mu}a \, F_{\nu\rho}\wtilde{F}^{\nu\rho}$ \\
$\cO_{\partial adu}^{VL8,TR}$ & $\partial_{\mu}a \, \( \bar{d}_L \gamma_{\nu} T^a d_L \) \(\bar{u}_L \sigma^{\mu\nu} T^a u_R\) $ & $\cO_{\partial a^2G}^{(1)}$ & $\partial_{\mu}a\partial^{\mu}a \, G_{\nu\rho}^aG^{a,\nu\rho}$ \\
$\cO_{\partial adu}^{VR1,SR}$ & $\partial_{\mu}a \, \( \bar{d}_R \gamma^{\mu} d_R \) \(\bar{u}_L u_R\)$ & $\cO_{\partial a^2G}^{(2)}$ & $\partial_{\mu}a\partial^{\nu}a\, G^{a,\mu\rho} G_{\nu\rho}^a$ \\
$\cO_{\partial adu}^{VR8,SR}$ & $\partial_{\mu}a \, \( \bar{d}_R \gamma^{\mu} T^a d_R \) \(\bar{u}_L T^a u_R\)$ & $\cO_{\partial a^2\tilde{G}}$ & $\partial_{\mu}a\partial^{\mu}a \, G_{\nu\rho}^a\wtilde{G}^{a,\nu\rho}$ \\ \cline{3-4}
$\cO_{\partial adu}^{VR1,TR}$ & $\partial_{\mu}a \, \( \bar{d}_R \gamma_{\nu} d_R \) \(\bar{u}_L \sigma^{\mu\nu} u_R\)$ & \multicolumn{2}{c|}{$(\partial a)^4$} \\ \cline{3-4}
$\cO_{\partial adu}^{VR8,TR}$ & $\partial_{\mu}a \, \( \bar{d}_R \gamma_{\nu} T^a d_R \) \(\bar{u}_L \sigma^{\mu\nu} T^a u_R\)$ & $\cO_{\partial a^4}$ & $\partial_{\mu}a \partial^{\mu}a\partial_{\nu}a \partial^{\nu}a$ \\
$\cO_{\partial aeu}^{VL,SR}$ & $\partial_{\mu}a \, \( \bar{e}_L \gamma^{\mu} e_L \) \(\bar{u}_L u_R\)$ & & \\
\hline
\end{tabular}
\caption{Operators in the aLEFT at mass dimension~8 with $\partial a$ as a building block. }
\label{tab:aLEFT_SS_dim8}
\end{table}

\begin{table}[h!]
\centering
\begin{tabular}{ |c|c|c|c| }
\hline
\multicolumn{2}{|c|}{$(\partial a)^2 \psi^2 D$} & \multicolumn{2}{c|}{$\partial a\, \psi^2 X D + \hc$~(cont.)} \\ \hline
$\cO_{\partial a^2eD}^{VL}$ & $\partial_{\mu}a\partial_{\nu}a \, \( \bar{e}_L \gamma^{\mu} \overleftrightarrow{D}^{\nu} e_L \)$ & $\cO_{\partial a FuD}^{TR}$ & $\partial_{\mu} a \, F_{\nu\rho} \( \bar{u}_L \sigma^{\mu\nu} D^{\rho} u_R \)$ \\
$\cO_{\partial a^2eD}^{VR}$ & $\partial_{\mu}a\partial_{\nu}a \, \( \bar{e}_R \gamma^{\mu} \overleftrightarrow{D}^{\nu} e_R \)$ & $\cO_{\partial a \tilde{F}uD}^{SR}$ & $\partial_{\mu} a \, \wtilde{F}^{\mu\nu} \( \bar{u}_L D_{\nu} u_R \)$ \\
$\cO_{\partial a^2\nu D}^{VL}$ & $\partial_{\mu}a\partial_{\nu}a \, \( \bar{\nu}_L \gamma^{\mu} \overleftrightarrow{D}^{\nu} \nu_L \)$ & $\cO_{\partial a FdD}^{SR}$ & $\partial_{\mu} a \, F^{\mu\nu} \( \bar{d}_L D_{\nu} d_R \)$ \\
$\cO_{\partial a^2uD}^{VL}$ & $\partial_{\mu}a\partial_{\nu}a \, \( \bar{u}_L \gamma^{\mu} \overleftrightarrow{D}^{\nu} u_L \)$ & $\cO_{\partial a FdD}^{TR}$ & $\partial_{\mu} a \, F_{\nu\rho} \( \bar{d}_L \sigma^{\mu\nu} D^{\rho} d_R \)$ \\
$\cO_{\partial a^2uD}^{VR}$ & $\partial_{\mu}a\partial_{\nu}a \, \( \bar{u}_R \gamma^{\mu} \overleftrightarrow{D}^{\nu} u_R \)$ & $\cO_{\partial a \tilde{F}dD}^{SR}$ & $\partial_{\mu} a \, \wtilde{F}^{\mu\nu} \( \bar{d}_L D_{\nu} d_R \)$ \\
$\cO_{\partial a^2dD}^{VL}$ & $\partial_{\mu}a\partial_{\nu}a \, \( \bar{d}_L \gamma^{\mu} \overleftrightarrow{D}^{\nu} d_L \)$ & $\cO_{\partial a GuD}^{SR}$ & $\partial_{\mu} a \, G^{a,\mu\nu} \( \bar{u}_L T^a D_{\nu} u_R \)$ \\
$\cO_{\partial a^2dD}^{VR}$ & $\partial_{\mu}a\partial_{\nu}a \, \( \bar{d}_R \gamma^{\mu} \overleftrightarrow{D}^{\nu} d_R \)$ & $\cO_{\partial a GuD}^{TR}$ & $\partial_{\mu} a \, G_{\nu\rho}^a \( \bar{u}_L \sigma^{\mu\nu} T^a D^{\rho} u_R \)$ \\ \cline{1-2}
\multicolumn{2}{|c|}{$\partial a\, \psi^2 X D + \hc$} & $\cO_{\partial a \tilde{G}uD}^{SR}$ & $\partial_{\mu} a \, \wtilde{G}^{a,\mu\nu} \( \bar{u}_L T^a D_{\nu} u_R \)$ \\ \cline{1-2}
$\cO_{\partial a FeD}^{SR}$ & $\partial_{\mu} a \, F^{\mu\nu} \( \bar{e}_L D_{\nu} e_R \)$ & $\cO_{\partial a GdD}^{SR}$ & $\partial_{\mu} a \, G^{a,\mu\nu} \( \bar{d}_L T^a D_{\nu} d_R \)$ \\
$\cO_{\partial a FeD}^{TR}$ & $\partial_{\mu} a \, F_{\nu\rho} \( \bar{e}_L \sigma^{\mu\nu} D^{\rho} e_R \)$ & $\cO_{\partial a GdD}^{TR}$ & $\partial_{\mu} a \, G_{\nu\rho}^a \( \bar{d}_L \sigma^{\mu\nu} T^a D^{\rho} d_R \)$ \\
$\cO_{\partial a \tilde{F}eD}^{SR}$ & $\partial_{\mu} a \, \wtilde{F}^{\mu\nu} \( \bar{e}_L D_{\nu} e_R \)$ & $\cO_{\partial a \tilde{G}dD}^{SR}$ & $\partial_{\mu} a \, \wtilde{G}^{a,\mu\nu} \( \bar{d}_L T^a D_{\nu} d_R \)$ \\
 $\cO_{\partial a FuD}^{SR}$ & $\partial_{\mu} a \, F^{\mu\nu} \( \bar{u}_L D_{\nu} u_R \)$  & & \\
\hline
\end{tabular}
\caption{Operators in the aLEFT at mass dimension~8 with $\partial a$ as a building block. [Tab.~\ref{tab:aLEFT_SS_dim8} continued.] }
\label{tab:aLEFT_SS_dim8_2}
\end{table}

\begin{table}[h!]
\centering
\begin{tabular}{ |c|c|c|c| }
\hline
\multicolumn{4}{|c|}{$\slashed{B}$ and $\slashed{L}$ terms} \\ \hline
\multicolumn{2}{|c|}{$\partial a \, \psi^4 + \hc$} & \multicolumn{2}{c|}{$\partial a \, \psi^4 + \hc$~(cont.)} \\ \hline
$\cO_{\partial ae\nu}^{VL,SR}$ & $\partial_{\mu}a \, \( \bar{e}_L \gamma^{\mu} e_L \) \( \bar{\nu}_L \nu_L^c \)$ & $\cO_{\partial aeddd}^{VL,SR}$ & $\partial_{\mu}a \, \epsilon^{\alpha\beta\gamma} \( \bar{e}_L \gamma^{\mu} d_{L,\alpha} \) \( \bar{d}_{R,\beta}^c d_{R,\gamma} \)$ \\
$\cO_{\partial ae\nu}^{VR,SR}$ & $\partial_{\mu}a \, \( \bar{e}_R \gamma^{\mu} e_R \) \( \bar{\nu}_L \nu_L^c \)$ & $\cO_{\partial addde}^{VL,SR}$ & $\partial_{\mu}a \, \epsilon^{\alpha\beta\gamma} \( \bar{d}_{L,\alpha} \gamma^{\mu} d_{R,\beta}^c \) \( \bar{d}_{L,\gamma} e_{R} \)$ \\
$\cO_{\partial au\nu}^{VL,SR}$ & $\partial_{\mu}a \, \( \bar{u}_L \gamma^{\mu} u_L \) \( \bar{\nu}_L \nu_L^c \)$ & $\cO_{\partial aeduu}^{VR,SR}$ & $\partial_{\mu}a \, \epsilon^{\alpha\beta\gamma} \( \bar{e}_{L}^c \gamma^{\mu} d_{R,\alpha} \) \( \bar{u}_{R,\beta}^c u_{R,\gamma} \)$ \\
$\cO_{\partial au\nu}^{VR,SR}$ & $\partial_{\mu}a \, \( \bar{u}_R \gamma^{\mu} u_R \) \( \bar{\nu}_L \nu_L^c \)$ & $\cO_{\partial adeuu}^{VR,SR}$ & $\partial_{\mu}a \, \epsilon^{\alpha\beta\gamma} \( \bar{d}_{L,\alpha}^c \gamma^{\mu} e_{R} \) \( \bar{u}_{R,\beta}^c u_{R,\gamma} \)$ \\
$\cO_{\partial ad\nu}^{VL,SR}$ & $\partial_{\mu}a \, \( \bar{d}_L \gamma^{\mu} d_L \) \( \bar{\nu}_L \nu_L^c \)$ & $\cO_{\partial aeudu}^{VL,SR}$ & $\partial_{\mu}a \, \epsilon^{\alpha\beta\gamma} \( \bar{e}_{R}^c \gamma^{\mu} u_{L,\alpha} \) \( \bar{d}_{R,\beta}^c u_{R,\gamma} \)$ \\
$\cO_{\partial ad\nu}^{VR,SR}$ & $\partial_{\mu}a \, \( \bar{d}_R \gamma^{\mu} d_R \) \( \bar{\nu}_L \nu_L^c \)$ & $\cO_{\partial adueu}^{VL,SR}$ & $\partial_{\mu}a \, \epsilon^{\alpha\beta\gamma} \( \bar{d}_{R,\alpha}^c \gamma^{\mu} u_{L,\beta} \) \( \bar{e}_{R}^c u_{R,\gamma} \)$ \\
$\cO_{\partial aude\nu}^{VL,SR}$ & $\partial_{\mu}a \, \( \bar{u}_L \gamma^{\mu} d_L \) \( \bar{e}_L \nu_L^c \)$ & $\cO_{\partial auude}^{VL,SR}$ & $\partial_{\mu}a \, \epsilon^{\alpha\beta\gamma} \( \bar{u}_{L,\alpha} \gamma^{\mu} u_{R,\beta}^c \) \( \bar{d}_{L,\beta} e_{L}^c \)$ \\
$\cO_{\partial aedu\nu}^{VL,SR}$ & $\partial_{\mu}a \, \( \bar{e}_L \gamma^{\mu} d_L \) \( \bar{u}_L \nu_L^c \)$ & $\cO_{\partial aduue}^{VL,SR}$ & $\partial_{\mu}a \, \epsilon^{\alpha\beta\gamma} \( \bar{d}_{L,\alpha} \gamma^{\mu} u_{R,\beta}^c \) \( \bar{u}_{L,\beta} e_{L}^c \)$ \\
$\cO_{\partial au\nu ed}^{VL,SR}$ & $\partial_{\mu}a \, \( \bar{u}_L \gamma^{\mu} \nu_L \) \( \bar{e}_L d_R \)$ & $\cO_{\partial aueud}^{VL,SR}$ & $\partial_{\mu}a \, \epsilon^{\alpha\beta\gamma} \( \bar{u}_{L,\alpha} \gamma^{\mu} e_{R}^c \) \( \bar{u}_{L,\beta} d_{L,\gamma}^c \)$ \\
$\cO_{\partial ae\nu ud}^{VL,SR}$ & $\partial_{\mu}a \, \( \bar{e}_L \gamma^{\mu} \nu_L \) \( \bar{u}_L d_R \)$ & $\cO_{\partial audue}^{VL,SR}$ & $\partial_{\mu}a \, \epsilon^{\alpha\beta\gamma} \( \bar{u}_{L,\alpha} \gamma^{\mu} d_{R,\beta}^c \) \( \bar{u}_{L,\beta} e_{L}^c \)$ \\
$\cO_{\partial ade\nu u}^{VL,SR}$ & $\partial_{\mu}a \, \( \bar{d}_L \gamma^{\mu} e_L \) \( \bar{\nu}_L u_R \)$ & $\cO_{\partial a\nu ddu}^{VL,SR}$ & $\partial_{\mu}a \, \epsilon^{\alpha\beta\gamma} \( \bar{\nu}_{L} \gamma^{\mu} d_{R,\alpha}^c \) \( \bar{d}_{L,\beta} u_{L,\gamma}^c \)$ \\
$\cO_{\partial a\nu ed u}^{VL,SR}$ & $\partial_{\mu}a \, \( \bar{\nu}_L \gamma^{\mu} e_L \) \( \bar{d}_L u_R \)$ & $\cO_{\partial add\nu u}^{VL,SR}$ & $\partial_{\mu}a \, \epsilon^{\alpha\beta\gamma} \( \bar{d}_{L,\alpha} \gamma^{\mu} d_{R,\beta}^c \) \( \bar{\nu}_{L} u_{L,\gamma}^c \)$ \\
$\cO_{\partial aeu\nu d}^{VL,SR}$ & $\partial_{\mu}a \, \( \bar{e}_L \gamma^{\mu} u_R^c \) \( \bar{\nu}_L d_R \)$ & $\cO_{\partial ad\nu ud}^{VL,SR}$ & $\partial_{\mu}a \, \epsilon^{\alpha\beta\gamma} \( \bar{d}_{L,\alpha} \gamma^{\mu} \nu_{L} \) \( \bar{u}_{L,\beta} d_{L,\gamma}^c \)$ \\
$\cO_{\partial a\nu ued}^{VL,SR}$ & $\partial_{\mu}a \, \( \bar{\nu}_L \gamma^{\mu} u_R^c \) \( \bar{e}_L d_R \)$ & $\cO_{\partial aud\nu d}^{VR,SR}$ & $\partial_{\mu}a \, \epsilon^{\alpha\beta\gamma} \( \bar{u}_{L,\alpha}^c \gamma^{\mu} d_{R,\beta} \) \( \bar{\nu}_{L} d_{R,\gamma} \)$ \\
$\cO_{\partial a\nu eud}^{VL,SR}$ & $\partial_{\mu}a \, \( \bar{\nu}_L \gamma^{\mu} e_R^c \) \( \bar{u}_L d_R \)$ & $\cO_{\partial ad\nu du}^{VL,SR}$ & $\partial_{\mu}a \, \epsilon^{\alpha\beta\gamma} \( \bar{d}_{R,\alpha}^c \gamma^{\mu} \nu_{L} \) \( \bar{d}_{R,\beta}^c u_{R,\gamma} \)$ \\
$\cO_{\partial aue\nu d}^{VL,SR}$ & $\partial_{\mu}a \, \( \bar{u}_L \gamma^{\mu} e_R^c \) \( \bar{\nu}_L d_R \)$ & $\cO_{\partial adud\nu}^{VL,SR}$ & $\partial_{\mu}a \, \epsilon^{\alpha\beta\gamma} \( \bar{d}_{L,\alpha} \gamma^{\mu} u_{R,\beta}^c \) \( \bar{d}_{L,\gamma} \nu_{L}^c \)$ \\
$\cO_{\partial adu\nu e}^{VL,SR}$ & $\partial_{\mu}a \, \( \bar{d}_L \gamma^{\mu} u_L \) \( \bar{\nu}_L e_R \)$ & $\cO_{\partial add\nu u}^{VR,SR}$ & $\partial_{\mu}a \, \epsilon^{\alpha\beta\gamma} \( \bar{d}_{L,\alpha}^c \gamma^{\mu} d_{R,\beta} \) \( \bar{\nu}_{L} u_{R,\gamma} \)$ \\ 
$\cO_{\partial a\nu ude}^{VL,SR}$ & $\partial_{\mu}a \, \( \bar{\nu}_L \gamma^{\mu} u_L \) \( \bar{d}_L e_R \)$ & $\cO_{\partial adu\nu d}^{VR,SR}$ & $\partial_{\mu}a \, \epsilon^{\alpha\beta\gamma} \( \bar{d}_{L,\alpha}^c \gamma^{\mu} u_{R,\beta} \) \( \bar{\nu}_{L} d_{R,\gamma} \)$ \\ 
$\cO_{\partial adu\nu e}^{VR,SR}$ & $\partial_{\mu} a \(\bar{d}_R \gamma^{\mu} u_R\) \(\bar{\nu}_L e_R \)$ & $\cO_{\partial adedd}^{VL,SR}~(\star)$ & $\partial_{\mu}a \, \epsilon^{\alpha\beta\gamma} \( \bar{d}_{L,\alpha}^c \gamma^{\mu} e_{L} \) \( \bar{d}_{L,\beta} d_{L,\gamma}^c \)$ \\
$\cO_{\partial ade\nu u}^{VR,SR}$ & $\partial_{\mu} a \(\bar{d}_R \gamma^{\mu} e_R\) \(\bar{\nu}_L u_R \)$ & $\cO_{\partial aeddd}^{VR,SR}~(\star)$ & $\partial_{\mu}a \, \epsilon^{\alpha\beta\gamma} \( \bar{e}_R \gamma^{\mu} d_{R,\alpha} \) \( \bar{d}_{R,\beta}^c d_{R,\gamma} \)$ \\ \cline{3-4}
$\cO_{\partial a\nu edu}^{VR,SR}$ & $\partial_{\mu} a \(\bar{\nu}_L^c \gamma^{\mu} e_R\) \(\bar{d}_L u_R \)$ & \multicolumn{2}{c|}{$\partial a\, \psi^2 X D + \hc$} \\ \cline{3-4}
$\cO_{\partial a\nu ude}^{VR,SR}$ & $\partial_{\mu} a \(\bar{\nu}_L^c \gamma^{\mu} u_R\) \(\bar{d}_L e_R \)$ & $\cO_{\partial aF\nu D}^{SR}$ & $\partial_{\mu}a \, F^{\mu\nu} \( \bar{\nu}_L \sigma_{\nu\rho} \partial^{\rho} \nu_L^c \)$ \\
$\cO_{\partial a\nu}^{VL,SR}~(\star)$ & $\partial_{\mu}a \, \( \bar{\nu}_{L} \gamma^{\mu} \nu_{L} \) \( \bar{\nu}_{L} \nu_{L}^c \)$ & & \\
\hline
\end{tabular}
\caption{B- and L-breaking operators in the aLEFT at mass dimension~8 with $\partial a$ as a building block. Note that the operators marked with $(\star)$ do not appear when fermion flavors are set to 1. [Tab.~\ref{tab:aLEFT_SS_dim8_2} continued].}
\label{tab:aLEFT_SS_dim8_3}
\end{table}

\clearpage
\subsection{Without shift symmetry} \label{app:aLEFTnonSSOpBasis}

For aLEFT without a shift symmetry, once again, we can use the PQ-breaking isolation condition Eq.~\eqref{eq:separation_aLEFT} to construct the operator basis easily. The operator bases at higher dimensions can be constructed with
\begin{equation}
    \cL_{n}^{\aLEFTnotPQ} = a \, \cL_{n-1}^{\aLEFTnotPQ} + a \, \cL_{n-1}^{\text{LEFT}} + \cL_{n}^{\aLEFTPQ}
\end{equation}
for $n>5$. As a start point, we show the operator basis at dimension~5 in Tab.~\ref{tab:aLEFT_nonSS_dim5}. The higher-dimensional operator bases can be easily constructed with the LEFT operator bases~\cite{Jenkins:2017jig,Liao:2020zyx} and the shift-symmetric operator bases. For completeness, the axion-dependent renormalizable operators are shown in Eq.~\eqref{eq:aLEFTnonPQreno}.

\begin{table}[h!]
\centering
\begin{tabular}{ |c|c|c|c| }
\hline
\multicolumn{2}{|c|}{$V(a,H)$} & \multicolumn{2}{c|}{$a^2 \psi^2 + \hc$} \\
\hline
$\cO_{a^5}$ & $a^5$ & $\cO_{a^2e}^{SR}$ & $a^2 \bar{e}_L e_R$ \\ \cline{1-2}
\multicolumn{2}{|c|}{$a X^2$} & $\cO_{a^2\nu}^{SR}$ $(\slashed{L})$ & $a^2 \bar{\nu}_L \nu_L^c$ \\ \cline{1-2}
$\cO_{aF}$ & $a F_{\mu\nu} F^{\mu\nu}$ & $\cO_{a^2u}^{SR}$ & $a^2 \bar{u}_L u_R$ \\
$\cO_{a\tilde{F}}$ & $a F_{\mu\nu} \wtilde{F}^{\mu\nu}$ & $\cO_{a^2d}^{SR}$ & $a^2 \bar{d}_L d_R$ \\ 
$\cO_{aG}$ & $a G_{\mu\nu}^a G^{a,\mu\nu}$ & & \\
$\cO_{a\tilde{G}}$ & $a G_{\mu\nu}^a \wtilde{G}^{a,\mu\nu}$ & & \\
\hline
\end{tabular}
\caption{Operators in the aLEFT at mass dimension~5 with $a$ as a building block. It is worth noting that the dimension-5 ALP Yukawa couplings in the aSMEFT in Tab.~\ref{tab:aSMEFT_nonSS_dim5} become dimension-4 ALP-dependent mass terms in the aLEFT. The lepton number violating operator $\cO_{a^2\nu}$ is marked with $(\slashed{L})$.
}
\label{tab:aLEFT_nonSS_dim5}
\end{table}

\section{Additional results for the Hilbert series and the operator counting}
\label{app:HS}

\subsection{aSMEFT} \label{app:HSaSMEFT}

In Section~\ref{sec:aSMEFT}, we have shown the Hilbert series for aSMEFT with (without) a shift symmetry up to dimension~8~(7). Due to the excessive length of the Hilbert series in higher dimensions, it cannot be included in this paper. Instead, we provide an ancillary Mathematica notebook to present all the Hilbert series from dimension~5 to dimension~15. By keeping the information of the baryon and lepton number violation in each term of the Hilbert series, and setting all spurions to unity, the number of operators at each mass dimension can be expanded in powers of the $B$ and $L$ violating unit $\epsilon_{B,L}$. The Hilbert series is calculated with $N_f$ flavors of the fermions, which allow us to count the operators with flavor dependence. The full results of the operator counting is also available in our ancillary notebook file. In this section, we will only show the results for number counting up to dimension~12.

For aSMEFT with a shift symmetry, the numbers of operators at each dimension are given by
\setcounter{equation}{1}
\begin{align*}
\#\,\cO_5^{\PQ} \,=\,& 2 - N_f + 5 N_f^2 \, , \\
\#\,\cO_6^{\PQ} \,=\,&  1 \, , \\
\#\,\cO_7^{\PQ} \,=\,&  5 + 39 N_f^2 \, , \\ 
\#\,\cO_8^{\PQ} \,=\,&  \left(13+11 N_f^2\right)+\left(-\frac{2 N_f^2}{3}+\frac{8 N_f^4}{3}\right) \epsilon_B \epsilon_L+\left(4 N_f^2+2 N_f^4\right) \epsilon_L^2 \, , \\
\#\,\cO_9^{\PQ} \,=\,&  \left(74+\frac{1799 N_f^2}{4}-\frac{N_f^3}{2}+\frac{847 N_f^4}{4}\right)+46 N_f^4 \epsilon_B \epsilon_L+\left(N_f+N_f^2\right) \epsilon_L^2 \, , \\ 
\#\,\cO_{10}^{\PQ} \,=\,&  \left(74+\frac{431 N_f^2}{2}+\frac{91 N_f^4}{2}\right)+\left(8 N_f^3+106 N_f^4\right) \epsilon_B \epsilon_L+\left(-N_f+75 N_f^2+N_f^3+121 N_f^4\right) \epsilon_L^2 \,,\\
\#\,\cO_{11}^{\PQ} \,=\,&  \left(799+\frac{67043 N_f^2}{12}-\frac{25 N_f^3}{2}+\frac{21687 N_f^4}{4}-N_f^5+\frac{2363 N_f^6}{6}\right) \numberthis\\
&+\left(\frac{259 N_f^3}{6}+\frac{5139 N_f^4}{4}+\frac{35 N_f^5}{6}+\frac{753 N_f^6}{4}\right) \epsilon_B \epsilon_L+\left(4 N_f+\frac{92 N_f^2}{3}+2 N_f^3+\frac{70 N_f^4}{3}\right) \epsilon_L^2\,,\\
\#\,\cO_{12}^{\PQ} \,=\,& \left(693+\frac{6629 N_f^2}{2}-5 N_f^3+\frac{3857 N_f^4}{2}\right)+\left(-\frac{4 N_f^2}{9}+\frac{13 N_f^3}{12}+\frac{5 N_f^4}{36}+\frac{71 N_f^5}{12}+\frac{2135 N_f^6}{36}\right) \epsilon_B^2\\
&+\left(-2 N_f^2+173 N_f^3+\frac{8450 N_f^4}{3}+74 N_f^5+\frac{2221 N_f^6}{3}\right) \epsilon_B \epsilon_L\\
&+\left(-21 N_f+1091 N_f^2-50 N_f^3+\frac{37349 N_f^4}{12}+7 N_f^5+\frac{8503 N_f^6}{12}\right) \epsilon_L^2\\
&+\left(-\frac{2 N_f^2}{9}-\frac{N_f^4}{18}+4 N_f^5+\frac{437 N_f^6}{18}\right) \epsilon_B \epsilon_L^3\,.
\end{align*}

For aSMEFT without a shift symmetry, the numbers of operators at each dimension are given by
\begin{align*}
 \#\,\cO_5^{\notPQ} \,=\,& 9 + 6 N_f^2 \, , \\
 \#\,\cO_6^{\notPQ} \,=\,& \left(10+6 N_f^2\right)+\left(N_f+N_f^2\right) \epsilon_L^2 \, , \\
 \#\,\cO_7^{\notPQ} \,=\,& \left(30+\frac{315 N_f^2}{4}+\frac{N_f^3}{2}+\frac{107 N_f^4}{4}\right)+\left(\frac{2 N_f^2}{3}+N_f^3+\frac{19 N_f^4}{3}\right) \epsilon_B \epsilon_L+\left(N_f+N_f^2\right) \epsilon_L^2 \, , \\
 \#\,\cO_8^{\notPQ} \,=\,& \left(43+\frac{359 N_f^2}{4}+\frac{N_f^3}{2}+\frac{107 N_f^4}{4}\right)+\left(3 N_f+\frac{41 N_f^2}{3}+N_f^3+\frac{37 N_f^4}{3}\right) \epsilon_L^2\\
 &+\left(2 N_f^3+16 N_f^4\right) \epsilon_B \epsilon_L\, ,\\
 \#\,\cO_9^{\notPQ} \,=\,& \left(206+934 N_f^2+650 N_f^4\right)+\left(\frac{2 N_f^2}{3}+3 N_f^3+\frac{475 N_f^4}{3}\right) \epsilon_B \epsilon_L+\left(4 N_f+\frac{44 N_f^2}{3}+N_f^3+\frac{37 N_f^4}{3}\right) \epsilon_L^2\,,\\
 \#\,\cO_{10}^{\notPQ} \,=\,& \left(280+\frac{2299 N_f^2}{2}+\frac{1391 N_f^4}{2}\right)+\left(\frac{N_f^2}{4}+\frac{61 N_f^3}{24}+\frac{29 N_f^4}{24}+\frac{11 N_f^5}{24}+\frac{85 N_f^6}{24}\right) \epsilon_B^2\\
 &+\left(-\frac{2 N_f^2}{3}+\frac{62 N_f^3}{3}+\frac{1256 N_f^4}{3}+\frac{4 N_f^5}{3}+40 N_f^6\right) \epsilon_B \epsilon_L\\
 &+\left(12 N_f+\frac{518 N_f^2}{3}+\frac{73 N_f^3}{12}+\frac{4187 N_f^4}{12}-\frac{N_f^5}{12}+\frac{437 N_f^6}{12}\right) \epsilon_L^2+\left(-N_f^5+N_f^6\right) \epsilon_B \epsilon_L^3\,,\\
 \#\,\cO_{11}^{\notPQ} \,=\,& \left(1609+\frac{22461 N_f^2}{2}-21 N_f^3+12961 N_f^4-7 N_f^5+\frac{3305 N_f^6}{2}\right)\numberthis\\
 &+\left(\frac{N_f^2}{4}+\frac{61 N_f^3}{24}+\frac{29 N_f^4}{24}+\frac{11 N_f^5}{24}+\frac{85 N_f^6}{24}\right) \epsilon_B^2\\
 &+\left(-\frac{4 N_f^2}{3}+\frac{695 N_f^3}{6}+\frac{40555 N_f^4}{12}+\frac{115 N_f^5}{6}+\frac{3331 N_f^6}{4}\right) \epsilon_B \epsilon_L\\
 &+\left(16 N_f+\frac{610 N_f^2}{3}+\frac{97 N_f^3}{12}+\frac{1489 N_f^4}{4}-\frac{N_f^5}{12}+\frac{437 N_f^6}{12}\right) \epsilon_L^2\\
 &+\left(-\frac{4 N_f^2}{9}-\frac{N_f^3}{3}-\frac{N_f^4}{9}-\frac{2 N_f^5}{3}+\frac{14 N_f^6}{9}\right) \epsilon_B \epsilon_L^3+\left(-\frac{N_f^2}{6}+\frac{N_f^4}{6}\right) \epsilon_L^4\,,\\
\#\,\cO_{12}^{\notPQ} \,=\,& \left(2302+14545 N_f^2-26 N_f^3+\frac{29779 N_f^4}{2}-7 N_f^5+\frac{3305 N_f^6}{2}\right)\\
&+\left(\frac{5 N_f^2}{9}+\frac{197 N_f^3}{12}+\frac{59 N_f^4}{36}+\frac{175 N_f^5}{12}+\frac{7661 N_f^6}{36}\right) \epsilon_B^2\\
&+\left(-\frac{16 N_f^2}{3}+\frac{871 N_f^3}{2}+\frac{109639 N_f^4}{12}+\frac{411 N_f^5}{2}+\frac{13823 N_f^6}{4}\right) \epsilon_B \epsilon_L\\
&+\left(13 N_f+\frac{6695 N_f^2}{3}-\frac{1111 N_f^3}{12}+7381 N_f^4-\frac{149 N_f^5}{12}+\frac{7786 N_f^6}{3}\right) \epsilon_L^2\\
&+\left(-\frac{2 N_f^2}{3}+\frac{2 N_f^3}{3}+\frac{17 N_f^4}{6}+\frac{25 N_f^5}{3}+\frac{569 N_f^6}{6}\right) \epsilon_B \epsilon_L^3+\left(-\frac{N_f^2}{6}+\frac{N_f^4}{6}\right) \epsilon_L^4\,.
\end{align*}

By setting $\epsilon_{B,L}\to 1$, the total number of operators can be easily obtained.

In Section~\ref{sec:aSMEFTCPV}, we have discussed the CP properties of the operators, the operators can be categorized into CP-even, CP-odd and CP-violating classes. In this section, we show additional results of the Hilbert series up to dimension 8. To reduce the length, we only show the reduced Hilbert series as we have done in Section~\ref{app:aSMEFTSSOpBasis}. For $\aSMEFTPQ$, the CP-even Hilbert series is given by
\begin{align}
\cH_{5,\text{even}}^{\PQ}\,=\,& 3a X^2 + \frac{1}{2} \left(-2+3 N_f+5 N_f^2\right)\partial a\, \psi ^2\,,\qquad
\cH_{6,\text{even}}^{\PQ}\,=\, (\partial a)^2 H^2\,,\nonumber\\
\cH_{7,\text{even}}^{\PQ}\,=\,& \partial a\, \cD H^4 + 2\partial a\, \cD H^2 X + \frac{7}{2} N_f \left(1+N_f\right)\partial a\, H^2 \psi ^2 + 6 N_f^2\partial a\, \cD H \psi ^2 + 10 N_f^2\partial a\, X \psi ^2\,,\nonumber\\
\cH_{8,\text{even}}^{\PQ}\,=\,& (\partial a)^4 + 2(\partial a)^2 \cD^2 H^2 + (\partial a)^2 H^4 + 6(\partial a)^2 X^2 + \frac{5}{2} N_f \left(1+N_f\right)(\partial a)^2 \cD \psi ^2 \\
&+ 3 N_f^2(\partial a)^2 H \psi ^2 + 2 N_f^2\partial a\, \cD H^2 \psi ^2 + \frac{1}{3} N_f^2 \left(-1+7 N_f^2\right)\partial a\, \psi ^4\,,\nonumber
\end{align}

The CP-odd Hilbert series is given by
\begin{equation}
\begin{split}
\cH_{5,\text{odd}}^{\PQ}\,=\,& \frac{5}{2} \left(-1+N_f\right) N_f\partial a\, \psi ^2\,,\qquad
\cH_{6,\text{odd}}^{\PQ}\,=\, 0\,,\\
\cH_{7,\text{odd}}^{\PQ}\,=\,& 2\partial a\, \cD H^2 X + \frac{7}{2} \left(-1+N_f\right) N_f\partial a\, H^2 \psi ^2 + 6 N_f^2\partial a\, \cD H \psi ^2 + 10 N_f^2\partial a\, X \psi ^2\,,\\
\cH_{8,\text{odd}}^{\PQ}\,=\,& 3(\partial a)^2 X^2 + \frac{5}{2} \left(-1+N_f\right) N_f(\partial a)^2 \cD \psi ^2 + 3 N_f^2(\partial a)^2 H \psi ^2 + 2 N_f^2\partial a\, \cD H^2 \psi ^2 \\
&+ \frac{1}{3} N_f^2 \left(-1+7 N_f^2\right)\partial a\, \psi ^4\,,
\end{split}
\end{equation}

The CP-violating Hilbert series is calculated with some specific $N_f$. For $\aSMEFTPQ$, we show the CP-violating Hilbert series with $N_f=3$ as follows
\begin{equation}
\begin{split}
\cH_{5,\text{CPV}}^{\PQ}\,=\,& 9\partial a\, \psi ^2\,,\qquad
\cH_{6,\text{CPV}}^{\PQ}\,=\, 0\,,\\
\cH_{7,\text{CPV}}^{\PQ}\,=\,& 2\partial a\, \cD H^2 X + 42\partial a\, \cD H \psi ^2 + 12\partial a\, H^2 \psi ^2 + 72\partial a\, X \psi ^2\,,\\
\cH_{8,\text{CPV}}^{\PQ}\,=\,& 3(\partial a)^2 X^2 + 9(\partial a)^2 \cD \psi ^2 + 21(\partial a)^2 H \psi ^2\,,
\end{split}
\end{equation}

For $\aSMEFTnotPQ$, The CP-even Hilbert series up to dimension 8 is given as
\begin{align*}
\cH_{5,\text{even}}^{\notPQ}\,=\,& 3a X^2 + 3 N_f^2a H \psi ^2\,,\\
\cH_{6,\text{even}}^{\notPQ}\,=\,& a^6 + a^4 H^2 + a^2 \cD^2 H^2 + a^2 H^4 + 3a^2 X^2 + \frac{1}{2} N_f \left(1+N_f\right)a H^2 \psi ^2 + 3 N_f^2a^2 H \psi ^2\,,\\
\cH_{7,\text{even}}^{\notPQ}\,=\,& a \cD^2 H^4 + 2a \cD^2 H^2 X + 3a^3 X^2 + 4a H^2 X^2 + 2a X^3 + \frac{1}{2} N_f \left(1+N_f\right)a^2 H^2 \psi ^2 \\
&+ 3 N_f^2a^3 H \psi ^2 + 6 N_f^2a \cD^2 H \psi ^2 + 8 N_f^2a \cD H^2 \psi ^2 + 3 N_f^2a H^3 \psi ^2 + 10 N_f^2a \cD X \psi ^2 \\
&+ 8 N_f^2a H X \psi ^2 + \frac{1}{24} N_f \left(-6-193 N_f+18 N_f^2+397 N_f^3\right)a \psi ^4\,,\\
\cH_{8,\text{even}}^{\notPQ}\,=\,& a^8 + a^4 \cD^4 + a^6 H^2 + a^4 \cD^2 H^2 + 2a^2 \cD^4 H^2 + a^4 H^4 + 3a^2 \cD^2 H^4 + a^2 H^6 \numberthis \\
&+ 2a^2 \cD^2 H^2 X + 3a^4 X^2 + 6a^2 \cD^2 X^2 + 4a^2 H^2 X^2 + 2a^2 X^3 \\
&+ \frac{5}{2} N_f \left(1+N_f\right)a^2 \cD^3 \psi ^2 + \frac{1}{2} N_f \left(1+N_f\right)a^3 H^2 \psi ^2 + N_f \left(1+3 N_f\right)a \cD^2 H^2 \psi ^2\\
&+ \frac{1}{2} N_f \left(1+N_f\right)a H^4 \psi ^2 + 3 N_f^2a^4 H \psi ^2 + 9 N_f^2a^2 \cD^2 H \psi ^2 + 8 N_f^2a^2 \cD H^2 \psi ^2 \\
&+ 3 N_f^2a^2 H^3 \psi ^2 + N_f^2a \cD H^3 \psi ^2 + \frac{1}{2} N_f \left(-1+3 N_f\right)a H^2 X \psi ^2 + 10 N_f^2a^2 \cD X \psi ^2 \\
&+ 8 N_f^2a^2 H X \psi ^2 + \frac{1}{24} N_f \left(6+275 N_f+18 N_f^2+397 N_f^3\right)a^2 \psi ^4\\
&+ \frac{1}{2} N_f^3 \left(3+7 N_f\right)a \cD \psi ^4 + \frac{1}{2} N_f^3 \left(-1+15 N_f\right)a H \psi ^4\,,
\end{align*}

The CP-odd Hilbert series is given by
\begin{align*}
\cH_{5,\text{odd}}^{\notPQ}\,=\,& a^5 + a^3 H^2 + a H^4 + 3a X^2 + 3 N_f^2a H \psi ^2\,,\\
\cH_{6,\text{odd}}^{\notPQ}\,=\,& 3a^2 X^2 + \frac{1}{2} N_f \left(1+N_f\right)a H^2 \psi ^2 + 3 N_f^2a^2 H \psi ^2\,,\\
\cH_{7,\text{odd}}^{\notPQ}\,=\,& a^7 + a^5 H^2 + a^3 \cD^2 H^2 + a^3 H^4 + 2a \cD^2 H^4 + a H^6 + 2a \cD^2 H^2 X + 3a^3 X^2\\
&+ 4a H^2 X^2 + 2a X^3 + \frac{1}{2} N_f \left(1+N_f\right)a^2 H^2 \psi ^2 + 3 N_f^2a^3 H \psi ^2 + 6 N_f^2a \cD^2 H \psi ^2\\
&+ 8 N_f^2a \cD H^2 \psi ^2 + 3 N_f^2a H^3 \psi ^2 + 10 N_f^2a \cD X \psi ^2 + 8 N_f^2a H X \psi ^2\\
&+ \frac{1}{24} N_f \left(6+275 N_f+18 N_f^2+397 N_f^3\right)a \psi ^4\,,\\
\cH_{8,\text{odd}}^{\notPQ}\,=\,& a^2 \cD^2 H^4 + 2a^2 \cD^2 H^2 X + 3a^4 X^2 + 3a^2 \cD^2 X^2 + 4a^2 H^2 X^2 + 2a^2 X^3 \numberthis\\
&+ \frac{5}{2} \left(-1+N_f\right) N_fa^2 \cD^3 \psi ^2 + \frac{1}{2} N_f \left(1+N_f\right)a^3 H^2 \psi ^2 + N_f \left(1+3 N_f\right)a \cD^2 H^2 \psi ^2\\
&+ \frac{1}{2} N_f \left(1+N_f\right)a H^4 \psi ^2 + 3 N_f^2a^4 H \psi ^2 + 9 N_f^2a^2 \cD^2 H \psi ^2 + 8 N_f^2a^2 \cD H^2 \psi ^2 \\
&+ 3 N_f^2a^2 H^3 \psi ^2 + N_f^2a \cD H^3 \psi ^2 + \frac{1}{2} N_f \left(-1+3 N_f\right)a H^2 X \psi ^2 + 10 N_f^2a^2 \cD X \psi ^2\\
&+ 8 N_f^2a^2 H X \psi ^2 + \frac{1}{24} N_f \left(-6-193 N_f+18 N_f^2+397 N_f^3\right)a^2 \psi ^4\\
&+ \frac{1}{2} N_f^3 \left(3+7 N_f\right)a \cD \psi ^4 + \frac{1}{2} N_f^3 \left(-1+15 N_f\right)a H \psi ^4 \,,
\end{align*}

The CP-violating Hilbert series with $N_f=3$ is given by
\begin{align*}
\cH_{5,\text{CPV}}^{\notPQ}\,=\,& a^5 + a^3 H^2 + a H^4 + 3a X^2 + 21a H \psi ^2\,,\qquad
\cH_{6,\text{CPV}}^{\notPQ}\,=\, 3a^2 X^2 + 21a^2 H \psi ^2\,,\\
\cH_{7,\text{CPV}}^{\notPQ}\,=\,& a^7 + a^5 H^2 + a^3 \cD^2 H^2 + a^3 H^4 + 2a \cD^2 H^4 + a H^6 + 2a \cD^2 H^2 X + 3a^3 X^2\\
&+ 4a H^2 X^2 + 2a X^3 + 21a^3 H \psi ^2 + 42a \cD^2 H \psi ^2 + 54a \cD H^2 \psi ^2 + 21a H^3 \psi ^2 \numberthis\\
&+ 72a \cD X \psi ^2 + 60a H X \psi ^2 + 774a \psi ^4\,,\\
\cH_{8,\text{CPV}}^{\notPQ}\,=\,& a^2 \cD^2 H^4 + 2a^2 \cD^2 H^2 X + 3a^4 X^2 + 3a^2 \cD^2 X^2 + 4a^2 H^2 X^2 + 2a^2 X^3 + 9a^2 \cD^3 \psi ^2\\
&+ 21a^4 H \psi ^2 + 63a^2 \cD^2 H \psi ^2 + 54a^2 \cD H^2 \psi ^2 + 21a^2 H^3 \psi ^2 + 72a^2 \cD X \psi ^2\\
&+ 60a^2 H X \psi ^2 + 597a^2 \psi ^4\,,
\end{align*}

\subsection{aLEFT} \label{app:HSaLEFT}

In Section~\ref{sec:aLEFT}, the Hilbert series for aLEFT with (without) a shift symmetry is shown up to dimension~7~(6). In this section, we provide more results for Hilbert series at higher dimensions. For the shift-symmetric aLEFT, the Hilbert series from dimension~5 to dimension~8 are given by
\begin{align*}
\cH_5^{\aLEFTPQ} \,=\, & \partial a\, u_L u_L^{\dagger} + \partial a\, u_R u_R^{\dagger} + \partial a\, d_L d_L^{\dagger} + \partial a\, d_R d_R^{\dagger} + \partial a\, \nu_L \nu_L^{\dagger} + \partial a\, e_L e_L^{\dagger} + \partial a\, e_R e_R^{\dagger} \\
&- \partial a\, F_L \cD - \partial a\, F_R \cD - \partial a\, \cD^3 \, , \\
\cH_6^{\aLEFTPQ} \,=\, & 0 \, , \\
\cH_7^{\aLEFTPQ} \,=\, & (\partial a)^2 u_L u_R + (\partial a)^2 u_L^{\dagger} u_R^{\dagger} + (\partial a)^2 d_L d_R + (\partial a)^2 d_L^{\dagger} d_R^{\dagger} + (\partial a)^2 \nu_L^2 + (\partial a)^2 \nu_L^{\dagger2}\\
&  + (\partial a)^2 e_L e_R + (\partial a)^2 e_L^{\dagger} e_R^{\dagger} + \partial a\, u_L u_L^{\dagger} F_L + \partial a\, u_L u_L^{\dagger} F_R + \partial a\, u_L u_L^{\dagger} G_L \\
& + \partial a\, u_L u_L^{\dagger} G_R + \partial a\, u_R u_R^{\dagger} F_L + \partial a\, u_R u_R^{\dagger} F_R + \partial a\, u_R u_R^{\dagger} G_L + \partial a\, u_R u_R^{\dagger} G_R \\
& + \partial a\, d_L d_L^{\dagger} F_L + \partial a\, d_R d_R^{\dagger} F_L + \partial a\, d_L d_L^{\dagger} F_R + \partial a\, d_R d_R^{\dagger} F_R + \partial a\, d_L d_L^{\dagger} G_L \\
& + \partial a\, d_R d_R^{\dagger} G_L + \partial a\, d_L d_L^{\dagger} G_R + \partial a\, d_R d_R^{\dagger} G_R + \partial a\, \nu_L \nu_L^{\dagger} F_L + \partial a\, \nu_L \nu_L^{\dagger} F_R \\
& + \partial a\, e_L e_L^{\dagger} F_L + \partial a\, e_R e_R^{\dagger} F_L + \partial a\, e_L e_L^{\dagger} F_R + \partial a\, e_R e_R^{\dagger} F_R \, ,  \\
\cH_8^{\aLEFTPQ}  \,=\,& (\partial a)^4 + (\partial a)^2 u_L u_L^{\dagger} \cD + (\partial a)^2 u_R u_R^{\dagger} \cD + (\partial a)^2 d_L d_L^{\dagger} \cD + (\partial a)^2 d_R d_R^{\dagger} \cD + (\partial a)^2 \nu_L \nu_L^{\dagger} \cD \\
& + (\partial a)^2 e_L e_L^{\dagger} \cD + (\partial a)^2 e_R e_R^{\dagger} \cD + (\partial a)^2 F_L^2 + (\partial a)^2 F_L F_R + (\partial a)^2 F_R^2 + (\partial a)^2 G_L G_R \\
& + (\partial a)^2 G_L^2 + (\partial a)^2 G_R^2 + 2 \partial a\, u_L^2 u_L^{\dagger} u_R + 2 \partial a\, u_L u_L^{\dagger2} u_R^{\dagger} + 2 \partial a\, u_L u_R^2 u_R^{\dagger} + 2 \partial a\, u_L^{\dagger} u_R u_R^{\dagger2} \\
& + 4 \partial a\, u_L u_L^{\dagger} d_L d_R + 4 \partial a\, u_L u_L^{\dagger} d_L^{\dagger} d_R^{\dagger} + 4 \partial a\, u_L u_R d_L d_L^{\dagger} + 4 \partial a\, u_L u_R d_R d_R^{\dagger} \\
& + 4 \partial a\, u_L^{\dagger} u_R^{\dagger} d_L d_L^{\dagger} + 4 \partial a\, u_L^{\dagger} u_R^{\dagger} d_R d_R^{\dagger} + 4 \partial a\, u_R u_R^{\dagger} d_L d_R + 4 \partial a\, u_R u_R^{\dagger} d_L^{\dagger} d_R^{\dagger} \\
& + 2 \partial a\, d_L^2 d_L^{\dagger} d_R + 2 \partial a\, d_L d_L^{\dagger2} d_R^{\dagger} + 2 \partial a\, d_L d_R^2 d_R^{\dagger} + 2 \partial a\, d_L^{\dagger} d_R d_R^{\dagger2} + \partial a\, u_L^2 d_R^{\dagger} e_L \\
& + \partial a\, u_L^2 d_L e_R^{\dagger} + \partial a\, u_L^{\dagger2} d_R e_L^{\dagger} + \partial a\, u_L^{\dagger2} d_L^{\dagger} e_R + 2 \partial a\, u_L^{\dagger} u_R d_L^{\dagger} e_L^{\dagger} + 2 \partial a\, u_L^{\dagger} u_R d_R e_R \\
& + \partial a\, u_R^2 d_R e_L^{\dagger} + \partial a\, u_R^2 d_L^{\dagger} e_R + 2 \partial a\, u_L u_R^{\dagger} d_L e_L + 2 \partial a\, u_L u_R^{\dagger} d_R^{\dagger} e_R^{\dagger} + \partial a\, u_R^{\dagger2} d_R^{\dagger} e_L \\
& + \partial a\, u_R^{\dagger2} d_L e_R^{\dagger} + 2 \partial a\, u_L d_L d_R^{\dagger} \nu_L + \partial a\, u_L d_L^2 \nu_L^{\dagger} + \partial a\, u_L d_R^{\dagger2} \nu_L^{\dagger} + \partial a\, u_L^{\dagger} d_L^{\dagger2} \nu_L \\
& + \partial a\, u_L^{\dagger} d_R^2 \nu_L + 2 \partial a\, u_L^{\dagger} d_L^{\dagger} d_R \nu_L^{\dagger} + 2 \partial a\, u_R d_L^{\dagger} d_R \nu_L + \partial a\, u_R d_L^{\dagger2} \nu_L^{\dagger} + \partial a\, u_R d_R^2 \nu_L^{\dagger} \\
& + \partial a\, u_R^{\dagger} d_L^2 \nu_L + \partial a\, u_R^{\dagger} d_R^{\dagger2} \nu_L + 2 \partial a\, u_R^{\dagger} d_L d_R^{\dagger} \nu_L^{\dagger} + \partial a\, d_L^{\dagger} d_R^2 e_L + \partial a\, d_L d_R^{\dagger2} e_L^{\dagger} \numberthis\\
& + \partial a\, d_L^2 d_R^{\dagger} e_R + \partial a\, d_L^{\dagger2} d_R e_R^{\dagger} + \partial a\, u_L u_L^{\dagger} \nu_L^2 + \partial a\, u_L u_L^{\dagger} \nu_L^{\dagger2} + 2 \partial a\, u_L u_R \nu_L \nu_L^{\dagger} \\
& + 2 \partial a\, u_L^{\dagger} u_R^{\dagger} \nu_L \nu_L^{\dagger} + \partial a\, u_R u_R^{\dagger} \nu_L^2 + \partial a\, u_R u_R^{\dagger} \nu_L^{\dagger2} + 2 \partial a\, u_L u_L^{\dagger} e_L e_R + 2 \partial a\, u_L u_L^{\dagger} e_L^{\dagger} e_R^{\dagger} \\
& + 2 \partial a\, u_L u_R e_L e_L^{\dagger} + 2 \partial a\, u_L u_R e_R e_R^{\dagger} + 2 \partial a\, u_L^{\dagger} u_R^{\dagger} e_L e_L^{\dagger} + 2 \partial a\, u_L^{\dagger} u_R^{\dagger} e_R e_R^{\dagger} \\
& + 2 \partial a\, u_R u_R^{\dagger} e_L e_R + 2 \partial a\, u_R u_R^{\dagger} e_L^{\dagger} e_R^{\dagger} + 2 \partial a\, u_L u_R F_L \cD + \partial a\, u_L u_R F_R \cD \\
& + 2 \partial a\, u_L u_R G_L \cD + \partial a\, u_L u_R G_R \cD + \partial a\, u_L^{\dagger} u_R^{\dagger} F_L \cD + 2 \partial a\, u_L^{\dagger} u_R^{\dagger} F_R \cD \\
& + \partial a\, u_L^{\dagger} u_R^{\dagger} G_L \cD + 2 \partial a\, u_L^{\dagger} u_R^{\dagger} G_R \cD + 2 \partial a\, u_L d_L^{\dagger} \nu_L e_L + 2 \partial a\, u_L d_R \nu_L e_R^{\dagger} \\
& + 2 \partial a\, u_L d_R \nu_L^{\dagger} e_L + 2 \partial a\, u_L d_L^{\dagger} \nu_L^{\dagger} e_R^{\dagger} + 2 \partial a\, u_L^{\dagger} d_R^{\dagger} \nu_L e_L^{\dagger} + 2 \partial a\, u_L^{\dagger} d_L \nu_L e_R \\
& + 2 \partial a\, u_L^{\dagger} d_L \nu_L^{\dagger} e_L^{\dagger} + 2 \partial a\, u_L^{\dagger} d_R^{\dagger} \nu_L^{\dagger} e_R + 2 \partial a\, u_R d_L \nu_L e_L^{\dagger} + 2 \partial a\, u_R d_R^{\dagger} \nu_L e_R \\
& + 2 \partial a\, u_R d_R^{\dagger} \nu_L^{\dagger} e_L^{\dagger} + 2 \partial a\, u_R d_L \nu_L^{\dagger} e_R + 2 \partial a\, u_R^{\dagger} d_R \nu_L e_L + 2 \partial a\, u_R^{\dagger} d_L^{\dagger} \nu_L e_R^{\dagger} \\
& + 2 \partial a\, u_R^{\dagger} d_L^{\dagger} \nu_L^{\dagger} e_L + 2 \partial a\, u_R^{\dagger} d_R \nu_L^{\dagger} e_R^{\dagger} + \partial a\, d_L d_L^{\dagger} \nu_L^2 + \partial a\, d_R d_R^{\dagger} \nu_L^2 + 2 \partial a\, d_L d_R \nu_L \nu_L^{\dagger} \\
& + 2 \partial a\, d_L^{\dagger} d_R^{\dagger} \nu_L \nu_L^{\dagger} + \partial a\, d_L d_L^{\dagger} \nu_L^{\dagger2} + \partial a\, d_R d_R^{\dagger} \nu_L^{\dagger2} + 2 \partial a\, d_L d_R e_L e_L^{\dagger} + 2 \partial a\, d_L^{\dagger} d_R^{\dagger} e_L e_L^{\dagger} \\
& + 2 \partial a\, d_L d_L^{\dagger} e_L e_R + 2 \partial a\, d_R d_R^{\dagger} e_L e_R + 2 \partial a\, d_L d_L^{\dagger} e_L^{\dagger} e_R^{\dagger} + 2 \partial a\, d_R d_R^{\dagger} e_L^{\dagger} e_R^{\dagger} \\
& + 2 \partial a\, d_L d_R e_R e_R^{\dagger} + 2 \partial a\, d_L^{\dagger} d_R^{\dagger} e_R e_R^{\dagger} + 2 \partial a\, d_L d_R F_L \cD + \partial a\, d_L^{\dagger} d_R^{\dagger} F_L \cD + \partial a\, d_L d_R F_R \cD \\
& + 2 \partial a\, d_L^{\dagger} d_R^{\dagger} F_R \cD + 2 \partial a\, d_L d_R G_L \cD + \partial a\, d_L^{\dagger} d_R^{\dagger} G_L \cD + \partial a\, d_L d_R G_R \cD + 2 \partial a\, d_L^{\dagger} d_R^{\dagger} G_R \cD \\
& + \partial a\, \nu_L^2 e_L e_L^{\dagger} + \partial a\, \nu_L^2 e_R e_R^{\dagger} + 2 \partial a\, \nu_L \nu_L^{\dagger} e_L e_R + 2 \partial a\, \nu_L \nu_L^{\dagger} e_L^{\dagger} e_R^{\dagger} + \partial a\, \nu_L^{\dagger2} e_L e_L^{\dagger} \\
& + \partial a\, \nu_L^{\dagger2} e_R e_R^{\dagger} + \partial a\, e_L^2 e_L^{\dagger} e_R + \partial a\, e_L e_L^{\dagger2} e_R^{\dagger} + \partial a\, e_L e_R^2 e_R^{\dagger} + \partial a\, e_L^{\dagger} e_R e_R^{\dagger2} + \partial a\, \nu_L^2 F_L \cD \\
& + \partial a\, \nu_L^{\dagger2} F_R \cD + 2 \partial a\, e_L e_R F_L \cD + \partial a\, e_L^{\dagger} e_R^{\dagger} F_L \cD + \partial a\, e_L e_R F_R \cD + 2 \partial a\, e_L^{\dagger} e_R^{\dagger} F_R \cD \, . \\
\end{align*}
Note that the number of quark flavors $N_{u,d}$ and lepton flavors $N_{e,\nu}$ in above Hilbert series are all set to 1. In the ancillary file, we provide the Hilbert series with all flavor dependence. Based on the full Hilbert series, and assume that $N_\nu=N_e=N_d$, we can obtain the operator counting with flavor independence, which are shown up to dimension~10 as follows,
\begin{align*}
 \#\,\cO_{5}^{\aLEFTPQ} \,=\,& 2-2N_d+5 N_d^2-N_u+2 N_u^2 \, , \\
 \#\,\cO_{6}^{\aLEFTPQ} \,=\,& 0 \, , \\
 \#\,\cO_{7}^{\aLEFTPQ} \,=\,& \left(18 N_d^2+10 N_u^2\right)+\left(N_d+N_d^2\right) \epsilon_L^2 \, , \\
 \#\,\cO_{8}^{\aLEFTPQ} \,=\,& \left(7+23 N_d^2+36 N_d^4+16 N_d^3 N_u+14 N_u^2+52 N_d^2 N_u^2+8 N_u^4\right)\\
 &+\left(-\frac{4 N_d^2}{3}+\frac{16 N_d^4}{3}+16 N_d^3 N_u+16 N_d^2 N_u^2\right) \epsilon_B \epsilon_L\\
 &+\left(-N_d+\frac{7 N_d^2}{3}+\frac{26 N_d^4}{3}+16 N_d^3 N_u+4 N_d^2 N_u^2\right) \epsilon_L^2 \, ,\\
\#\,\cO_{9}^{\aLEFTPQ} \,=\,& \left(6+\frac{429 N_d^2}{4}-\frac{5 N_d^3}{2}+\frac{393 N_d^4}{4}+44 N_d^3 N_u+66 N_u^2+142 N_d^2 N_u^2+22 N_u^4\right)\\
&+\left(\frac{4 N_d^2}{3}+4 N_d^3+\frac{44 N_d^4}{3}+8 N_d^2 N_u+44 N_d^3 N_u+44 N_d^2 N_u^2\right) \epsilon_B \epsilon_L \numberthis\\
&+\left(4 N_d^2-4 N_d^3+24 N_d^4+44 N_d^3 N_u-2 N_d N_u^2+12 N_d^2 N_u^2\right) \epsilon_L^2\\
&+\left(-\frac{N_d}{2}+\frac{5 N_d^2}{12}-\frac{N_d^3}{2}+\frac{7 N_d^4}{12}\right) \epsilon_L^4 \, , \\
\#\,\cO_{10}^{\aLEFTPQ} \,=\,& \left(17+\frac{539 N_d^2}{2}-2 N_d^3+\frac{865 N_d^4}{2}+192 N_d^3 N_u+186 N_u^2+704 N_d^2 N_u^2+116 N_u^4\right)\\
&+\left(-\frac{4 N_d^2}{3}+4 N_d^3+\frac{232 N_d^4}{3}+8 N_d^2 N_u+232 N_d^3 N_u+232 N_d^2 N_u^2\right) \epsilon_B \epsilon_L\\
&+\left(-N_d+\frac{79 N_d^2}{3}-7 N_d^3+\frac{275 N_d^4}{3}+192 N_d^3 N_u-4 N_d N_u^2+48 N_d^2 N_u^2\right) \epsilon_L^2\\
&+\left(-\frac{N_d^2}{6}+\frac{N_d^4}{6}\right) \epsilon_L^4 \, .
\end{align*}

For aLEFT without a shift symmetry, the Hilbert series from dimension~5 to dimension~7 are shown as follows
\begin{align*}
\cH_{5}^{\aLEFTnotPQ} \,=\, & a^5 + a^2 u_L u_R + a^2 u_L^{\dagger} u_R^{\dagger} + a^2 d_L d_R + a^2 d_L^{\dagger} d_R^{\dagger} + a^2 \nu_L^2 + a^2 \nu_L^{\dagger2} + a^2 e_L e_R + a^2 e_L^{\dagger} e_R^{\dagger} \\
&+ a F_L^2 + a F_R^2 + a G_L^2 + a G_R^2 \, , \\
\cH_{6}^{\aLEFTnotPQ} \,=\, & a^6 + a^3 u_L u_R + a^3 u_L^{\dagger} u_R^{\dagger} + a^3 d_L d_R + a^3 d_L^{\dagger} d_R^{\dagger} + a^3 \nu_L^2 + a^3 \nu_L^{\dagger2} + a^3 e_L e_R + a^3 e_L^{\dagger} e_R^{\dagger} \\
& + a^2 F_L^2 + a^2 F_R^2 + a^2 G_L^2 + a^2 G_R^2 + a u_L u_R F_L + a u_L u_R G_L + a u_L^{\dagger} u_R^{\dagger} F_R + a u_L^{\dagger} u_R^{\dagger} G_R \\
& + a d_L d_R F_L + a d_L^{\dagger} d_R^{\dagger} F_R + a d_L d_R G_L + a d_L^{\dagger} d_R^{\dagger} G_R + a e_L e_R F_L + a e_L^{\dagger} e_R^{\dagger} F_R \, , \\
\cH_{7}^{\aLEFTnotPQ} \,=\, & a^7 + a^4 u_L u_R + a^4 u_L^{\dagger} u_R^{\dagger} + a^4 d_L d_R + a^4 d_L^{\dagger} d_R^{\dagger} + a^4 \nu_L^2 + a^4 \nu_L^{\dagger2} + a^4 e_L e_R + a^4 e_L^{\dagger} e_R^{\dagger} \\
&+ a^3 F_L^2 + a^3 F_R^2 + a^3 G_L^2 + a^3 G_R^2 + a^2 u_L u_R F_L + a^2 u_L u_R G_L + a^2 u_L^{\dagger} u_R^{\dagger} F_R \\
& + a^2 u_L^{\dagger} u_R^{\dagger} G_R + a^2 u_L u_R \cD^2 + a^2 u_L^{\dagger} u_R^{\dagger} \cD^2 + a^2 d_L d_R F_L + a^2 d_L^{\dagger} d_R^{\dagger} F_R + a^2 d_L d_R G_L \\
& + a^2 d_L^{\dagger} d_R^{\dagger} G_R + a^2 d_L d_R \cD^2 + a^2 d_L^{\dagger} d_R^{\dagger} \cD^2 + a^2 \nu_L^2 \cD^2 + a^2 \nu_L^{\dagger2} \cD^2 + a^2 e_L e_R F_L \\
& + a^2 e_L^{\dagger} e_R^{\dagger} F_R + a^2 e_L e_R \cD^2 + a^2 e_L^{\dagger} e_R^{\dagger} \cD^2 + a u_L^2 u_L^{\dagger2} + 2 a u_L^2 u_R^2 + 2 a u_L u_L^{\dagger} u_R u_R^{\dagger} \\
& + 2 a u_L^{\dagger2} u_R^{\dagger2} + a u_R^2 u_R^{\dagger2} + 2 a u_L u_L^{\dagger} d_L d_L^{\dagger} + 2 a u_L u_L^{\dagger} d_R d_R^{\dagger} + 4 a u_L u_R d_L d_R \\
& + 2 a u_L u_R d_L^{\dagger} d_R^{\dagger} + 2 a u_L^{\dagger} u_R^{\dagger} d_L d_R + 4 a u_L^{\dagger} u_R^{\dagger} d_L^{\dagger} d_R^{\dagger} + 2 a u_R u_R^{\dagger} d_L d_L^{\dagger} + 2 a u_R u_R^{\dagger} d_R d_R^{\dagger} \\
& + a d_L^2 d_L^{\dagger2} + 2 a d_L^2 d_R^2 + 2 a d_L d_L^{\dagger} d_R d_R^{\dagger} + 2 a d_L^{\dagger2} d_R^{\dagger2} + a d_R^2 d_R^{\dagger2} + a u_L^2 d_L e_L + a u_L^{\dagger2} d_L^{\dagger} e_L^{\dagger} \\
& + a u_L^{\dagger} u_R d_R e_L^{\dagger} + a u_L^{\dagger} u_R d_L^{\dagger} e_R + a u_R^2 d_R e_R + a u_L u_R^{\dagger} d_R^{\dagger} e_L + a u_L u_R^{\dagger} d_L e_R^{\dagger} + a u_R^{\dagger2} d_R^{\dagger} e_R^{\dagger} \\
& + a u_L d_L^2 \nu_L + a u_L d_L d_R^{\dagger} \nu_L^{\dagger} + a u_L^{\dagger} d_L^{\dagger} d_R \nu_L + a u_L^{\dagger} d_L^{\dagger2} \nu_L^{\dagger} + a u_R d_R^2 \nu_L + a u_R d_L^{\dagger} d_R \nu_L^{\dagger} \\
& + a u_R^{\dagger} d_L d_R^{\dagger} \nu_L + a u_R^{\dagger} d_R^{\dagger2} \nu_L^{\dagger} + a u_L u_L^{\dagger} \nu_L \nu_L^{\dagger} + a u_L u_R \nu_L^2 + a u_L u_R \nu_L^{\dagger2} + a u_L^{\dagger} u_R^{\dagger} \nu_L^2 \\
& + a u_L^{\dagger} u_R^{\dagger} \nu_L^{\dagger2} + a u_R u_R^{\dagger} \nu_L \nu_L^{\dagger} + a u_L u_L^{\dagger} e_L e_L^{\dagger} + a u_L u_L^{\dagger} e_R e_R^{\dagger} + 2 a u_L u_R e_L e_R \numberthis\\
& + a u_L u_R e_L^{\dagger} e_R^{\dagger} + a u_L^{\dagger} u_R^{\dagger} e_L e_R + 2 a u_L^{\dagger} u_R^{\dagger} e_L^{\dagger} e_R^{\dagger} + a u_R u_R^{\dagger} e_L e_L^{\dagger} + a u_R u_R^{\dagger} e_R e_R^{\dagger} \\
& + a u_L u_L^{\dagger} F_L \cD + a u_L u_L^{\dagger} F_R \cD + a u_L u_L^{\dagger} G_L \cD + a u_L u_L^{\dagger} G_R \cD + a u_R u_R^{\dagger} F_L \cD \\
& + a u_R u_R^{\dagger} F_R \cD + a u_R u_R^{\dagger} G_L \cD + a u_R u_R^{\dagger} G_R \cD + 2 a u_L d_R \nu_L e_L + a u_L d_L^{\dagger} \nu_L e_R^{\dagger} \\
& + a u_L d_L^{\dagger} \nu_L^{\dagger} e_L + a u_L d_R \nu_L^{\dagger} e_R^{\dagger} + a u_L^{\dagger} d_L \nu_L e_L^{\dagger} + a u_L^{\dagger} d_R^{\dagger} \nu_L e_R + 2 a u_L^{\dagger} d_R^{\dagger} \nu_L^{\dagger} e_L^{\dagger} \\
& + a u_L^{\dagger} d_L \nu_L^{\dagger} e_R + a u_R d_R^{\dagger} \nu_L e_L^{\dagger} + 2 a u_R d_L \nu_L e_R + a u_R d_L \nu_L^{\dagger} e_L^{\dagger} + a u_R d_R^{\dagger} \nu_L^{\dagger} e_R \\
& + a u_R^{\dagger} d_L^{\dagger} \nu_L e_L + a u_R^{\dagger} d_R \nu_L e_R^{\dagger} + a u_R^{\dagger} d_R \nu_L^{\dagger} e_L + 2 a u_R^{\dagger} d_L^{\dagger} \nu_L^{\dagger} e_R^{\dagger} + a d_L d_R \nu_L^2 + a d_L^{\dagger} d_R^{\dagger} \nu_L^2 \\
& + a d_L d_L^{\dagger} \nu_L \nu_L^{\dagger} + a d_R d_R^{\dagger} \nu_L \nu_L^{\dagger} + a d_L d_R \nu_L^{\dagger2} + a d_L^{\dagger} d_R^{\dagger} \nu_L^{\dagger2} + a d_L d_L^{\dagger} e_L e_L^{\dagger} + a d_R d_R^{\dagger} e_L e_L^{\dagger} \\
& + 2 a d_L d_R e_L e_R + a d_L^{\dagger} d_R^{\dagger} e_L e_R + a d_L d_R e_L^{\dagger} e_R^{\dagger} + 2 a d_L^{\dagger} d_R^{\dagger} e_L^{\dagger} e_R^{\dagger} + a d_L d_L^{\dagger} e_R e_R^{\dagger} \\
& + a d_R d_R^{\dagger} e_R e_R^{\dagger} + a d_L d_L^{\dagger} F_L \cD + a d_R d_R^{\dagger} F_L \cD + a d_L d_L^{\dagger} F_R \cD + a d_R d_R^{\dagger} F_R \cD + a d_L d_L^{\dagger} G_L \cD \\
& + a d_R d_R^{\dagger} G_L \cD + a d_L d_L^{\dagger} G_R \cD + a d_R d_R^{\dagger} G_R \cD + a \nu_L^2 \nu_L^{\dagger2} + a \nu_L^2 e_L e_R + a \nu_L^2 e_L^{\dagger} e_R^{\dagger} \\
& + a \nu_L \nu_L^{\dagger} e_L e_L^{\dagger} + a \nu_L \nu_L^{\dagger} e_R e_R^{\dagger} + a \nu_L^{\dagger2} e_L e_R + a \nu_L^{\dagger2} e_L^{\dagger} e_R^{\dagger} + a e_L^2 e_L^{\dagger2} + a e_L^2 e_R^2 + a e_L e_L^{\dagger} e_R e_R^{\dagger} \\
& + a e_L^{\dagger2} e_R^{\dagger2} + a e_R^2 e_R^{\dagger2} + a \nu_L \nu_L^{\dagger} F_L \cD + a \nu_L \nu_L^{\dagger} F_R \cD + a e_L e_L^{\dagger} F_L \cD + a e_R e_R^{\dagger} F_L \cD \\
& + a e_L e_L^{\dagger} F_R \cD + a e_R e_R^{\dagger} F_R \cD + a G_L^3 + a G_R^3 \, ,
\end{align*}
where the flavors of fermions are set to 1, and the general Hilbert series with flavor dependence is shown in the ancillary file. The numbers of operators from dimension~5 to dimension~10 are presented in powers of $\epsilon_{B,L}$, which are shown as follows,
\begin{align*}
\#\,\cO_{5}^{\aLEFTnotPQ} \,=\,& \left(5+4 N_d^2+2 N_u^2\right)+\left(N_d+N_d^2\right) \epsilon_L^2 \, , \\
\#\,\cO_{6}^{\aLEFTnotPQ} \,=\,& \left(5+10 N_d^2+6 N_u^2\right)+2 N_d^2 \epsilon_L^2 \, , \\
\#\,\cO_{7}^{\aLEFTnotPQ} \,=\,& \left(7+\frac{131 N_d^2}{4}+\frac{3 N_d^3}{2}+\frac{87 N_d^4}{4}+10 N_d^3 N_u+19 N_u^2+32 N_d^2 N_u^2+5 N_u^4\right)\\
&+\left(-\frac{4 N_d^2}{3}-2 N_d^3+\frac{10 N_d^4}{3}-4 N_d^2 N_u+10 N_d^3 N_u+10 N_d^2 N_u^2\right) \epsilon_B \epsilon_L\\
&+\left(N_d+3 N_d^2+2 N_d^3+6 N_d^4+10 N_d^3 N_u+N_d N_u^2+3 N_d^2 N_u^2\right) \epsilon_L^2\\
&+\left(-\frac{N_d^2}{6}+\frac{N_d^4}{6}\right) \epsilon_L^4 \, , \\
\#\,\cO_{8}^{\aLEFTnotPQ} \,=\,&\left(14+\frac{335 N_d^2}{4}-\frac{N_d^3}{2}+\frac{303 N_d^4}{4}+34 N_d^3 N_u+53 N_u^2+110 N_d^2 N_u^2+17 N_u^4\right)\\
&+\left(-\frac{4 N_d^2}{3}+2 N_d^3+\frac{34 N_d^4}{3}+4 N_d^2 N_u+34 N_d^3 N_u+34 N_d^2 N_u^2\right) \epsilon_B \epsilon_L\\
&+\left(4 N_d+10 N_d^2-3 N_d^3+19 N_d^4+34 N_d^3 N_u-N_d N_u^2+9 N_d^2 N_u^2\right) \epsilon_L^2\\
&+\left(-\frac{N_d^2}{6}+\frac{N_d^4}{6}\right) \epsilon_L^4 \, ,\\
\#\,\cO_{9}^{\aLEFTnotPQ} \,=\,& \left(43+220 N_d^2-3 N_d^3+294 N_d^4+132 N_d^3 N_u+141 N_u^2+462 N_d^2 N_u^2+75 N_u^4\right)\\
&+\left(6 N_d^3+50 N_d^4+12 N_d^2 N_u+150 N_d^3 N_u+150 N_d^2 N_u^2\right) \epsilon_B \epsilon_L\\
&+\left(4 N_d+14 N_d^2-8 N_d^3+68 N_d^4+132 N_d^3 N_u-4 N_d N_u^2+36 N_d^2 N_u^2\right) \epsilon_L^2\\
&+\left(\frac{N_d}{2}+\frac{3 N_d^2}{4}-\frac{N_d^3}{2}+\frac{5 N_d^4}{4}\right) \epsilon_L^4 \, , \numberthis\\
\#\,\cO_{10}^{\aLEFTnotPQ} \,=\,& \left(60+\frac{11435 N_d^2}{18}-\frac{25 N_d^3}{3}+\frac{18407 N_d^4}{18}+\frac{10 N_d^5}{3}+\frac{865 N_d^6}{9}+452 N_d^3 N_u+3 N_d^4 N_u\right.\\
&+113 N_d^5 N_u+\frac{1309 N_u^2}{3}+\frac{3409}{2} N_d^2 N_u^2+2 N_d^3 N_u^2+\frac{623}{2} N_d^4 N_u^2+70 N_d^3 N_u^3\\
&\left.+286 N_u^4+148 N_d^2 N_u^4+\frac{35 N_u^6}{3}\right)+\left(\frac{N_d N_u}{2}+\frac{97}{12} N_d^2 N_u+\frac{3}{2} N_d^3 N_u-\frac{1}{12} N_d^4 N_u\right.\\
&\left.+\frac{1}{2} N_d N_u^2+\frac{41}{12} N_d^2 N_u^2-\frac{1}{2} N_d^3 N_u^2+\frac{175}{12} N_d^4 N_u^2\right) \epsilon_B^2+\left(-\frac{8 N_d^2}{3}+13 N_d^3\right.\\
&\left.+178 N_d^4-2 N_d^5+\frac{113 N_d^6}{3}+\frac{82}{3} N_d^2 N_u+\frac{3331}{6} N_d^3 N_u-\frac{9}{2} N_d^4 N_u+148 N_d^5 N_u\right.\\
&\left.+\frac{1123}{2} N_d^2 N_u^2-3 N_d^3 N_u^2+\frac{471}{2} N_d^4 N_u^2-\frac{17}{6} N_d^2 N_u^3+\frac{665}{6} N_d^3 N_u^3+70 N_d^2 N_u^4\right) \epsilon_B \epsilon_L\\
&+\left(5 N_d+\frac{175 N_d^2}{3}-\frac{46 N_d^3}{3}+\frac{641 N_d^4}{3}+\frac{7 N_d^5}{3}+36 N_d^6+9 N_d^2 N_u+\frac{1357}{3} N_d^3 N_u\right.\\
&\left.+2 N_d^4 N_u+\frac{350}{3} N_d^5 N_u-\frac{15}{2} N_d N_u^2+\frac{697}{6} N_d^2 N_u^2+4 N_d^3 N_u^2+\frac{280}{3} N_d^4 N_u^2\right.\\
&\left.+70 N_d^3 N_u^3+\frac{1}{2} N_d N_u^4+\frac{19}{2} N_d^2 N_u^4\right) \epsilon_L^2+\left(-\frac{2 N_d^3}{3}-\frac{13 N_d^4}{6}-\frac{N_d^5}{3}+\frac{19 N_d^6}{6}\right.\\
&\left.+\frac{2}{3} N_d^2 N_u-\frac{13}{2} N_d^3 N_u-\frac{11}{6} N_d^4 N_u+\frac{11}{3} N_d^5 N_u-\frac{3}{2} N_d^2 N_u^2+\frac{19}{2} N_d^4 N_u^2\right.\\
&\left.+\frac{1}{6} N_d^2 N_u^3+\frac{35}{6} N_d^3 N_u^3\right) \epsilon_B \epsilon_L^3+\left(\frac{N_d}{2}+\frac{7 N_d^2}{12}+\frac{N_d^3}{2}+\frac{N_d^4}{4}-N_d^5+\frac{7 N_d^6}{6}\right.\\
&\left.-\frac{8}{3} N_d^3 N_u-N_d^4 N_u+\frac{11}{3} N_d^5 N_u+\frac{1}{2} N_d N_u^2-\frac{7}{12} N_d^2 N_u^2-\frac{1}{2} N_d^3 N_u^2+\frac{7}{12} N_d^4 N_u^2\right) \epsilon_L^4\\
&+\left(-\frac{N_d^2}{36}+\frac{N_d^3}{24}+\frac{N_d^4}{72}-\frac{N_d^5}{24}+\frac{N_d^6}{72}\right) \epsilon_L^6 \, .
\end{align*}

In the following, we show the reduced Hilbert series for CP-even, CP-odd and CP-violating operators up to dimension~8. For $\aLEFTPQ$, the CP-even Hilbert series is given by
\begin{align*}
\cH_{5,\text{even}}^{\aLEFTPQ} \,=\, & 2a X^2 + \frac{1}{2}\left(N_d+5N_d^2+2N_u^2\right)\partial a\, \psi ^2 \, , \\
\cH_{6,\text{even}}^{\aLEFTPQ} \,=\, & 0 \, , \\
\cH_{7,\text{even}}^{\aLEFTPQ} \,=\, & \frac{1}{2}\left(N_d+5N_d^2+2N_u^2\right)(\partial a)^2 \psi ^2 + \left(7N_d^2+4N_u^2\right)\partial a\, X \psi ^2 \, , \numberthis \\
\cH_{8,\text{even}}^{\aLEFTPQ}  \,=\,&  (\partial a)^4 + 4(\partial a)^2 X^2 + \left(25N_d^4+24N_d^3N_u+4N_u^4+N_d^2\left(-1+36N_u^2\right)\right)\partial a\, \psi ^4\\
& + \left(\frac{5N_d}{2}+\frac{5N_d^2}{2}+N_u+N_u^2\right)(\partial a)^2 \cD \psi ^2+ \left(-\frac{N_d}{2}+\frac{21N_d^2}{2}+6N_u^2\right)\partial a\, \cD X \psi ^2\, . \\
\end{align*}

The CP-odd Hilbert series of $\aLEFTPQ$ is shown as
\begin{align*}
\cH_{5,\text{odd}}^{\aLEFTPQ} \,=\, & \left(-\frac{5N_d}{2}+\frac{5N_d^2}{2}+\left(-1+N_u\right)N_u\right)\partial a\, \psi ^2 \, , \\
\cH_{6,\text{odd}}^{\aLEFTPQ} \,=\, & 0 \, , \\
\cH_{7,\text{odd}}^{\aLEFTPQ} \,=\, & \frac{1}{2}\left(N_d+5N_d^2+2N_u^2\right)(\partial a)^2 \psi ^2 + \left(7N_d^2+4N_u^2\right)\partial a\, X \psi ^2 \, , \numberthis \\
\cH_{8,\text{odd}}^{\aLEFTPQ}  \,=\,& 2(\partial a)^2 X^2  + \left(25N_d^4+24N_d^3N_u+4N_u^4+N_d^2\left(-1+36N_u^2\right)\right)\partial a\, \psi ^4\\
& + \left(-\frac{5N_d}{2}+\frac{5N_d^2}{2}+\left(-1+N_u\right)N_u\right)(\partial a)^2 \cD \psi ^2 + \left(-\frac{N_d}{2}+\frac{21N_d^2}{2}+6N_u^2\right)\partial a\, \cD X \psi ^2 \, . \\
\end{align*}

The CP-violating Hilbert series of $\aLEFTPQ$ with $N_u=2$ and $N_d=N_e=N_\nu=3$ is shown as
\begin{align*}
\cH_{5,\text{CPV}}^{\aLEFTPQ} \,=\, & 3\partial a\, \psi ^2 \, , \\
\cH_{6,\text{CPV}}^{\aLEFTPQ} \,=\, & 0 \, , \\
\cH_{7,\text{CPV}}^{\aLEFTPQ} \,=\, & 14(\partial a)^2 \psi ^2 + 35\partial a\, X \psi ^2 \, , \numberthis \\
\cH_{8,\text{CPV}}^{\aLEFTPQ}  \,=\,& 2(\partial a)^2 X^2 + 3(\partial a)^2 \cD \psi ^2 + 51\partial a\, \cD X \psi ^2 + 642\partial a\, \psi ^4 \, . \\
\end{align*}

For the $\aLEFTnotPQ$, the CP-even Hilbert series is given by
\begin{align*}
\cH_{5,\text{even}}^{\aLEFTnotPQ} \,=\, & 2a X^2 + \frac{1}{2}\left(N_d+5N_d^2+2N_u^2\right)a^2 \psi ^2 \, , \\
\cH_{6,\text{even}}^{\aLEFTnotPQ} \,=\, & a^6 + 2a^2 X^2 + \frac{1}{2}\left(N_d+5N_d^2+2N_u^2\right)a^3 \psi ^2 + \frac{1}{2}\left(-N_d+7N_d^2+4N_u^2\right)a X \psi ^2 \, , \\
\cH_{7,\text{even}}^{\aLEFTnotPQ} \,=\, & 2a^3 X^2 + a X^3 + \frac{1}{2}\left(N_d+5N_d^2+2N_u^2\right)a^4 \psi ^2 + \frac{1}{2}\left(N_d+5N_d^2+2N_u^2\right)a^2 \cD^2 \psi ^2\\
&+ \frac{1}{2}\left(-N_d+7N_d^2+4N_u^2\right)a^2 X \psi ^2 + \left(7N_d^2+4N_u^2\right)a \cD X \psi ^2\\
&+ \frac{1}{8}\left(125N_d^4+6N_d^3\left(1+20N_u\right)+N_d\left(-6-56N_u+4N_u^2\right)+4N_u^2\left(-1+5N_u^2\right)\right.\\
&\left.+N_d^2\left(-45-16N_u+180N_u^2\right)\right)a \psi ^4 \, ,  \\
\cH_{8,\text{even}}^{\aLEFTnotPQ}  \,=\,& a^8 + a^4 \cD^4 + 2a^4 X^2 + 4a^2 \cD^2 X^2 + a^2 X^3 + \frac{1}{2}\left(N_d+5N_d^2+2N_u^2\right)a^5 \psi ^2 \numberthis\\
&+ \frac{1}{2}\left(N_d+5N_d^2+2N_u^2\right)a^3 \cD^2 \psi ^2 + \left(\frac{5N_d}{2}+\frac{5N_d^2}{2}+N_u+N_u^2\right)a^2 \cD^3 \psi ^2\\
&+ \frac{1}{2}\left(-N_d+7N_d^2+4N_u^2\right)a^3 X \psi ^2 + \left(-\frac{N_d}{2}+\frac{21N_d^2}{2}+6N_u^2\right)a \cD^2 X \psi ^2\\
&+ \left(7N_d^2+4N_u^2\right)a^2 \cD X \psi ^2 + 2\left(N_d+8N_d^2+5N_u^2\right)a X^2 \psi ^2\\
&+ \frac{1}{8}\left(125N_d^4+6N_d^3\left(1+20N_u\right)+N_d\left(6+56N_u+4N_u^2\right)+4N_u^2\left(7+5N_u^2\right)\right.\\
&\left.+N_d^2\left(71-16N_u+180N_u^2\right)\right)a^2 \psi ^4 \\
&+ \left(\frac{75N_d^4}{2}-N_dN_u^2+6N_u^4+2N_d^2N_u\left(2+27N_u\right)+N_d^3\left(-\frac{3}{2}+36N_u\right)\right)a \cD \psi ^4 \, . \\
\end{align*}

The CP-odd Hilbert series of $\aLEFTnotPQ$ is given as
\begin{align*}
\cH_{5,\text{odd}}^{\aLEFTnotPQ} \,=\, & a^5 + 2a X^2 + \frac{1}{2}\left(N_d+5N_d^2+2N_u^2\right)a^2 \psi ^2 \, , \\
\cH_{6,\text{odd}}^{\aLEFTnotPQ} \,=\, & 2a^2 X^2 + \frac{1}{2}\left(N_d+5N_d^2+2N_u^2\right)a^3 \psi ^2 + \frac{1}{2}\left(-N_d+7N_d^2+4N_u^2\right)a X \psi ^2 \, , \\
\cH_{7,\text{odd}}^{\aLEFTnotPQ} \,=\, & a^7 + 2a^3 X^2 + a X^3 + \frac{1}{2}\left(N_d+5N_d^2+2N_u^2\right)a^4 \psi ^2 + \frac{1}{2}\left(N_d+5N_d^2+2N_u^2\right)a^2 \cD^2 \psi ^2 \\
&+ \frac{1}{2}\left(-N_d+7N_d^2+4N_u^2\right)a^2 X \psi ^2 + \left(7N_d^2+4N_u^2\right)a \cD X \psi ^2\\
&+ \frac{1}{8}\left(125N_d^4+6N_d^3\left(1+20N_u\right)+N_d\left(6+56N_u+4N_u^2\right)+4N_u^2\left(7+5N_u^2\right)\right.\\
&\left.+N_d^2\left(71-16N_u+180N_u^2\right)\right)a \psi ^4 \, ,  \\
\cH_{8,\text{odd}}^{\aLEFTnotPQ}  \,=\,& 2a^4 X^2 + 2a^2 \cD^2 X^2 + a^2 X^3 + \frac{1}{2}\left(N_d+5N_d^2+2N_u^2\right)a^5 \psi ^2 \numberthis\\
&+ \frac{1}{2}\left(N_d+5N_d^2+2N_u^2\right)a^3 \cD^2 \psi ^2 + \left(-\frac{5N_d}{2}+\frac{5N_d^2}{2}+\left(-1+N_u\right)N_u\right)a^2 \cD^3 \psi ^2 \\
&+ \frac{1}{2}\left(-N_d+7N_d^2+4N_u^2\right)a^3 X \psi ^2 + \left(-\frac{N_d}{2}+\frac{21N_d^2}{2}+6N_u^2\right)a \cD^2 X \psi ^2\\
&+ \left(7N_d^2+4N_u^2\right)a^2 \cD X \psi ^2 + 2\left(N_d+8N_d^2+5N_u^2\right)a X^2 \psi ^2\\
&+ \frac{1}{8}\left(125N_d^4+6N_d^3\left(1+20N_u\right)+N_d\left(-6-56N_u+4N_u^2\right)+4N_u^2\left(-1+5N_u^2\right)\right.\\
&\left.+N_d^2\left(-45-16N_u+180N_u^2\right)\right)a^2 \psi ^4 + \left(\frac{75N_d^4}{2}-N_dN_u^2+6N_u^4+2N_d^2N_u\left(2+27N_u\right)\right.\\
&\left.+N_d^3\left(-\frac{3}{2}+36N_u\right)\right)a \cD \psi ^4 \, . \\
\end{align*}

The CP-violating Hilbert series of $\aLEFTnotPQ$ with $N_u=2$, $N_d=N_e=N_\nu=3$ is given by
\begin{align*}
\cH_{5,\text{CPV}}^{\aLEFTnotPQ} \,=\, & a^5 + 2a X^2 + 14a^2 \psi ^2 \, , \\
\cH_{6,\text{CPV}}^{\aLEFTnotPQ} \,=\, & 2a^2 X^2 + 14a^3 \psi ^2 + 16a X \psi ^2 \, , \numberthis\\
\cH_{7,\text{CPV}}^{\aLEFTnotPQ} \,=\, & a^7 + 2a^3 X^2 + a X^3 + 14a^4 \psi ^2 + 14a^2 \cD^2 \psi ^2 + 16a^2 X \psi ^2 + 35a \cD X \psi ^2 + 551a \psi ^4 \, ,  \\
\cH_{8,\text{CPV}}^{\aLEFTnotPQ}  \,=\,& 2a^4 X^2 + 2a^2 \cD^2 X^2 + a^2 X^3 + 14a^5 \psi ^2 + 14a^3 \cD^2 \psi ^2 + 3a^2 \cD^3 \psi ^2 + 16a^3 X \psi ^2\\
&+ 35a^2 \cD X \psi ^2 + 51a \cD^2 X \psi ^2 + 86a X^2 \psi ^2 + 316a^2 \psi ^4 + 927a \cD \psi ^4 \, . \\
\end{align*}

\section{Details on the basis change from the derivative to the Yukawa basis}
\label{app:ShiftFieldRedef}

The discussion about the shift symmetry in the presence of the EOM redundancy at dimension-5 usually does not take the effect of the field redefinition on higher order operators into consideration. Furthermore, the effect on ALP-independent effective operators built from SM fields are also ignored. In this section, we will study the effect of the field redefinition on those operators.

\subsection{ALP-dependent operators}
We will first ignore the SMEFT operators and start with the full derivatively coupled Lagrangian up to dimension-7, i.e., all operators in Tabs.~\ref{tab:aSMEFT_SS_dim5}-\ref{tab:aSMEFT_SS_dim7}. To keep the discussion concise, we only show the calculations for one higher order operator, while the calculations for the other operators follow in a straightforward way. Furthermore, we ignore the bosonic operators here, as they are irrelevant to the discussion. The first derivatively coupled fermionic operators beyond the leading order appears at dimension-7. Eventually, we consider the following Lagrangian
\begin{equation}
\begin{split}
    \cL = & \sum_{\psi \in \text{SM}} \bar{\psi} i \slashed{D} \psi - \( \bar{Q} Y_u \tilde{H} u + \bar{Q} Y_d H d + \bar{L} Y_e H e + \hc \) + \frac{\partial_{\mu} a}{f} \sum_{\psi \in \text{SM}} \bar{\psi} \gamma^{\mu} C_{\psi} \psi \\
    & + \frac{\partial^{\mu} a}{f^3} \(\bar{L} \gamma^{\nu} C_{\partial aLB}  L\) B_{\mu\nu} + \dots
\end{split}
\end{equation}
where the dots collect all the other terms in the derivatively coupled EFT that follow the same discussion. Redefining the fermion fields by $\psi \to \exp\( i C_{\psi} \frac{a}{f} \) \psi$, trades the derivatively coupled operators at dimension-5 for the ALP-Yukawa couplings but also generates more operators at higher dimensions. We find
\begin{align}
    & \cL \to \sum_{\psi \in \text{SM}} \bar{\psi} i \slashed{D} \psi - \cancel{ \frac{\partial_{\mu} a}{f} \sum_{\psi \in \text{SM}} \bar{\psi} \gamma^{\mu} C_{\psi} \psi } - \( \bar{Q} e^{-iC_Q \frac{a}{f}} Y_u e^{iC_u \frac{a}{f}} \tilde{H} u + \bar{Q} e^{-iC_Q \frac{a}{f}} Y_d e^{iC_d \frac{a}{f}} H d \right. \\
    & \left. + \bar{L} e^{-iC_L \frac{a}{f}} Y_e e^{iC_e \frac{a}{f}}  H e + \hc \) + \cancel{ \frac{\partial_{\mu} a}{f} \sum_{\psi \in \text{SM}} \bar{\psi} \gamma^{\mu} C_{\psi} \psi } + \frac{\partial^{\mu} a}{f^3} \(\bar{L} e^{-iC_L \frac{a}{f}} \gamma^{\nu} C_{\partial aLB} e^{iC_L \frac{a}{f}} L\) B_{\mu\nu} + \dots\nonumber
\end{align}
Expanding these exponentials in the SM Yukawa couplings to leading order yields the usual relations at dimension-5. However, we will also study how they alter the dimension-7 operators. 

Focusing only on the leptonic terms, we have after expanding the exponentials
\begin{align}
    \cL \to & \sum_{\psi \in \text{SM}} \bar{\psi} i \slashed{D} \psi - \bar{L} \[  Y_e + \frac{a}{f} i \( Y_e C_e - C_L Y_e \) + \frac{a^2}{f^2} \( C_L Y_e C_e - \frac{1}{2} \( C_L^2 Y_e  + Y_e C_e^2 \) \) + \dots \] H e \nonumber\\
    & + \frac{\partial^{\mu} a}{f^3} \(\bar{L} \gamma^{\nu} \[ C_{\partial aLB} + \frac{a}{f} i \( C_{\partial aLB} C_L - C_L C_{\partial aLB} \) + \frac{a^2}{f^2} \( C_L C_{\partial aLB} C_L \vphantom{\frac{1}{2}}\right.\right.\right. \\
    & \left.\left.\left. \qquad\qquad\qquad -\frac{1}{2} \( C_L^2 C_{\partial aLB} + C_{\partial aLB} C_L^2 \) \)  + \dots  \] L\) B_{\mu\nu} + \dots \, . \nonumber
\end{align}
Notice that expanding the exponential introduces more shift-breaking interactions beyond what is usually shown in the literature. We have constructed these terms previously with the help of the PQ-breaking isolation condition in App.~\ref{app:aSMEFTnonSSOpBasis}. As for the dimension-5 Yukawa couplings, the Wilson coefficients of those shift-breaking operators have to fulfill relations dictated by the exponentiated form of the ALP interactions. We find for the operators shown in the Lagrangian
\begin{equation}
\begin{split}
    C_{ae} & = i \( C_L Y_e - Y_e C_e \) \, , \\
    C_{a^2e} & = \( \frac{1}{2} \( C_L^2 Y_e  + Y_e C_e^2 \) - C_L Y_e C_e  \) \, , \\
    C_{a\partial aLB} & = i \( C_{\partial aLB} C_L - C_L C_{\partial aLB} \) \, , \\
    C_{a^2\partial aLB} & = \( C_L C_{\partial aLB} C_L -\frac{1}{2} \( C_L^2 C_{\partial aLB} + C_{\partial aLB} C_L^2 \) \) \, ,
\end{split}
\end{equation}
where we have used the notation introduced in App.~\ref{app:aSMEFTnonSSOpBasis} for the shift-breaking operators. 

Note that the parameter counting in the EFT before and after the field redefinition is still consistent after including the higher order operators as well. The relations for the dimension-5 ALP-Yukawa operators remove exactly the difference in physical parameters between the ALP-Yukawa and the derivatively coupled basis as was shown in Ref.~\cite{Bonnefoy:2022rik}. At higher mass dimension, the relations fully saturate the freedom in the Wilson coefficients of the shift-breaking operators and no new parameters are added, as expected. This will always happen, since the exponential generates interactions proportional to the Wilson coefficient of the shift-symmetric operator that is affected by the chiral rotation. The main difference to the dimension-5 case is that, there, the derivatively coupled operator and the operator generated by the field redefinition are connected by an EOM redundancy. Then, the field content of the operator is changed by the field redefinition and a different amount of degrees of freedom is captured if the relations are disregarded.\footnote{To be precise, the same is true for the dimension-5 couplings if one counts the physical parameters in the SM Yukawa couplings as well. Then, one can also explain the CP parities of the flavor invariants in Ref.~\cite{Bonnefoy:2022rik}; there are 3 CP-odd relations in the lepton sector and 9 CP-odd relations and 1 CP-even relations in the quark sector corresponding to the numbers of physical parameters in the SM Yukawas. The CP parities are flipped here with respect to the SM parameters because the ALP is assigned to be odd under parity. This however does not imply that imposing CP on the full theory (i.e. the vanishing of the CKM phase implies a vanishing of the single CP-even relation in the ALP EFT) makes all order parameters of shift-symmetry vanish. This is because the CP-even parameter in the Wilson coefficient of the dimension-5 operator is independent and the rephasing properties of the CKM matrix are unchanged by setting the CKM phase to zero. Hence, the single CP-even invariant remains untouched.}

In order to have a shift-symmetric ALP EFT in the Yukawa basis beyond dimension-5, one needs to include these additional interactions with their constrained Wilson coefficients. Otherwise one will run into shift-breaking results while doing computations. Only when all the additional diagrams from the operators generated by the field redefinitions are considered, one will recover a shift-symmetric result.

In particular, if one allows for shift-breaking effects and wants to understand the shift-symmetric limit it is practical to work in the Yukawa basis instead of the basis with the derivatively coupled fermionic operators at dimension-5. Then, to take the shift-symmetric limit, on has to impose the relations discussed in this section on the generic Wilson coefficients of the shift-breaking operators in order to get consistent results.

\subsection{SMEFT operators}
The ALP-dependent chiral transformation also affects SMEFT operators, introducing a new source of ALP-dependent operators. We will give some examples here, working with the following Lagrangian\footnote{For simplicity we have taken the operators with and without an ALP to be suppressed by the same UV scale $f$. Depending on the structure of the UV theory and the details of PQ-breaking the operators can also come with a suppression of different scales corresponding to different UV sectors.}
\begin{align*}
     \cL_{\text{SMEFT}} = & \frac{1}{f^2} |H|^2 \bar{L} C_{eH} H e + \frac{1}{f^2} C_{lequ,ijkl}^{(1)} \(\bar{L}_i e_j \) \epsilon \(\bar{Q}_k u_l \) \\
     \longrightarrow & \frac{|H|^2}{f^2} \bar{L} \[ C_{eH} + \frac{a}{f} i \( C_{eH} C_e - C_L C_{eH} \) + \frac{a^2}{f^2} \( C_L C_{eH} C_e - \frac{1}{2} \( C_L^2 C_{eH}  + C_{eH} C_e^2 \) \) \] H e \\
     & +  \frac{1}{f^2} \[ C_{lequ,ijkl}^{(1)} + \frac{a}{f} i\( C_{lequ,ij^{\prime}kl}^{(1)} C_{e,j^{\prime}j} + C_{lequ,ijkl^{\prime}}^{(1)} C_{u,l^{\prime}l} - C_{lequ,i^{\prime}jkl}^{(1)} C_{L,ii^{\prime}} - C_{lequ,ijk^{\prime}l}^{(1)} C_{Q,kk^{\prime}} \) \right. \\
     & \left. + \frac{a^2}{f^2} \( C_{lequ,i^{\prime}j^{\prime}kl}^{(1)}  C_{L,ii^{\prime}} C_{e,j^{\prime}j} - C_{lequ,i^{\prime}jk^{\prime}l}^{(1)}  C_{L,ii^{\prime}} C_{Q,kk^{\prime}} + C_{lequ,i^{\prime}jkl^{\prime}}^{(1)}  C_{L,ii^{\prime}} C_{u,l^{\prime}l} \right.\right. \\
     & \left.\left. + C_{lequ,ij^{\prime}k^{\prime}l}^{(1)} C_{e,j^{\prime}j} C_{Q,kk^{\prime}} - C_{lequ,ij^{\prime}kl^{\prime}}^{(1)} C_{e,j^{\prime}j} C_{u,l^{\prime}l} + C_{lequ,ijk^{\prime}l^{\prime}}^{(1)} C_{Q,kk^{\prime}} C_{u,l^{\prime}l} \right.\right. \\
     & \left.\left. - \frac{1}{2} \( C_{lequ,i^{\prime}jkl}^{(1)}  (C_{L}^2)_{ii^{\prime}} + C_{lequ,ij^{\prime}kl}^{(1)} (C_{e}^2)_{j^{\prime}j}+ C_{lequ,ijk^{\prime}l}^{(1)}  (C_{Q}^2)_{kk^{\prime}} + C_{lequ,ijkl^{\prime}}^{(1)}  (C_{u}^2)_{l^{\prime}l} \) \) \vphantom{\frac{1}{f^2}} \] \times \\
     & \times \(\bar{L}_i e_j \) \epsilon \(\bar{Q}_k u_l \) \, .
\end{align*}
Comparing to the generic shift-breaking Lagrangian, similar relations are found as before. They become more and more complicated as more fermions appear in the operators. One can simply read of the relations from the Lagrangian and we will not give them explicitly again.

Instead of starting with the derivatively coupled EFT, one can also start at the opposite end and consider the $\aSMEFTnotPQ$ which is constructed without imposing a shift symmetry. Then, shifting the ALP $a \to a+c$ and demanding that the shift $c$ vanishes in the EFT after performing field redefinitions, while staying in the same operator basis, leads to the same relations as we have just derived. We have checked this explicitly with our operator basis up to dimension-8. To this end, a field redefinition should be used that allows to remove the shift $c$ in the Lagrangian while keeping the kinetic terms of the fermions invariant and at the same time not generating new operators outside of the operator basis we start with. The only such transformation is given by redefining the fermion fields with powers of the shift $\frac{c}{f}$ as $\psi \to \psi + i \sum_{k=1}^{\infty} c_{\psi}^{(k)} \(\frac{c}{f}\)^k \psi$, where the $c_{\psi}^{(k)}$ are generic hermitian matrices. We keep the terms in this expansion up to the order that is relevant for the EFT expansion in each step of the discussion. To get consistent relations for all terms that are proportional to the shift for operators with more than one power of an ALP, the following choice for the coefficients in the field redefinition $c_{\psi}^{(n)} = i^{n-1} C_{\psi}/(n!)$ has to be made, i.e. $\psi \to \exp\(i C_{\psi} \frac{c}{f}\) \psi$. This is reminiscent of the chiral transformation we have started with in the purely derivatively coupled EFT.

\subsection{List of additional relations in Yukawa basis} \label{app:ShiftRelations}
In this appendix we list all the constrained Wilson coefficients of operators that have to be added in the Yukawa basis up to dimension~8. Here, we only restrict to those operators which already have an ALP field before performing the chiral rotation, i.e., we ignore the contributions from the SMEFT operators due to their length. Those relations can be straightforwardly constructed as shown in the previous section.

At dimension-5, we find the well-known relations
\begin{equation}
    C_{ae} = i \(C_L Y_e - Y_e C_e \), \quad C_{au} = i \( C_Q Y_u - Y_u C_u \), \quad C_{ad} = i \( C_Q Y_d - Y_d C_d \) \, .
\end{equation}
Since at dimension-6 the only existing operator is bosonic, the only relations at this mass-dimension come again from the Yukawa-like operators. They read
\begin{equation}
\begin{split}
    C_{a^2e} = & \( \frac{1}{2} \( C_L^2 Y_e  + Y_e C_e^2 \) - C_L Y_e C_e  \), \quad C_{a^2u} = \( \frac{1}{2} \( C_Q^2 Y_u  + Y_u C_u^2 \) - C_Q Y_u C_u  \), \\
    &\qquad\qquad\qquad C_{a^2d} = \( \frac{1}{2} \( C_Q^2 Y_d  + Y_d C_d^2 \) - C_Q Y_d C_d  \) \, .
\end{split}
\end{equation}
The same is true at dimension-7 
\begin{equation}
\begin{split}
    C_{a^3e} & = \frac{i}{6} \( Y_e C_e^3 - C_L^3 Y_e \) + \frac{i}{2} \( C_L^2 Y_e C_e - C_L Y_e C_e^2 \), \\
    C_{a^3u} & = \frac{i}{6} \( Y_u C_u^3 - C_Q^3 Y_u \) + \frac{i}{2} \( C_Q^2 Y_u C_u - C_Q Y_u C_u^2 \), \\
    C_{a^3d} & = \frac{i}{6} \( Y_d C_d^3 - C_Q^3 Y_d \) + \frac{i}{2} \( C_Q^2 Y_d C_d - C_Q Y_d C_d^2 \)\, .
\end{split}
\end{equation}
Only at dimension-8 there are relations introduced by new fermionic operators at dimension-7 reading as follows
\begin{equation}
\begin{split}
    C_{a^4e} & = \frac{1}{6} \(C_L^3 Y_e C_e + C_L Y_e C_e^3\) - \frac{1}{4} C_L^2 Y_e C_e^2 - \frac{1}{24} \(C_L^4 Y_e + Y_e C_e^4\) \, , \\
    C_{a^4u} & = \frac{1}{6} \(C_Q^3 Y_u C_u + C_Q Y_u C_u^3\) - \frac{1}{4} C_Q^2 Y_u C_u^2 - \frac{1}{24} \(C_Q^4 Y_u + Y_u C_u^4\) \, , \\
    C_{a^4d} & = \frac{1}{6} \(C_Q^3 Y_d C_d + C_Q Y_d C_d^3\) - \frac{1}{4} C_Q^2 Y_d C_d^2 - \frac{1}{24} \(C_Q^4 Y_d + Y_d C_d^4\) \, , \\
    C_{a \partial a e HD}^{(1,2)} & = i \( C_{\partial a eHD}^{(1,2)} C_e - C_L C_{\partial a eHD}^{(1,2)} \) \, , \\ 
    C_{a \partial a u HD}^{(1,2)} & = i \( C_{\partial a uHD}^{(1,2)} C_u - C_Q C_{\partial a uHD}^{(1,2)} \) \, , \\
    C_{a \partial a d HD}^{(1,2)} & = i \( C_{\partial a dHD}^{(1,2)} C_d - C_L C_{\partial a dHD}^{(1,2)} \) \, , \\
    C_{a \partial a \psi H^2} & = i \( C_{\partial a \psi H^2} C_{\psi} - C_{\psi} C_{\partial a \psi H^2} \) \, , \\
    C_{a \partial a \psi V} & = i \( C_{\partial a \psi V} C_{\psi} - C_{\psi} C_{\partial a \psi V} \) \, .
\end{split}
\end{equation}
For the last relation, the same relation holds true for the operators with the dual field strength. Furthermore, some shift-symmetric operator at dimension-8 get shifted as follows
\begin{equation}
\begin{split}
    C_{\partial a^2eH } & \to C_{\partial a^2eH } + i \( C_{\partial a eHD}^{(2)} C_e - C_L C_{\partial a eHD}^{(1)} \) , \\
    C_{\partial a^2uH} & \to C_{\partial a^2uH } + i \( C_{\partial a uHD}^{(2)} C_u - C_Q C_{\partial a uHD}^{(1)} \) , \\
    C_{\partial a^2dH} & \to C_{\partial a^2dH } + i \( C_{\partial a dHD}^{(2)} C_d - C_Q C_{\partial a dHD}^{(1)} \) , \\
\end{split}
\end{equation}
due to the derivative acting on the fermions in the operator corresponding to the last relation in the previous equation.

\clearpage
\bibliographystyle{apsrev4-1_title}
\bibliography{biblio.bib}

\end{document}